\pdfoutput=1

\documentclass[11pt,twoside,a4paper,cmspaper,final,collab]{cms-tdr}
\def\svnVersion{215791}\def\cmsCernNo{2012-373}\def\cmsCernDate{\today}\def\cmsMessage{Published in Physics Letters B as \href{http://dx.doi.org/10.1016/j.physletb.2013.10.033}{\doi{10.1016/j.physletb.2013.10.033.}}}

\begin{document}\cmsNoteHeader{BPH-11-001}

\hyphenation{had-ron-i-za-tion}
\hyphenation{cal-or-i-me-ter}
\hyphenation{de-vices}

\RCS$Revision: 210824 $
\RCS$HeadURL: svn+ssh://svn.cern.ch/reps/tdr2/papers/BPH-11-001/trunk/BPH-11-001.tex $
\RCS$Id: BPH-11-001.tex 210824 2013-10-09 08:50:04Z nuno $

\newcommand{\upsn}  {\ensuremath{\PgU\cmsSymbolFace{(nS)}}\xspace}
\newcommand{\upsm}  {\ensuremath{\PgU\cmsSymbolFace{(mS)}}\xspace}
\newcommand{\sqrts} {\sqrt{s}}
\newcommand{\eff}{\varepsilon}
\newcommand{\acc}{\ensuremath{\mathcal{A}}}
\newcommand{\CASCADE}{{\textsc{cascade}}\xspace}
\newcommand{\ups}{\ensuremath{\PgU}\xspace}

\newcommand{\samplelumi} {\ensuremath{35.8 \pm 1.4}\xspace}
\newcommand{\triggerpathA} {{HLT\_L1DoubleMuOpen\_Tight\;}}
\newcommand{\triggerpathB} {{HLT\_DoubleMu0\;}}
\newcommand{\triggerpathC} {{HLT\_DoubleMu0\_Quarkonium\_v1\;}}
\ifthenelse{\boolean{cms@external}}{\providecommand{\cmsLeft}{top}}{\providecommand{\cmsLeft}{left}}
\ifthenelse{\boolean{cms@external}}{\providecommand{\cmsRight}{bottom}}{\providecommand{\cmsRight}{right}}
\ifthenelse{\boolean{cms@external}}{\providecommand{\suppMaterial}{the supplemental material [URL will be inserted by publisher]}} {\providecommand{\suppMaterial}{Appendix~\ref{app:suppMat}}}

\cmsNoteHeader{BPH-11-001}
\title{\texorpdfstring{Measurement of the \PgUa, \PgUb, and \PgUc\ cross sections  in $\Pp\Pp$ collisions at $\sqrt{s}=7\TeV$ }{Measurement of the Y(1S), Y(2S), and Y(3S) cross sections in pp collisions at sqrt(s) = 7 TeV}}

\date{\today}

\abstract{
   The \PgUa, \PgUb, and \PgUc\ production cross sections are measured using a data sample corresponding to an integrated luminosity of $35.8\pm 1.4\pbinv$ of proton-proton collisions at $\sqrt{s} = 7$\TeV,  collected with the CMS detector at the LHC. The \PgU\ resonances are identified through their decays to dimuons. Integrated over the \ensuremath{\PgU} transverse momentum range $\pt^{\PgU}< 50\GeVc$ and rapidity range $\abs{y^{\PgU}} < 2.4$, and assuming unpolarized \PgU\ production, the products of the \PgU\ production cross sections and dimuon branching fractions are
\begin{align*}
\sigma(\Pp\Pp \to \PgUa X ) \cdot \mathcal{B} (\PgUa \to \Pgmp \Pgmm) &= (8.55 \pm 0.05 {\,}^{+0.56}_{-0.50}\pm 0.34)\unit{nb}, \\
\sigma(\Pp\Pp \to \PgUb X ) \cdot \mathcal{B} (\PgUb \to \Pgmp \Pgmm) &= (2.21 \pm 0.03 {\,}^{+0.16}_{-0.14}\pm 0.09)\unit{nb}, \\
\sigma(\Pp\Pp \to \PgUc X ) \cdot \mathcal{B} (\PgUc \to \Pgmp \Pgmm) &= (1.11 \pm 0.02 {\,}^{+0.10}_{-0.08}\pm 0.04)\unit{nb},
\end{align*}
where the first  uncertainty is statistical, the second is systematic, and the third is from the uncertainty in the integrated luminosity. The differential cross sections in bins of transverse momentum and rapidity, and the cross section ratios are presented.  Cross section measurements performed within a restricted muon kinematic range and not corrected for acceptance are also provided. These latter measurements are independent of \PgU\ polarization assumptions. The results are compared to theoretical predictions and previous measurements.}

\hypersetup{%
pdfauthor={CMS Collaboration},%
pdftitle={Measurement of the Y(1S), Y(2S), and Y(3S) cross sections in pp collisions at sqrt(s) = 7 TeV},%
pdfsubject={CMS},%
pdfkeywords={CMS, physics, quarkonia, upsilon, dimuons}}

\maketitle 

\section{Introduction}
\label{sec:introduction}

No existing theoretical approach successfully reproduces both the differential cross section and the polarization measurements of the $\PJGy$ or $\PgU$ states~\cite{yellow} in hadron collisions.
Studying quarkonium hadroproduction at high center-of-mass energies and over a wide rapidity and transverse momentum range will facilitate significant improvements in our understanding of the processes involved.

Measurements of \ups production have been performed by several experiments~\cite{yellow,bib-cdfups,bib-d0ups,bib-lhcb-paper,yplb_ATLAS}.
The first measurement at $\sqrt{s} = 7 \TeV$ at the Large Hadron Collider (LHC) was reported by the Compact Muon Solenoid (CMS) Collaboration~\cite{yprd10}, using a data sample corresponding to an integrated luminosity of $3\pbinv$.
This paper constitutes an extension of that first cross section measurement, using a larger, independent sample, corresponding to an integrated luminosity of $35.8 \pm 1.4\pbinv$ collected in 2010.

Two different approaches to the measurement of the \upsn\ production cross sections, where n\,=\,1--3, are pursued in this paper. In each approach, the $\PgU$
is reconstructed in the decay $\PgU \to \Pgmp \Pgmm$.
In the first approach, a cross section measurement corrected for detector acceptance and efficiencies is presented, as in Ref.~\cite{yprd10}. This cross section measurement depends on the spin alignment of the $\PgU$. No net polarization is assumed for the main results. To show the sensitivity of the results to the polarization and to allow for interpolation, we provide measurements for other polarization assumptions. Recently, the CMS Collaboration has measured the polarizations of the \upsn in \Pp\Pp\ collisions at $\sqrt{s} = 7\TeV$, which are found to be small~\cite{cms_ypol}. Cross section measurements are also provided in the $\PgU$ transverse momentum ($\pt^{\PgU}$) and rapidity ($y^{\PgU}$) ranges matching those of the polarization measurement, and these polarization results are used to estimate the associated systematic uncertainty.
The motivation for the second approach, also used by the ATLAS Collaboration~\cite{yplb_ATLAS}, is to eliminate the dependence of the measured cross sections on the spin alignment of the $\PgU$.
In this second approach, a fiducial cross section measurement, corrected for detector efficiencies but not for acceptance, is presented. This cross section is defined within a muon kinematic range.

The paper is organized as follows. Section~\ref{detector} contains a short description of the CMS detector. Section~\ref{selection} presents the data collection, the trigger and offline event selections, and the reconstruction of the \ups resonances. Section~\ref{crosssection} describes the measurement technique. The detector acceptance and efficiencies to reconstruct \ups resonances that decay to two muons are discussed in Sections~\ref{sec:acceptance} and~\ref{efficiency}. The evaluation of systematic uncertainties in the measurements is described in Section~\ref{systematics}. In Sections~\ref{crosssection_noAcc_results} and~\ref{crosssection_results}, the \upsn fiducial and acceptance-corrected cross section results and comparisons to other experiments and to theoretical predictions are presented.

\section{CMS detector}
\label{detector}

The central feature of the CMS apparatus is a superconducting solenoid, of 6\unit{m} inner diameter, producing a magnetic field of 3.8\unit{T}. Within the superconducting solenoid volume are a silicon pixel and strip tracker, a lead tungstate crystal electromagnetic calorimeter, and a brass/scintillator hadron calorimeter. Muons are detected by three types of gas-ionization detectors embedded in the magnet steel return yoke surrounding the solenoid: drift tubes, cathode strip chambers, and resistive-plate chambers. The muon measurement covers the pseudorapidity range $\abs{\eta^{\mu}}<2.4$, where $\eta = -\ln[\tan(\theta/2)]$ and the polar angle $\theta$ is measured from the axis pointing along the counterclockwise-beam direction. The muon transverse momentum measurement, $\pt^{\mu}$, based on information from the silicon tracker alone, has a resolution of about 1\% for a typical muon in this analysis. The two-level CMS trigger system selects events of interest for permanent storage. The first trigger level, composed of custom hardware processors, uses information from the calorimeter and muon detectors to select events in less than 3.2\mus. The high-level trigger software algorithms, executed on a farm of commercial processors, further reduce the event rate using information from all detector subsystems. A detailed description of the CMS detector can be found in Ref.~\cite{JINST}.

\section{Data selection and event reconstruction}
\label{selection}

The data sample was collected in 2010, in low instantaneous luminosity conditions, allowing a less restrictive selection at the trigger level in comparison to subsequent data taking periods.
Data are included in the analysis for all periods where the silicon tracker, the muon detectors, and the trigger were performing well and the luminosity information was available.
In the first data-taking period, the trigger requires the detection of two muons without an explicit $\pt^{\mu}$ requirement.  The minimum distance
between each reconstructed muon trajectory and the average proton-proton interaction point in the transverse plane must be less than 2\cm.  In the second data-taking period, characterized by higher LHC instantaneous luminosities, additional requirements are imposed at trigger level: the two muons must have opposite charge and an invariant mass in the mass range $1.5 < M_{\mu\mu} < 14.5$\GeVcc.  All three muon systems take part in the trigger decision. In the first (second) data-taking period the trigger selected about 2 (5) million events.

Simulation is employed to design the offline selection, assess the detector acceptance, and study systematic effects. The \upsn events are simulated using \PYTHIA 6.412~\cite{bib-PYTHIA}, which generates events based on the leading-order color-singlet and color-octet mechanisms, with nonrelativistic quantum chromodynamics (QCD) matrix elements, tuned by comparing calculations with CDF data~\cite{bib-cdfPythia}, and applying the normalization and wave functions recommended in Ref.~\cite{bib-PYTHIA-tuning}. The underlying-event simulation uses the CTEQ6L1 parton distribution functions~\cite{bib-cteq6l1}.
Since \PYTHIA does not provide a simulation of \PgUb\ and \PgUc, the predictions for these states are obtained by replacing the \PgUa\ mass in the simulation with the \PgUb\ and \PgUc\ masses, respectively. Contributions from the decays of higher-mass bottomonium states (feed-down) are included in the simulation. For simulating the Y(2S) feed-down component, the masses of the 2P states replace the corresponding 1P states. For the \PgUc\, the feed-down is assumed to be small and is not simulated.
Final-state radiation (FSR) is implemented using \PHOTOS~\cite{bib-photos1,bib-photos2}. The response of the CMS detector is simulated with a \GEANTfour-based~\cite{bib-GEANT4} Monte Carlo (MC) simulation program. Simulated events are processed with the same reconstruction and trigger algorithms used for data.

The offline selection starts from $\PgU$ candidates reconstructed from pairs of oppositely charged muons with invariant mass between 7 and 14\GeVcc. The muons are required to have one or more reconstructed track segments in the muon systems that are well matched to the extrapolated position of a track reconstructed in the silicon tracker.
Quality criteria are applied to the tracks to reject muons from kaon and pion decays. Tracks are required to have at least 11 hits in the silicon tracker, at least one of which must be in the pixel detector, and a track-fit $\chi^2$ per degree of freedom smaller than 5. In addition, tracks are required to extrapolate back to a cylindrical volume of radius $2\mm$ and length $25\cm$, centered on the \Pp\Pp\ interaction region and parallel to the beam line. After offline confirmation of the trigger selection, muons are required to satisfy a kinematic threshold that depends on pseudorapidity
\begin{linenomath}
\begin{equation}
\begin{aligned}
 \label{eq:selection}
  &\pt^\mu > 3.75\GeVc &\qquad\text{if}\quad &&\abs{\eta^\mu}<0.8,\\
  &\pt^\mu > 3.5\GeVc  &\qquad\text{if}\quad &0.8<&\abs{\eta^\mu}<1.6, \\
  &\pt^\mu > 3.0\GeVc  &\qquad\text{if}\quad &1.6<&\abs{\eta^\mu}<2.4.
 \end{aligned}
\end{equation}
\end{linenomath}
These kinematic acceptance criteria are chosen to ensure that the
trigger and muon reconstruction efficiencies are high and not rapidly
changing within the phase space of the analysis.
The longitudinal separation between the two muons along the beam axis is required to be less than 2\cm.
The two muon helices are fit with a common vertex constraint, and
events are retained if the fit $\chi^2$ probability is larger than 0.1\%.
If multiple dimuon candidates are found in the same event, the candidate with the smallest vertex-fit $\chi^{2}$ probability is retained;
the fraction of \PgU\ candidates rejected by this requirement is about 0.6\%.

\section{Measurement of the inclusive differential cross section}
\label{crosssection}

The product of the \upsn differential cross section, $\sigma$, and the dimuon branching fraction, $\mathcal{B}$, is determined from the signal yield $N^{\text{cor}}_{\upsn}$, corrected by the acceptance $\mathcal{A}$ and the efficiency $\epsilon$, using
\ifthenelse{\boolean{cms@external}}
{
\begin{multline}
\frac{\rd\sigma\left(\Pp\Pp \to \upsn X\right)}{\rd\pt^{\PgU}\,\rd y^{\PgU}} \cdot \mathcal{B} \left(\upsn\to\Pgmp\Pgmm\right) =\\
\frac{ N^{\text{cor}}_{\upsn} (\pt^{\PgU},y^{\PgU}; {\acc},\epsilon)}{\lumi \ \cdot \Delta \pt^{\PgU} \cdot \Delta y^{\PgU} },
\label{eqn:xsection}
\end{multline}
}{
\begin{linenomath}
\begin{equation}
\frac{\rd\sigma\left(\Pp\Pp \to \upsn X\right)}{\rd\pt^{\PgU}\,\rd y^{\PgU}} \cdot \mathcal{B} \left(\upsn\to\Pgmp\Pgmm\right) =
\frac{ N^{\text{cor}}_{\upsn} (\pt^{\PgU},y^{\PgU}; {\acc},\epsilon)}{\lumi \ \cdot \Delta \pt^{\PgU} \cdot \Delta y^{\PgU} },
\label{eqn:xsection}
\end{equation}
\end{linenomath}
}
where $\mathcal{L}$ is the integrated luminosity of the data set, and $\Delta \pt^{\PgU}$ and $\Delta y^{\PgU}$ are the bin widths of the $\PgU$ transverse momentum and rapidity, respectively. The rapidity is defined as
$y=\frac{1}{2}\ln(\frac{E+p_z c}{E-p_z c})$, where $E$ is the energy and $p_z$ is the momentum component parallel to the beam axis of the muon pair.

\begin{figure}[!ht]
  \centering
  \includegraphics[width=0.49\textwidth]{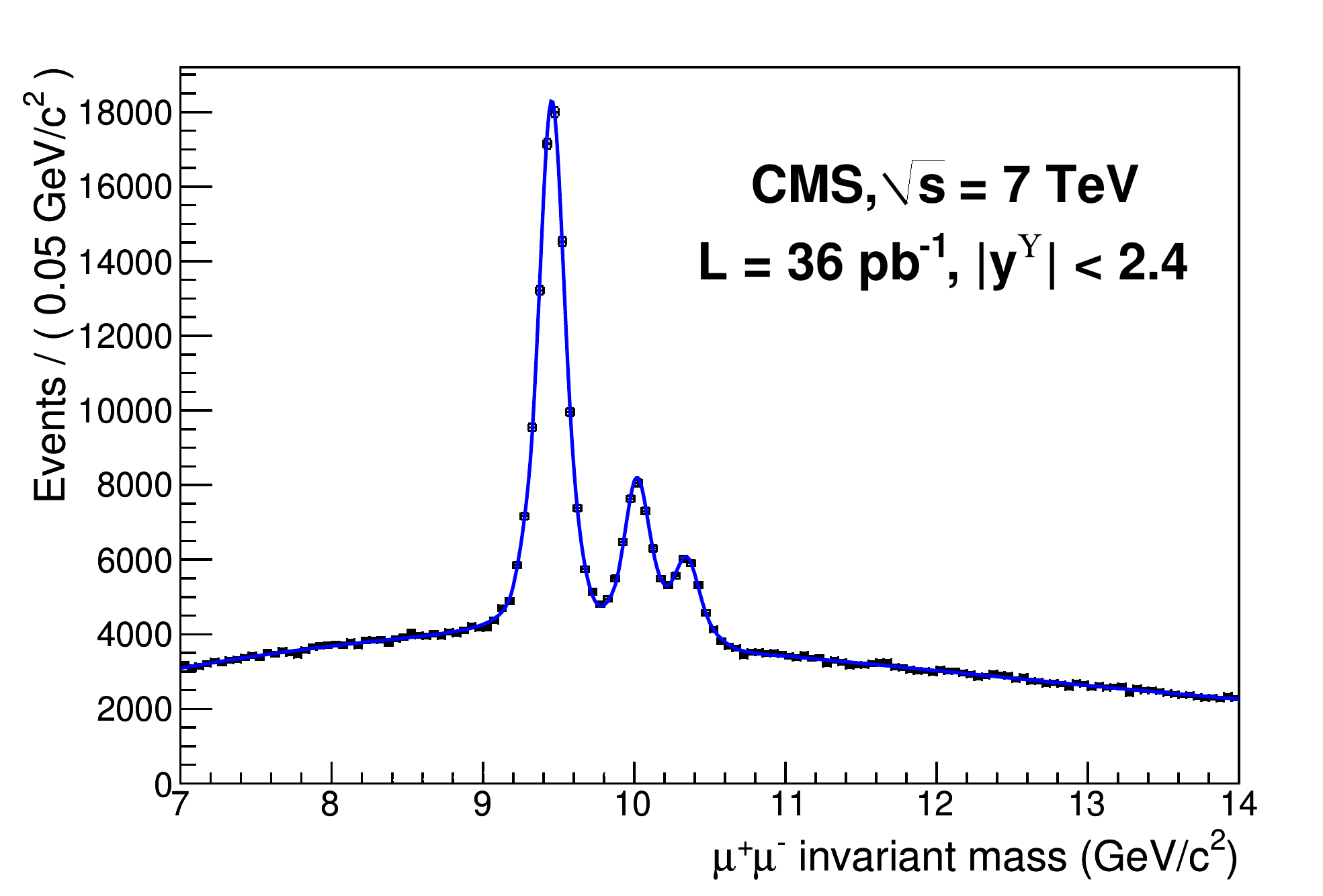}\label{fig:massfit_all}
  \includegraphics[width=0.49\textwidth]{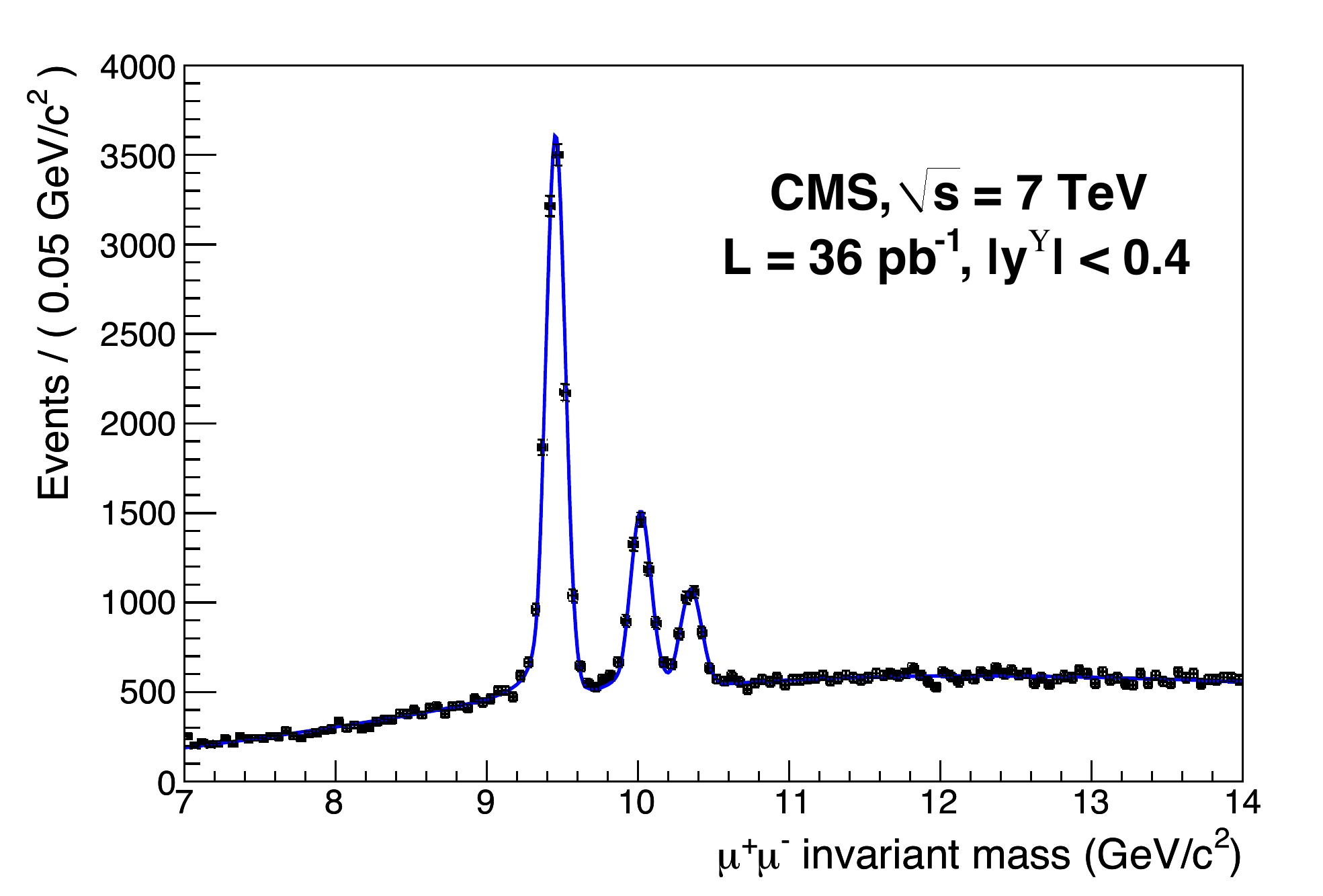}\label{fig:massfit_raw_0_04_y}
  \caption{The dimuon invariant-mass distribution in the vicinity of
    the \upsn resonances for $\abs{y^{\PgU}}<2.4$ (\cmsLeft) and for the subset of events where the rapidity of the \upsn satisfies $\abs{y^{\PgU}}<0.4$ (\cmsRight). The solid lines represent the results of the fits to the signal-plus-background functions described in the text.}
  \label{fig:massFit-raw}
\end{figure}

The \upsn yields are extracted via an extended unbinned maximum-likelihood fit to the dimuon invariant-mass spectrum. The measured mass line shape of each \ups state is parametrized by a ``Crystal Ball" (CB)~\cite{bib-crystalball} function, which consists of a Gaussian core portion and a power-law low-side tail to allow for FSR, with the low-mass tail parameters fixed from MC simulation~\cite{yprd10}. The three \upsn states are fitted simultaneously since the three resonances overlap in the measured dimuon mass range. The resolution, given by the standard deviation of the Gaussian component of the CB, is a free parameter in the fit, but is constrained to scale with the ratios of the resonance masses. However, the mass resolution varies with \ups rapidity. Consequently, a single resolution term in the Gaussian component of the CB is not sufficient to describe the data. For this reason, in the $\pt^{\PgU}$ intervals with sufficient statistical precision, the sum of two CBs with the same mean and FSR tail parameters, but different resolutions, is used for each \ups state. The fitted resolution is consistent with expectation from MC at the few percent level. The \upsn mass ratios are fixed to their world-average values~\cite{bib-pdg}. The background in the 7--14\GeVcc mass-fit range is nonpeaking and in some kinematic bins has a turn-on caused by the trigger and offline requirements. In general, the product of an error function and an exponential is chosen to describe the background~\cite{PbPb}, except when, for bins with poor statistical precision, a single exponential function is used. The dimuon invariant-mass spectra in the \upsn region, before accounting for acceptance and efficiencies, are shown in Fig.~\ref{fig:massFit-raw} and in \suppMaterial.

Following Ref.~\cite{yprd10}, given the significant $\pt^{\mu}$ and $\eta^{\mu}$ dependencies of the acceptances and efficiencies of the muons from \upsn decays, we correct for them on a candidate-by-candidate basis before performing the mass fit to obtain $N^{\text{cor}}_{\upsn}$ used in Eq.~(\ref{eqn:xsection}).
The fiducial differential cross section is determined from the efficiency-corrected signal yield within the kinematic region defined in Eq.~(\ref{eq:selection}).

\section{Acceptance}
\label{sec:acceptance}

The $\PgU \to \Pgmp \Pgmm$ acceptance of the CMS detector is the
product of two terms. The first is, for a given $\pt^{\PgU}$ and $y^{\PgU}$, the fraction of dimuon decays in which both muons are within
the phase space specified in Eq.~(\ref{eq:selection}).  The second is the probability that when
there are only two muons in the event both can be reconstructed in the tracker without requiring the quality criteria.
Both components are evaluated by simulation and parametrized as a function of $\pt^{\PgU}$ and $y^{\PgU}$.
The second component is close to unity, as verified in simulation and data.

Following Ref.~\cite{yprd10}, the acceptance is defined by the ratio
\begin{linenomath}
\begin{equation}
  {\acc}\left(p_T^\PgU,y^\PgU \right) =
  \frac{N^\text{reco}\left( \left. \pt^\PgU,y^\PgU \right| \text{Si tracks satisfying Eq.~(\ref{eq:selection})} \right)}
  {N^\text{gen}\left({p}_\mathrm{T}^\PgU,y^\PgU \right)},
\label{eq:acceptance}
\end{equation}
\end{linenomath}
and is computed in small bins in $(\pt^\PgU, y^\PgU)$. 
The parameter $N^\text{gen}$ is the number of $\PgU$ particles generated within a given $({p}_\mathrm{ T}^\PgU,y^\PgU)$ bin, while $N^\text{reco}$ is the number of $\PgU$ particles with reconstructed $(\pt^\PgU,y^\PgU)$ values within that bin, and having the silicon tracks satisfying Eq.~(\ref{eq:selection}). 
The $(\pt^\PgU,y^\PgU)$ values represent the generated and reconstructed values, respectively in the denominator and the numerator, thus accounting also for the effect of detector resolution in the definition of $\acc$. 
In addition the numerator requires the two tracks to be reconstructed with opposite charges and have an invariant mass within the \ups mass-fit range of 7--14\GeVcc.

The acceptance is evaluated with a signal MC simulation sample in which the \ups decay to two muons is generated with the \EVTGEN~\cite{bib-evtgen}
package, including FSR. There are no particles in the event besides the \PgU, its daughter muons, and the FSR photons. The \ups mesons are generated uniformly in $\pt^{\PgU}$ and $y^{\PgU}$. This sample is then simulated and reconstructed with
the CMS detector simulation software to assess the effects of multiple scattering and finite resolution of the detector.  An acceptance map with the assumption of zero \PgU\ polarization can be found in Ref.~\cite{yprd10}.  Systematic uncertainties
arising from the dependence of the cross section measurement on the MC simulation description of the \pt spectrum and resolution are evaluated in Section~\ref{systematics}.
The acceptance is calculated as a two-dimensional grid in $\pt^{\PgU}$ and $\abs{y^{\PgU}}$ using bin sizes of 0.1 in rapidity and 0.5\GeVc in $\pt^\PgU$ for $0 < \pt^\PgU < 2\GeVc$ and 1\GeVc for $2 < \pt^\PgU < 50\GeVc$. The corresponding correction is then performed on a candidate-by-candidate basis. The acceptance depends on the resonance mass; the \PgUc\ gives rise to higher-momenta muons which results in a roughly 10\% larger acceptance for the \PgUc\ than for the \PgUa. Consequently, the corrected yield for each of the \upsn resonances is obtained from a fit in which the corresponding \upsn acceptance is employed.
The acceptance decreases with rapidity, and there are no accepted events beyond $\abs{y^\PgU}=2.4$.  The acceptance has a minimum near $\pt^{\PgU}=5\GeVc$, as a result of the softer
muon failing the $\pt^{\Pgm}$ cut.
The polarization of the \ups strongly influences the muon angular distributions and could be a function of $\pt^{\PgU}$.
In order to show the sensitivity of the result to the \upsn polarization and to allow for interpolation, we provide cross section measurements
for unpolarized (default) and 6 polarization scenarios in which the polar anisotropy parameter $\lambda_\theta$~\cite{cms_ypol} is changed from fully longitudinal to fully transverse polarization, corresponding to $\lambda_\theta = -1,$ $-0.5$, $-0.25$, 0.25, 0.5, 1, in both the center-of-mass helicity and Collins--Soper~\cite{bib:CS} reference frames.
Cross section measurements for the $\pt^{\PgU}$ and $y^{\PgU}$ ranges used in Ref.~\cite{cms_ypol} are also provided in Fig.~\ref{fig:xsec_overlay_cmspol}. In that case, the polarization results from Ref.~\cite{cms_ypol} are used to estimate the corresponding systematic uncertainty.

\section{Efficiency}
\label{efficiency}

The total muon efficiency is factorized into the three conditional terms,
\ifthenelse{\boolean{cms@external}}{
\begin{equation}
\begin{aligned}
  \eff = & \eff(\text{trig} | \text{id}) \times \eff(\text{id} | \text{track}) \times \eff(\text{track} | \text{accepted})\\ \equiv & \eff_\text{trig} \times \eff_\text{id} \times \eff_\text{track}.
\end{aligned}
\end{equation}
}{
\begin{linenomath}
\begin{equation}
  \eff = \eff(\text{trig} | \text{id}) \times \eff(\text{id} | \text{track}) \times \eff(\text{track} | \text{accepted}) \equiv \eff_\text{trig} \times \eff_\text{id} \times \eff_\text{track}.
\end{equation}
\end{linenomath}
}
The tracking efficiency, $\eff_\text{track}$, combines the efficiency that the accepted track of a muon from a \upsn decay is reconstructed in the presence of additional particles in the silicon tracker, as determined with a track-embedding technique~\cite{bib-trackingefficiency}, and the efficiency for the track to satisfy the track-quality criteria.
The efficiency of the track-quality criteria~\cite{bib-trackingefficiency} is nearly uniform in \pt and $\eta$ and has an average value of $(98.66\pm0.05)\%$, as measured in Ref.~\cite{yprd10}, with negligible dependence on instantaneous luminosity.
The muon identification efficiency, $\eff_\text{id}$, is the probability that the silicon track caused by a muon is correctly identified as a muon. The efficiency that an identified muon satisfies the trigger is denoted by $\eff_\text{trig}$. The track quality, muon trigger, and muon identification efficiencies are determined using the tag-and-probe (T\&P) technique. The T\&P implementation follows Ref.~\cite{yprd10}, and utilizes a $\PJGy$ data sample as it provides a statistically independent, large-yield dimuon sample.

The \PgU\ efficiency is estimated from the product of the single-muon efficiencies. A factor, $\rho$, is used as a correction to this factorization hypothesis, and to account for possible biases introduced by the T\&P efficiency measurement with the $\PJGy$ sample. We define $\rho$ as
\begin{linenomath}
 \begin{align}
 \label{eq:rho}
	\rho(\pt^{\PgU}, \abs{y^{\PgU}}) = \frac{\epsilon(\PgU)}{\epsilon(\Pgmp_{\PJGy})\cdot\epsilon(\Pgmm_{\PJGy})},
 \end{align}
\end{linenomath}
where $\epsilon(\PgU)$ is the efficiency for a \ups to pass the trigger and muon identification selections, and $\epsilon(\Pgmp_{\PJGy})$ and $\epsilon(\Pgmm_{\PJGy})$ are the corresponding efficiencies for positively and negatively charged muons from a $\PJGy$ decay with the same \pt and $\eta$ as a muon in the \ups decay. The \ups efficiency is taken from MC simulation generator-level matching, which is performed by associating the two generated muons from the \ups with the reconstructed muons or trigger objects. The single-muon efficiencies are from the T\&P method utilizing a $\PJGy$ MC simulation sample.
Finally, the efficiency of the vertex-fit $\chi^{2}$ probability requirement is determined from data to be $(99.16\pm0.09)\%$ and constant over the entire kinematic range.

\section{Systematic uncertainties}
\label{systematics}

Systematic uncertainties in the cross section measurement stem from variations in the acceptance determination, potential residual inaccuracies in the efficiency measurement, the method of yield extraction, and the integrated luminosity. For each uncertainty, we give below in parentheses a representative range of values corresponding to the variation with $\pt^{\PgU}$. The acceptance is varied in the dimuon invariant-mass fit coherently by $\pm$1 standard deviation, reflecting the uncertainty from the finite MC simulation statistics (0.3--1\%). The acceptance is sensitive to biases in track momentum and differences in resolution between simulation and data. To determine the effect on the \ups acceptance, we introduce a track $\pt$ bias of 0.2\%, chosen based on the momentum scale biases seen in simulation and data~\cite{bib-trackermomentum}. We also vary the transverse momentum resolution by $\pm$10\%, corresponding to the uncertainty in the resolution measurement using $\PJGy$ in data. This reflects a conservative estimation of resolution effects. The acceptance map as a function of $\pt^{\PgU}$ and $\abs{y^{\PgU}}$ is then recalculated, and the systematic uncertainty is the difference in the resulting cross sections when using the perturbed acceptance map rather than the nominal one (0.0--0.7\%).
Imperfect knowledge of the production \pt spectrum of the $\PgU$ resonances at $\sqrt{s} = 7\TeV$ contributes a systematic uncertainty.
Using either a flat \pt distribution or the \pt distribution from \PYTHIA, which is found to be consistent with the previously
measured \pt distribution~\cite{yprd10}, gives rise to a systematic uncertainty (0.2$\%$).
FSR is incorporated into the simulation using the \PHOTOS algorithm.
To estimate the systematic uncertainty associated with this procedure, the acceptance is calculated without FSR, and 20\% of the difference is taken as the uncertainty (0.1--0.8\%), based on a study in Ref.~\cite{bib-photos2}.

Variation of the measured factorized efficiencies within their uncertainties also gives rise to a systematic uncertainty. The systematic uncertainties for the tracking efficiency (0.3--0.4\%), muon identification efficiency (2--4\%), and trigger efficiency (1--5\%) are evaluated conservatively by coherently varying all bins by $\pm$1 standard deviation. 
The systematic uncertainty arising from the choice of bin size for the efficiencies is determined by fitting the efficiency turn-on curves as a function of muon $\pt^{\mu}$ in different $|\eta^{\mu}|$ regions using a hyperbolic tangent function and taking the muon efficiencies from the function instead of the binned value to compute the cross section (1--4\%).
The intrinsic bias from the T\&P method, including possible bias in the T\&P technique and differences in the $\PJGy$ and $\PgU$ kinematics, as well as the possible misestimation of the double-muon $\PgU$ efficiency as the product of the single-muon efficiencies, are all included in the correction factor $\rho$. The average rho factor value is 1.07 and the full range of variation is from 0.92 to 1.20. As a conservative estimate of the systematic uncertainty associated with $\rho$, the measurements are repeated with a correction factor of unity and half of the variation is taken as the systematic uncertainty (2--5\%).

In addition, systematic uncertainties may arise from differences between the dimuon invariant-mass distribution in the data and the probability density functions (PDF) chosen for the signal and background components in the fit. Since the CB parameters, which describe the radiative tail of each signal resonance, are fixed from MC simulation in the fit to the data, we fit the full data set with free tail parameters and use the values obtained to fix the tail parameters for the yield extraction in the ($\Delta \pt^{\PgU}$, $\Delta y^{\PgU}$) bins. The difference in the fit yield is taken as a systematic uncertainty (1--4\%). We vary the background PDF by replacing the product of the exponential and error function by a polynomial function, while restricting the fit to the mass range 8--12\GeVcc (1--5\%). The determination of the integrated luminosity is made with an uncertainty of 4\%~\cite{bib-lumi-pas}.
A summary of systematic uncertainties for the \upsn production cross section, integrated over the full transverse momentum ($\pt^{\PgU}$) and rapidity ($y^{\PgU}$) ranges, is shown in Table~\ref{tab:results-xsecns-diff-y02}. The largest sources of systematic uncertainty arise from the statistical precision of the efficiency measurements determined from data, the efficiency correction factor $\rho$, and from the measurement of the integrated luminosity.

The cross section measurement uses acceptance maps corresponding to different \ups polarization scenarios. The values of the resulting cross sections vary approximately linearly by about $\pm5\%$, $\pm10\%$, and $\pm20\%$, respectively, assuming $\lambda_\theta = \pm0.25$, $\pm 0.5$, and $\pm$1,
as shown in Table~\ref{tab:polar_upsns}. The cross sections are also measured for $10< \pt^{\PgU}<50\GeVc$ and $\abs{y^{\PgU}}<1.2$ using the measured $\upsn$ polarizations~\cite{cms_ypol} to compute the acceptance corrections. The three anisotropy parameters in the center-of-mass helicity and Collins--Soper frames are varied coherently by $\pm$1 standard deviation, and the largest positive and negative variations with respect to the nominal (no polarization) case are taken as systematic uncertainties. These are listed in Table~\ref{tab:cmspol_upsns}. They are comparable to, or smaller than, the result of varying the longitudinal or transverse polarizations by setting $\lambda_\theta$ to $\pm$0.25
for the \PgUa\ case, while they are between the results obtained by setting $\lambda_\theta$ to ${\pm}0.25$ and ${\pm}0.5$
for the \PgUb\ and \PgUc. The fiducial cross sections do not depend on the acceptance, the assumed \PgU\ polarization, or the associated uncertainties. The definition of the acceptance in Eq.~(\ref{eq:acceptance}) includes reconstructed quantities. The variation in the cross section using only generator-level quantities is less than 1\%.

\begin{table*}[!htb]
\topcaption{Relative systematic uncertainties in the \upsn production cross section, integrated over the rapidity range $\abs{y^{\PgU}}<2.4$, times the dimuon branching fraction, in percent. The symbols $\acc$, $\epsilon_\text{T\&P}$, $\epsilon_{\rho}$, and PDF refer to the systematic uncertainties arising from the acceptance, tag-and-probe efficiencies, correction factor $\rho$, and signal-and-background PDF. The remaining systematic uncertainties are summed in the ``other" category. The integrated luminosity uncertainty of 4\% is not shown. The numbers in parentheses are negative variations.}
\centering
{
\renewcommand{\arraystretch}{1.2}
\begin{tabular}{c|c|c c c c c}
\hline
 & \pt (\GeVcns{})  &  $\acc$ & $\epsilon_\text{T\&P}$ & $\epsilon_{\rho}$ & PDF  & other \\ \hline
$\PgUa$ &  0--50     &$ 1.0\,(1.0)$ & $ 5.2\,( 4.3)$ & $3.4$ & 1.8 & $0.4\,(0.3)$\\
$\PgUb$ &  0--42     &$ 1.1\,(1.1)$ & $ 5.5\,( 4.1)$ & $3.7$ & $2.6$ & $0.4\,(0.4)$\\
$\PgUc$ &  0--38     &$ 1.2\,(1.1)$ & $ 6.7\,( 4.9)$ & $4.0$ & $3.8$ & $0.6\,(0.5)$\\
\hline
\end{tabular}
}
\label{tab:results-xsecns-diff-y02}%
\end{table*}

\begin{table*}[!ht]
\topcaption{The fractional change in percent to the central value of the \upsn cross section integrated over the rapidity range $\abs{y^{\PgU}} < 2.4$, relative to the unpolarized value, for six polarization scenarios in the center-of-mass helicity and Collins--Soper frames. The polarization assumption changes from fully longitudinal to fully transverse polarization as $\lambda_\theta$ changes from $-1$ to 1.}
\centering
{\footnotesize
\renewcommand{\arraystretch}{1.2}
\begin{tabular}{c|c|c c c c c c |c c c c c c}
\hline
\multicolumn{2}{c|}{}  &\multicolumn{6}{c|}{Helicity Frame $\lambda_{\theta}$} & \multicolumn{6}{c}{ Collins--Soper Frame $\lambda_{\theta}$} \\ \hline
& \pt (\GeVcns{}) & $ 1$ & $ 0.5$ & $ 0.25$ & $ -0.25$ & $ -0.5$ & $ -1$ & $ 1$ & $ 0.5$ & $ 0.25$ & $ -0.25$ & $ -0.5$ & $ -1$\\ \hline
$\PgUa$ &  0--50     &$ +19$ & $+10$ & $+5$ & $-5$ & $ -11$ & $-24$  & $+16$ & $+8$ &$ +4$ & $-5$ & $-9$ & $-19$\\
$\PgUb$ &  0--42     &$ +14$ & $+5$ & $+3$ & $-7$ & $ -12$ & $-24$ & $+13$ & $+6$ &$ +2$ & $-6$ & $-10$ & $-20$\\
$\PgUc$ &  0--38     &$ +16$ & $+9$ & $+5$ & $-4$ & $ -9$ & $-21$ & $+14$ & $+8$ &$ +5$ & $-3$ & $-7$ & $-17$\\
\hline
\end{tabular}
}
\label{tab:polar_upsns}%
\end{table*}

\section{Differential fiducial cross section measurement and comparison to theory}
\label{crosssection_noAcc_results}

The fiducial \upsn cross sections are determined from the efficiency-corrected signal yields within the muon kinematic range specified by Eq.~(\ref{eq:selection}), using Eq.~(\ref{eqn:xsection}) with the acceptance term set to unity. The resulting total fiducial \upsn cross sections times dimuon branching fractions at $\sqrt{s} = 7\TeV$ for $\abs{y^{\PgU}}<2.4$ are
\ifthenelse{\boolean{cms@external}}{
\begin{equation*}
\begin{aligned}
&\sigma(\Pp\Pp \to \PgUa X ) \cdot \mathcal{B} (\PgUa \to \Pgmp \Pgmm) &=\\ &\qquad (3.06 \, \pm 0.02\,^{+ 0.20}_{- 0.18}\pm 0.12)\unit{nb}, \\
&\sigma(\Pp\Pp \to \PgUb X ) \cdot \mathcal{B} (\PgUb \to \Pgmp \Pgmm) &=\\ &\qquad (0.910 \pm 0.011\,^{+ 0.055}_{- 0.046}\pm 0.036)\unit{nb}, \\
&\sigma(\Pp\Pp \to \PgUc X ) \cdot \mathcal{B} (\PgUc \to \Pgmp \Pgmm) &=\\ &\qquad (0.490 \pm 0.010\,^{+ 0.029}_{- 0.029}\pm 0.020)\unit{nb},
\end{aligned}
\end{equation*}
}{
\begin{equation*}
\begin{aligned}
\sigma(\Pp\Pp \to \PgUa X ) \cdot \mathcal{B} (\PgUa \to \Pgmp \Pgmm) &= (3.06 \, \pm 0.02\,^{+ 0.20}_{- 0.18}\pm 0.12)\unit{nb}, \\
\sigma(\Pp\Pp \to \PgUb X ) \cdot \mathcal{B} (\PgUb \to \Pgmp \Pgmm) &= (0.910 \pm 0.011\,^{+ 0.055}_{- 0.046}\pm 0.036)\unit{nb}, \\
\sigma(\Pp\Pp \to \PgUc X ) \cdot \mathcal{B} (\PgUc \to \Pgmp \Pgmm) &= (0.490 \pm 0.010\,^{+ 0.029}_{- 0.029}\pm 0.020)\unit{nb},
\end{aligned}
\end{equation*}
}
where the first uncertainty is statistical, the second is systematic, and the third is associated with the estimation of the integrated luminosity of the data sample. The integrated results are obtained from the sum of the differential $\pt^{\PgU}$ results. The measured cross sections include feed-down from higher-mass bottomonium states.

 \begin{figure*}[thbp]
   \centering
    \includegraphics[width=0.45\textwidth]{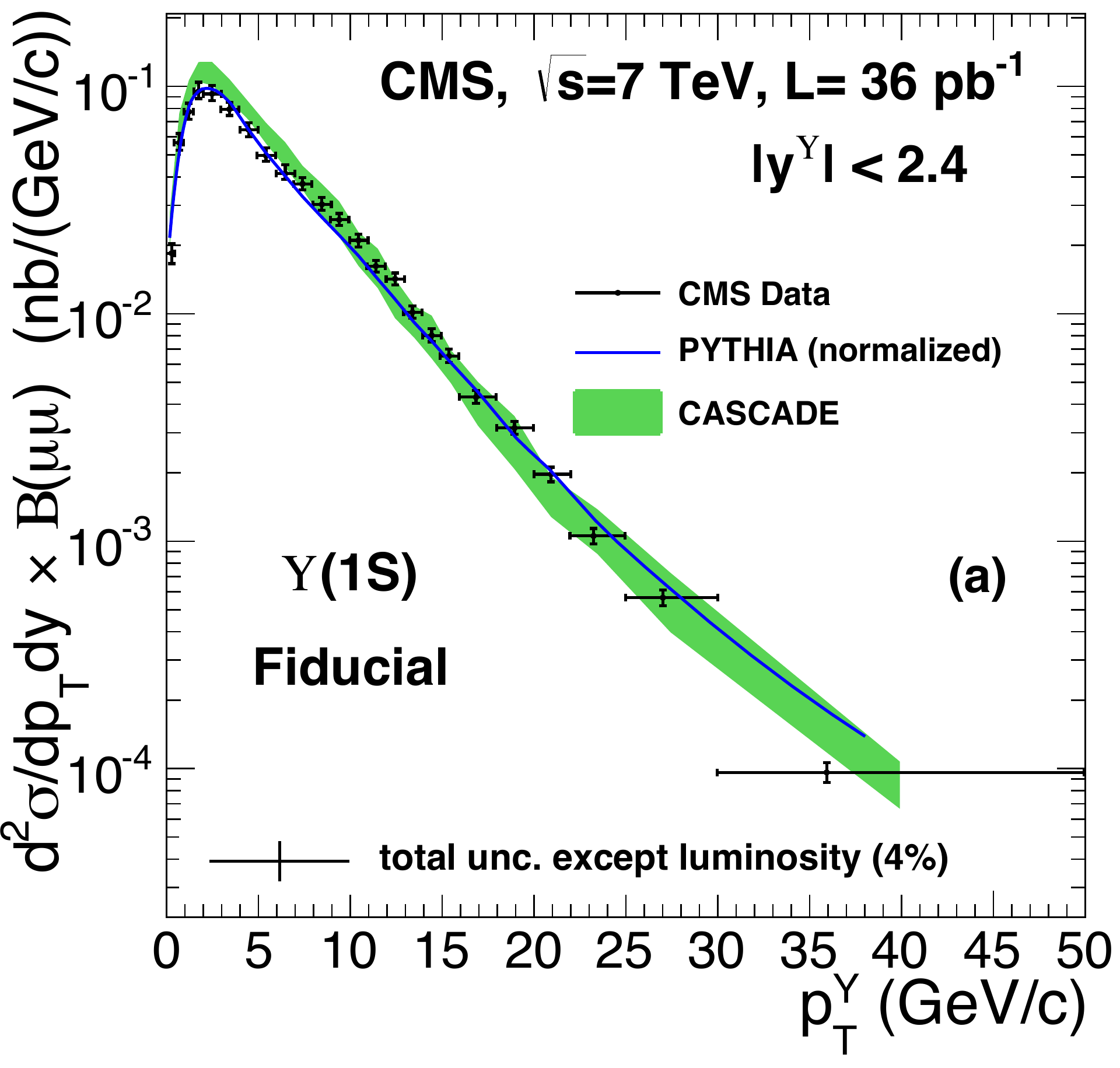}\label{fig:theory_pt_fiducial_1S}
    \includegraphics[width=0.45\textwidth]{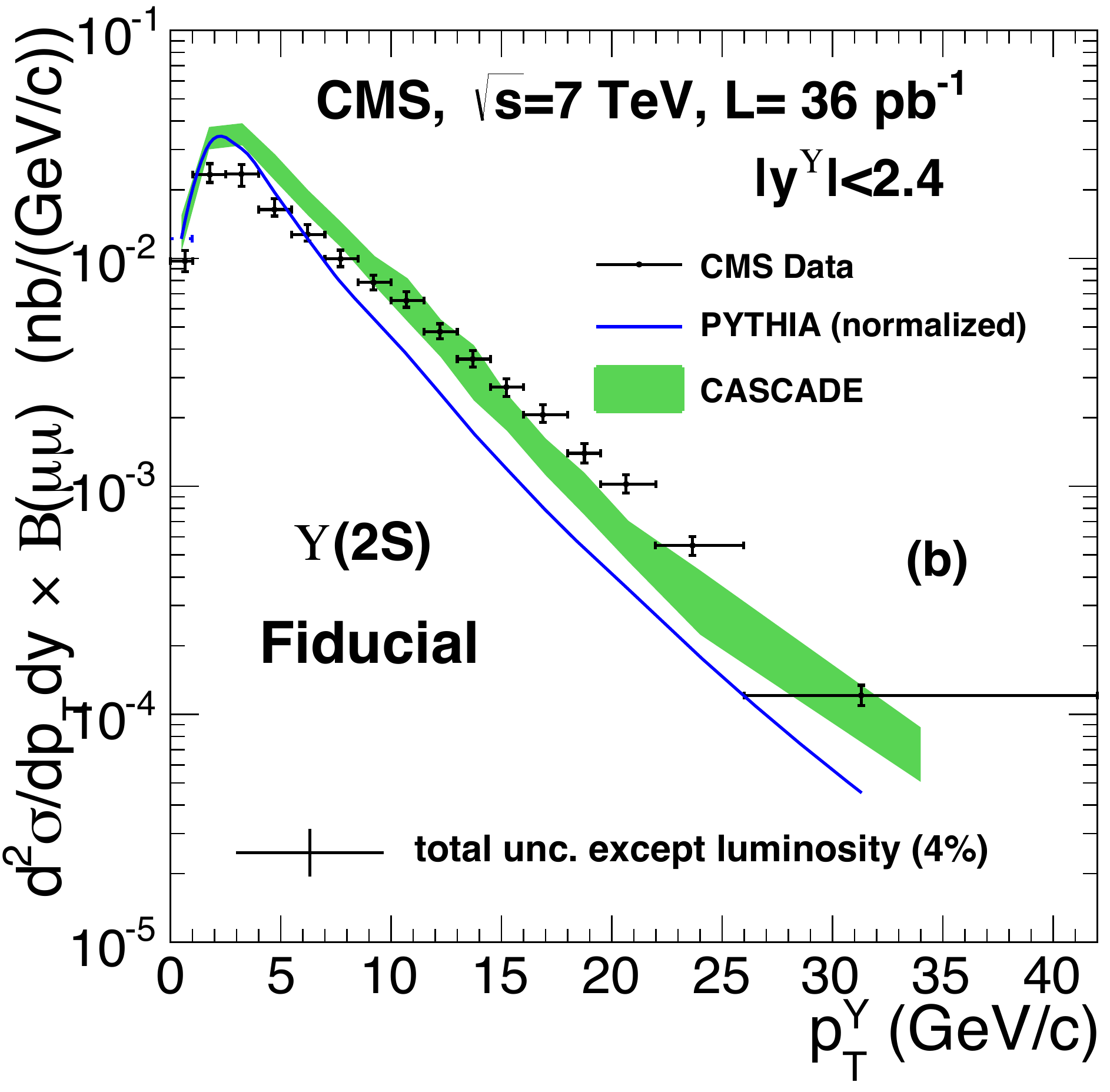}\label{fig:theory_pt_fiducial_2S}\\
    \includegraphics[width=0.45\textwidth]{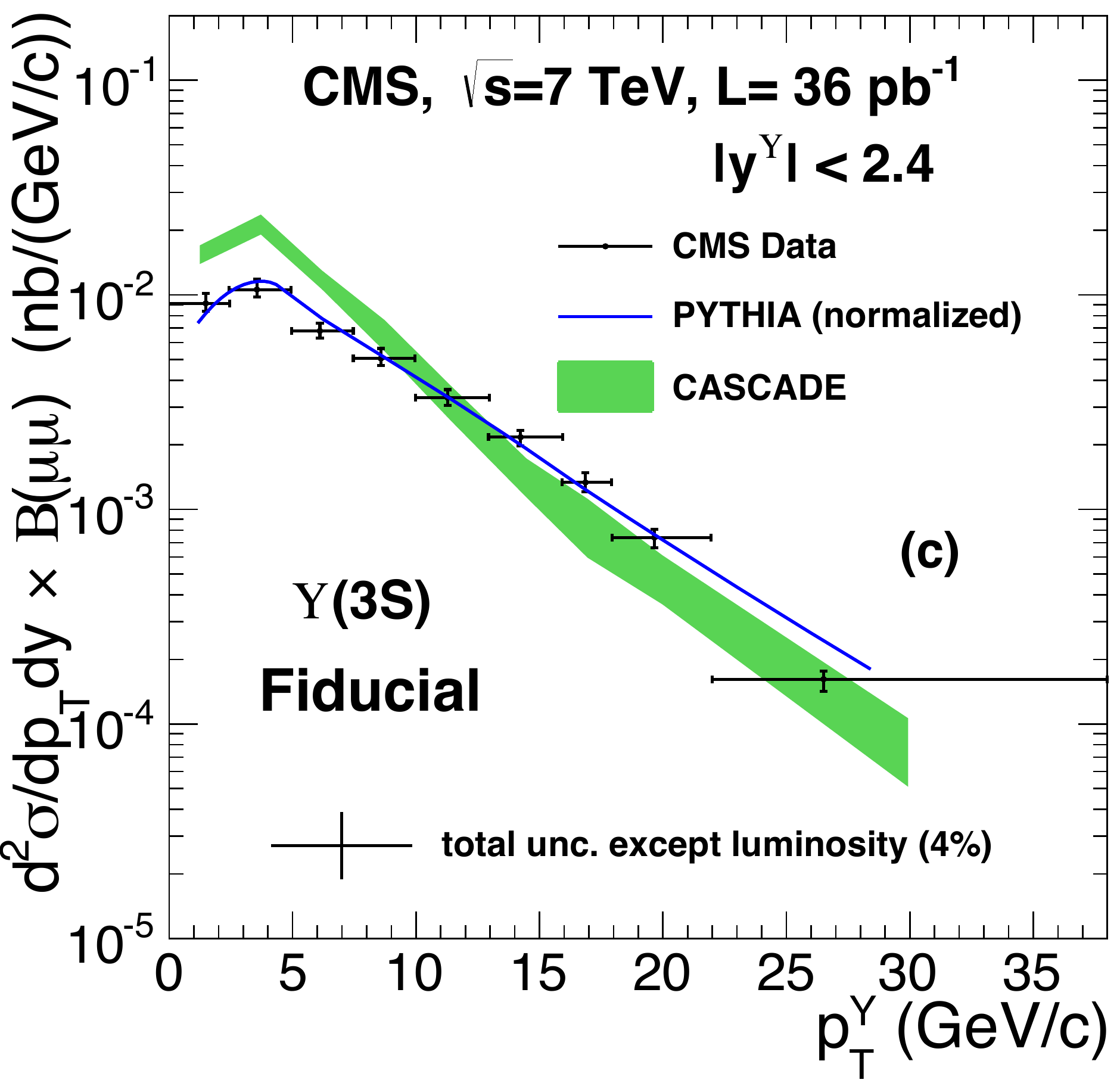}\label{fig:theory_pt_fiducial_3S}
    \includegraphics[width=0.45\textwidth]{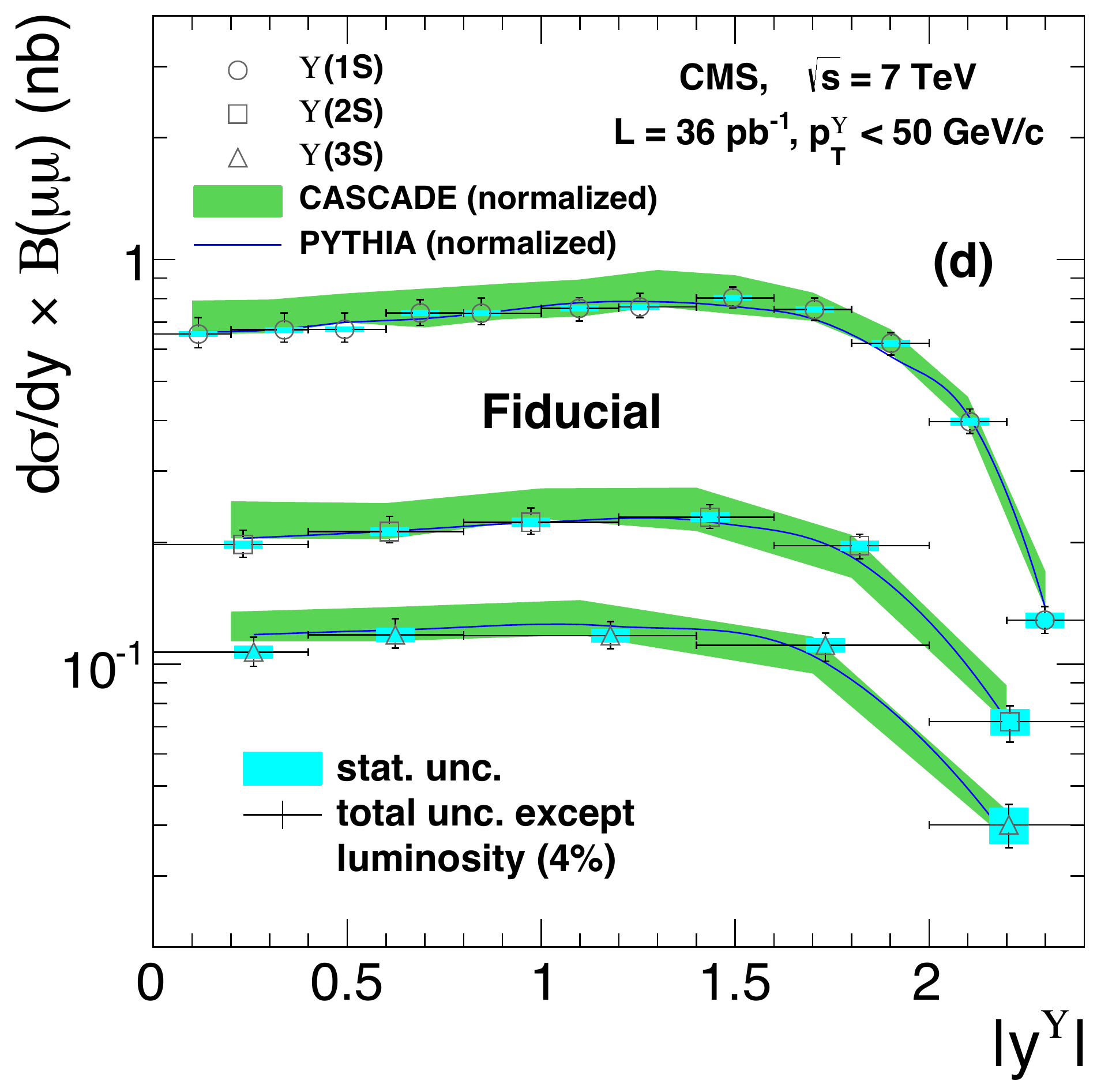}\label{fig:theory_y_fiducial_nS}
   \caption{
  Differential fiducial cross section of (a) \PgUa, (b) \PgUb, and (c) \PgUc\ as a function of $\pt^{\PgU}$ in the rapidity range $\abs{y^{\PgU}}<2.4$, and comparison to the predictions from \CASCADE and \PYTHIA. (d) Differential fiducial cross section of the \upsn as a function of rapidity and comparison to the predictions from \CASCADE and \PYTHIA. The \PYTHIA prediction is normalized to the measured total cross section, in order to facilitate the comparison of the shape of the dependences. The full \CASCADE prediction is shown in (a), (b), and (c); the normalised \CASCADE prediction is shown in (d). The bands indicate the estimated uncertainties in the \CASCADE prediction.} 
   \label{fig:CASCADE_fiducial}
 \end{figure*}

The \upsn differential \pt fiducial cross sections are summarized in Table~\ref{table:cross section-y-ns} and plotted in Fig.~\ref{fig:CASCADE_fiducial}~(a,b,c) and \suppMaterial. 
 In the figures, $\mathcal{B}(\upsn\to \Pgmp \Pgmm)$ is denoted as B$(\mu\mu)$. The results are also given for six rapidity intervals in \suppMaterial. 
Here, and throughout the paper, in figures illustrating differential cross sections, the data points are plotted at the average \pt (or rapidity) of the data in each bin. The $\pt^{\PgU}$ dependence of the cross sections has the same trend for all six rapidity intervals.
The \upsn \pt-integrated, differential rapidity fiducial cross sections, plotted in Fig.~\ref{fig:CASCADE_fiducial}~(d) and \suppMaterial, 
are all roughly constant from $\abs{y^{\PgU}} = 0$ to about 1.6, where they then fall quickly.
The ratios of the \upsn differential \pt fiducial cross sections, also shown in \suppMaterial, 
increase with $\pt^{\PgU}$.

\begin{table*}[!ht]
\topcaption{The product of the fiducial or acceptance-corrected \upsn production cross sections, $\sigma$, integrated and differential in $\pt^{\PgU}$, and the respective dimuon branching fraction, $\mathcal{B}$, integrated over the rapidity range $\abs{y^{\PgU}} < 2.4$. The cross sections assume the \upsn are unpolarized. The fiducial \upsn cross sections are independent of the \upsn polarization. The statistical uncertainty (stat.), the sum of the systematic uncertainties in quadrature ($\sum_\text{syst.}$), and the total uncertainty ($\Delta\sigma$; including stat., $\sum_\text{syst.}$, and the uncertainty in the integrated luminosity) are in percent. The numbers in parentheses are negative variations.}
  \centering
{\scriptsize
\begin{tabular}{c| c D{.}{.}{2.2}  | D{.}{.}{1.4} c c c |  D{.}{.}{1.4} c c c}
\hline
\multirow{2}{*}{ }&\multicolumn{2}{c|}{}	     & \multicolumn{4}{c|}{Fiducial Cross Section} & \multicolumn{4}{c}{Cross Section}\\\hline
&\pt(\GeVcns{}) & \multicolumn{1}{c}{mean}& \multicolumn{1}{|c}{$\sigma \cdot \mathcal{B}\unit{(nb)}$} & 	$\frac{\text{stat.}}{\sigma}$ & $\frac{\sum_{\text{syst.}}}{\sigma}$ & $\frac{\Delta\sigma}{\sigma}$
&   \multicolumn{1}{c}{$\sigma \cdot \mathcal{B}\unit{(nb)}$} &       $\frac{\text{stat.}}{\sigma}$ & $\frac{\sum_\text{{syst.}}}{\sigma}$ & $\frac{\Delta\sigma}{\sigma}$ \\ \hline
\multirow{24}{*}{$\PgUa$}
&0--0.5	& 0.33  &   0.0440 & 	${ 5.4}$ &$   8\,(   8)$ & $  11\,(  11)$& 0.0859&	${ 5.4}$ &$   8\,(   7)$ & $  11\,(  10)$\\
&0.5--1	& 0.77  &   0.133 & 	${ 3.1}$ &$   8\,(   8)$ & $  10\,(  10)$& 0.263&	${ 3.3}$ &$   8\,(   7)$ & $   9\,(   9)$\\
&1--1.5	& 1.26  &   0.182 & 	${ 2.5}$ &$   8\,(   8)$ & $   9\,(   9)$& 0.374&	${ 2.6}$ &$   8\,(   8)$ & $   9\,(   9)$\\
&1.5--2	& 1.75  &   0.228 & 	${ 2.4}$ &$   8\,(   8)$ & $  10\,(   9)$& 0.505&	${ 2.4}$ &$   9\,(   8)$ & $  10\,(   9)$\\
&2--3	& 2.49  &   0.442 & 	${ 1.6}$ &$   8\,(   7)$ & $   9\,(   8)$& 1.16&	${ 1.6}$ &$   8\,(  10)$ & $   9\,(  11)$\\
&3--4	& 3.48  &   0.374 & 	${ 1.8}$ &$   6\,(   6)$ & $   8\,(   7)$& 1.21&	${ 2.1}$ &$   7\,(   6)$ & $   9\,(   8)$\\
&4--5	& 4.48  &   0.302 & 	${ 1.8}$ &$   7\,(   7)$ & $   8\,(   8)$& 1.084&	${ 2.1}$ &$   7\,(   6)$ & $   8\,(   8)$\\
&5--6	& 5.49  &   0.236 & 	${ 2.0}$ &$   7\,(   6)$ & $   8\,(   7)$& 0.879&	${ 1.9}$ &$   7\,(   9)$ & $   8\,(  10)$\\
&6--7	& 6.49  &   0.195 & 	${ 2.0}$ &$   8\,(   7)$ & $   9\,(   9)$& 0.680&	${ 2.6}$ &$   6\,(   6)$ & $   8\,(   7)$\\
&7--8	& 7.49  &   0.174 & 	${ 2.1}$ &$   5\,(   5)$ & $   7\,(   6)$& 0.556&	${ 2.0}$ &$   6\,(   5)$ & $   7\,(   7)$\\
&8--9	& 8.48  &   0.144 & 	${ 2.3}$ &$   6\,(   5)$ & $   7\,(   7)$& 0.419&	${ 2.2}$ &$   5\,(   5)$ & $   7\,(   7)$\\
&9--10	& 9.48  &   0.1235 & 	${ 2.4}$ &$   5\,(   4)$ & $   7\,(   6)$& 0.331&	${ 2.3}$ &$   5\,(   4)$ & $   7\,(   6)$\\
&10--11	& 10.48 &   0.0988 & 	${ 2.5}$ &$   6\,(   5)$ & $   8\,(   7)$ &0.238&	${ 2.5}$ &$   5\,(   4)$ & $   7\,(   6)$\\
&11--12	& 11.49 &   0.0759 & 	${ 2.8}$ &$   4\,(   4)$ & $   7\,(   6)$ &0.179&	${ 2.9}$ &$   5\,(   4)$ & $   7\,(   6)$\\
&12--13	& 12.49 &   0.0670 & 	${ 2.9}$ &$   4\,(   4)$ & $   7\,(   6)$ &0.145&	${ 2.9}$ &$   5\,(   4)$ & $   7\,(   7)$\\
&13--14	& 13.47 &   0.0477 & 	${ 3.3}$ &$   5\,(   4)$ & $   7\,(   7)$ &0.0990&	${ 3.2}$ &$   4\,(   5)$ & $   7\,(   7)$\\
&14--15	& 14.49 &   0.0381 & 	${ 3.6}$ &$   5\,(   5)$ & $   7\,(   7)$ &0.0750&	${ 3.6}$ &$   5\,(   5)$ & $   8\,(   7)$\\
&15--16	& 15.48 &   0.0312 & 	${ 4.0}$ &$   5\,(   4)$ & $   8\,(   7)$ &0.0595&	${ 3.8}$ &$   5\,(   5)$ & $   7\,(   7)$\\
&16--18	& 16.91 &   0.0412 & 	${ 3.5}$ &$   5\,(   5)$ & $   7\,(   7)$ &0.0732&	${ 3.4}$ &$   5\,(   5)$ & $   7\,(   7)$\\
&18--20	& 18.98 &   0.0296 & 	${ 4.0}$ &$   5\,(   4)$ & $   7\,(   7)$ &0.0500&	${ 3.8}$ &$   5\,(   4)$ & $   7\,(   7)$\\
&20--22	& 20.94 &   0.0187 & 	${ 5.1}$ &$   4\,(   4)$ & $   8\,(   8)$ &0.0302&	${ 5.1}$ &$   5\,(   4)$ & $   8\,(   8)$\\
&22--25	& 23.30 &   0.0148 & 	${ 5.8}$ &$   4\,(   4)$ & $   8\,(   8)$ &0.0237&	${ 5.6}$ &$   5\,(   4)$ & $   8\,(   8)$\\
&25--30	& 27.03 &   0.0133 & 	${ 6.1}$ &$   4\,(   4)$ & $   8\,(   8)$ &0.0205&	${ 6.0}$ &$   5\,(   4)$ & $   9\,(   8)$\\
&30--50	& 35.97 &   0.00923 & 	${ 7.8}$ &$   6\,(   6)$ & $  11\,(  10)$ &0.0123&	${ 7.4}$ &$   6\,(   6)$ & $  10\,(  10)$\\
& 0--50   & 5.34  &   3.06 &        ${ 0.6}$ &$   6\,(   6)$ & $  8\,(   7)$& 8.55 &    ${ 0.6}$ &$   7\,(   6)$ & $  8\,(   7)$\\\hline
\multirow{16}{*}{$\PgUb$}
&0--1     &0.66  &  0.0467&	        ${ 6.3}$ &$   7\,(   8)$ & $  10\,(  11)$& 0.0829&	     ${ 5.9}$ &$   9\,(   8)$ & $  11\,(  11)$\\
&1--2.5     &1.79  &  0.168&	        ${ 3.4}$ &$   8\,(   8)$ & $  10\,(  10)$& 0.331&	     ${ 3.3}$ &$  11\,(  10)$ & $  12\,(  12)$\\
&2.5--4     &3.21  &  0.169&	        ${ 3.1}$ &$   8\,(  11)$ & $   9\,(  12)$& 0.409&	     ${ 3.1}$ &$   9\,(   8)$ & $  10\,(   9)$\\
&4--5.5     &4.71  &  0.118&	        ${ 3.3}$ &$   8\,(   7)$ & $  10\,(   9)$& 0.362&	     ${ 3.3}$ &$   8\,(   7)$ & $   9\,(   9)$\\
&5.5--7     &6.22  &  0.0917&	        ${ 3.6}$ &$   6\,(   5)$ & $   8\,(   8)$& 0.286&	     ${ 3.6}$ &$   7\,(   6)$ & $   9\,(   8)$\\
&7--8.5     &7.71  &  0.0716&	        ${ 3.4}$ &$   7\,(   7)$ & $   9\,(   9)$& 0.212&	     ${ 3.9}$ &$   7\,(   7)$ & $   9\,(   9)$\\
&8.5--10     &9.21  &  0.0564&	        ${ 4.0}$ &$   5\,(   5)$ & $   8\,(   8)$& 0.146&	     ${ 4.0}$ &$   6\,(   6)$ & $   9\,(   8)$\\
&10--11.5     &10.69 &  0.0470&	        ${ 4.1}$ &$   6\,(   5)$ & $   8\,(   8)$ &0.1123&	     ${ 4.1}$ &$   6\,(   6)$ & $   9\,(   8)$\\
&11.5--13     &12.21 &  0.0343&	        ${ 4.6}$ &$   4\,(   4)$ & $   7\,(   8)$ &0.0765&	     ${ 4.6}$ &$   5\,(   5)$ & $   8\,(   8)$\\
&13--14.5     &13.70 &  0.0260&	        ${ 5.2}$ &$   5\,(   5)$ & $   8\,(   8)$ &0.0519&	     ${ 5.1}$ &$   5\,(   5)$ & $   8\,(   8)$\\
&14.5--16     &15.22 &  0.0196&	        ${ 5.7}$ &$   4\,(   6)$ & $   8\,(   9)$ &0.0376&	     ${ 5.7}$ &$   5\,(   7)$ & $   9\,(  10)$\\
&16--18     &16.88 &  0.0198&	        ${ 5.5}$ &$   6\,(   5)$ & $   9\,(   8)$ &0.0373&	     ${ 5.3}$ &$   6\,(   5)$ & $   9\,(   8)$\\
&18--19.5     &18.76 &  0.01005&	        ${ 7.5}$ &$   4\,(   5)$ & $   9\,(  10)$ &0.0159&	     ${ 7.4}$ &$   5\,(   4)$ & $  10\,(   9)$\\
&19.5--22     &20.65 &  0.0123&	        ${ 6.8}$ &$   5\,(   5)$ & $   9\,(   9)$ &0.0204&	     ${ 6.6}$ &$   5\,(   5)$ & $   9\,(   9)$\\
&22--26     &23.69 &  0.0104&	        ${ 7.4}$ &$   4\,(   5)$ & $   9\,(  10)$ &0.0158&	     ${ 7.2}$ &$   5\,(   4)$ & $  10\,(   9)$\\
&26--42     &31.30 &  0.00930&	        ${ 8.0}$ &$   5\,(   5)$ & $  10\,(  10)$ &0.0126&	     ${ 7.7}$ &$   6\,(   5)$ & $  10\,(  10)$\\
&0--42     &5.32  &   0.910  &        ${ 1.2}$ &$   6\,(   5)$ & $   7\,(   7)$& 2.21 &      ${ 1.2}$ &$  7\,(   6)$ & $  8\,(  7)$\\\hline
\multirow{9}{*}{\PgUc}
&0--2.5     &1.54  & 0.107&	         ${ 5.3}$ &$   7\,(   7)$ & $  10\,(  10)$&0.203&	    ${ 5.3}$ &$   8\,(   8)$ & $  11\,(  10)$\\
&2.5--5     &3.62  & 0.125&	         ${ 4.5}$ &$   8\,(   8)$ & $  10\,(  10)$&0.287&	    ${ 4.5}$ &$  10\,(  11)$ & $  12\,(  12)$\\
&5--7.5     &6.15  & 0.0801&	         ${ 4.7}$ &$   6\,(   6)$ & $   9\,(   8)$&0.227&	    ${ 4.6}$ &$   9\,(   8)$ & $  11\,(  10)$\\
&7.5--10     &8.62  & 0.0604&	         ${ 4.8}$ &$   9\,(   8)$ & $  11\,(  10)$&0.157&	    ${ 4.8}$ &$  11\,(  10)$ & $  12\,(  12)$\\
&10--13     &11.31 & 0.0476&	         ${ 4.5}$ &$   6\,(   7)$ & $   8\,(   9)$&0.113&	    ${ 4.3}$ &$   7\,(   5)$ & $   9\,(   8)$\\
&13--16     &14.30 & 0.0308&	         ${ 5.1}$ &$   5\,(   6)$ & $   8\,(   9)$&0.0617&	    ${ 5.0}$ &$   5\,(   5)$ & $   8\,(   8)$\\
&16--18     &16.94 & 0.0127&	         ${ 7.5}$ &$   6\,(   5)$ & $  10\,(  10)$&0.0227&	    ${ 7.4}$ &$   6\,(   5)$ & $  10\,(  10)$\\
&18--22     &19.72 & 0.0140&	         ${ 6.9}$ &$   7\,(   7)$ & $  11\,(  11)$&0.0229&	    ${ 7.0}$ &$   7\,(   6)$ & $  10\,(  10)$\\
&22--38     &26.51 & 0.0124&	         ${ 7.4}$ &$   9\,(   9)$ & $  12\,(  12)$&0.0185&	    ${ 7.6}$ &$  13\,(  13)$ & $  15\,(  15)$\\
&0--38     &5.31  &  0.490 &         ${ 2.0}$ &$   6\,(   6)$ & $  8\,(   7)$& 1.11 &    ${ 2.0}$ &$  9\,(  8)$ & $  10\,(  9)$\\
\hline
\end{tabular}
}
  \label{table:cross section-y-ns}
\end{table*}

A comparison between the fiducial cross section measurement and theoretical predictions is shown in Fig.~\ref{fig:CASCADE_fiducial}.
 Each of the predictions is made with the assumption of unpolarized \upsn production.
The comparison is made to the \CASCADE~\cite{bib-CASCADE} MC generator in the fixed-order-plus-next-to-leading-log (FONLL) framework, including feed-down
from $\chi_b$(1P), $\chi_b$(2P), $\chi_b$(3P)~\cite{PhysRevD.86.054015}, and other higher-mass \ups states, and to \PYTHIA~\cite{bib-PYTHIA-tuning} including feed-down for the \PgUa\ and \PgUb\ from the P-wave states with the same principal quantum
number.
The \pt dependence of the cross section predicted by \CASCADE agrees with the data for the \PgUa, is marginally consistent for the \PgUb\, but does not describe the \PgUc\ spectrum, where it predicts a softer \pt spectrum.
For each resonance, the total cross section predicted by \PYTHIA is higher, by factors of about 2, than the measured cross section. In Fig.~\ref{fig:CASCADE_fiducial}, 
for each resonance the \PYTHIA prediction is normalized to the measured total cross section, in order to facilitate the comparison of the cross section dependences with the predictions.
The \PYTHIA prediction of the \pt dependence agrees with data for the \PgUa\ and \PgUc, but not for the \PgUb. Both \CASCADE and \PYTHIA provide a good description of the shape of the rapidity dependence for the three states. Complete tables of results for the differential fiducial cross sections for the three \ups states are available in \suppMaterial.

\section{Acceptance-corrected differential cross section measurement and comparison to theory}
\label{crosssection_results}

The acceptance-corrected \upsn production cross sections times the dimuon branching fractions at $\sqrt{s} = 7\TeV$ for $\abs{y^{\PgU}}<2.4$ 
are measured to be
\ifthenelse{\boolean{cms@external}}{
\begin{equation}
\begin{aligned}
&\sigma(\Pp\Pp \to \PgUa X ) \cdot \mathcal{B} (\PgUa \to \Pgmp \Pgmm) &= \\ &\qquad(8.55 \pm 0.05\,^{+0.56}_{-0.50}\pm 0.34)\unit{nb}, \\
&\sigma(\Pp\Pp \to \PgUb X ) \cdot \mathcal{B} (\PgUb \to \Pgmp \Pgmm) &= \\ &\qquad(2.21 \pm 0.03\,^{+0.16}_{-0.14}\pm 0.09)\unit{nb}, \\
&\sigma(\Pp\Pp \to \PgUc X ) \cdot \mathcal{B} (\PgUc \to \Pgmp \Pgmm) &= \\ &\qquad(1.11 \pm 0.02\,^{+0.10}_{-0.08}\pm 0.04)\unit{nb},
\end{aligned}
\end{equation}
}{
\begin{equation}
\begin{aligned}
\sigma(\Pp\Pp \to \PgUa X ) \cdot \mathcal{B} (\PgUa \to \Pgmp \Pgmm) &= (8.55 \pm 0.05\,^{+0.56}_{-0.50}\pm 0.34)\unit{nb}, \\
\sigma(\Pp\Pp \to \PgUb X ) \cdot \mathcal{B} (\PgUb \to \Pgmp \Pgmm) &= (2.21 \pm 0.03\,^{+0.16}_{-0.14}\pm 0.09)\unit{nb}, \\
\sigma(\Pp\Pp \to \PgUc X ) \cdot \mathcal{B} (\PgUc \to \Pgmp \Pgmm) &= (1.11 \pm 0.02\,^{+0.10}_{-0.08}\pm 0.04)\unit{nb},
\end{aligned}
\end{equation}
}

where the first uncertainty is statistical, the second is systematic, and the third is from the estimation of the integrated luminosity. These results assume unpolarized \upsn production. The \PgUa\ integrated production cross section in the restricted rapidity range $\abs{y^{\PgU}}<2.0$ is $7.496 \pm 0.052\stat\unit{nb}$, which is consistent with the previous CMS result of $7.37\pm0.13\stat\unit{nb}$ ~\cite{yprd10}, measured in the same rapidity range. The results of the \upsn production cross sections for the same $\pt^{\PgU}$ and $y^{\PgU}$ ranges used for the measurement of the \upsn polarizations in Ref.~\cite{cms_ypol} are shown in Table~\ref{tab:cmspol_upsns}.

\begin{table}[!ht]
\topcaption{The product of the acceptance-corrected \upsn production cross sections, $\sigma$, and the dimuon branching fraction, $\mathcal{B}$, integrated over the rapidity range $\abs{y^{\PgU}} < 1.2$, and the $\pt^{\PgU}$ range from 10 to 50\GeVc, as used in Ref.~\cite{cms_ypol} for the measurement of the \ups polarizations. The cross sections assume the \upsn are unpolarized. The statistical uncertainty (stat.), the sum in quadrature of the systematic uncertainties ($\sum_\text{syst.}$), excluding the contribution from the polarization uncertainty, the systematic uncertainties from the polarization (pol.), and the total uncertainty ($\Delta\sigma$; including stat., $\sum_\text{syst.}$, pol., and the uncertainty in the integrated luminosity) are in percent. The numbers in parentheses are negative variations.}
\centering
\begin{tabular}{c  | c c c c c }
\hline
	     & $\sigma \cdot \mathcal{B}\unit{(nb)}$ & $\frac{\text{stat.}}{\sigma}$ & $\frac{\sum_{\text{syst.}}}{\sigma}$ & $\frac{\text{pol.}}{\sigma}$ & $\frac{\Delta\sigma}{\sigma}$ \\\hline
$\PgUa$ & ${0.558}$ & ${1.3}$ & $6\,(5)$ & $4\,(2)$ & $8\,(7)$\\
$\PgUb$ & ${0.213}$ & ${2.4}$ & $5\,(5)$ & $7\,(3)$ & $10\,(8)$\\
$\PgUc$ & ${0.127}$ & ${3.2}$ & $7\,(5)$ & $7\,(3)$ & $11\,(8)$\\
\hline
\end{tabular}

  \label{tab:cmspol_upsns}
\end{table}

\begin{figure}[htb]
  \centering
  \includegraphics[width=0.45\textwidth]{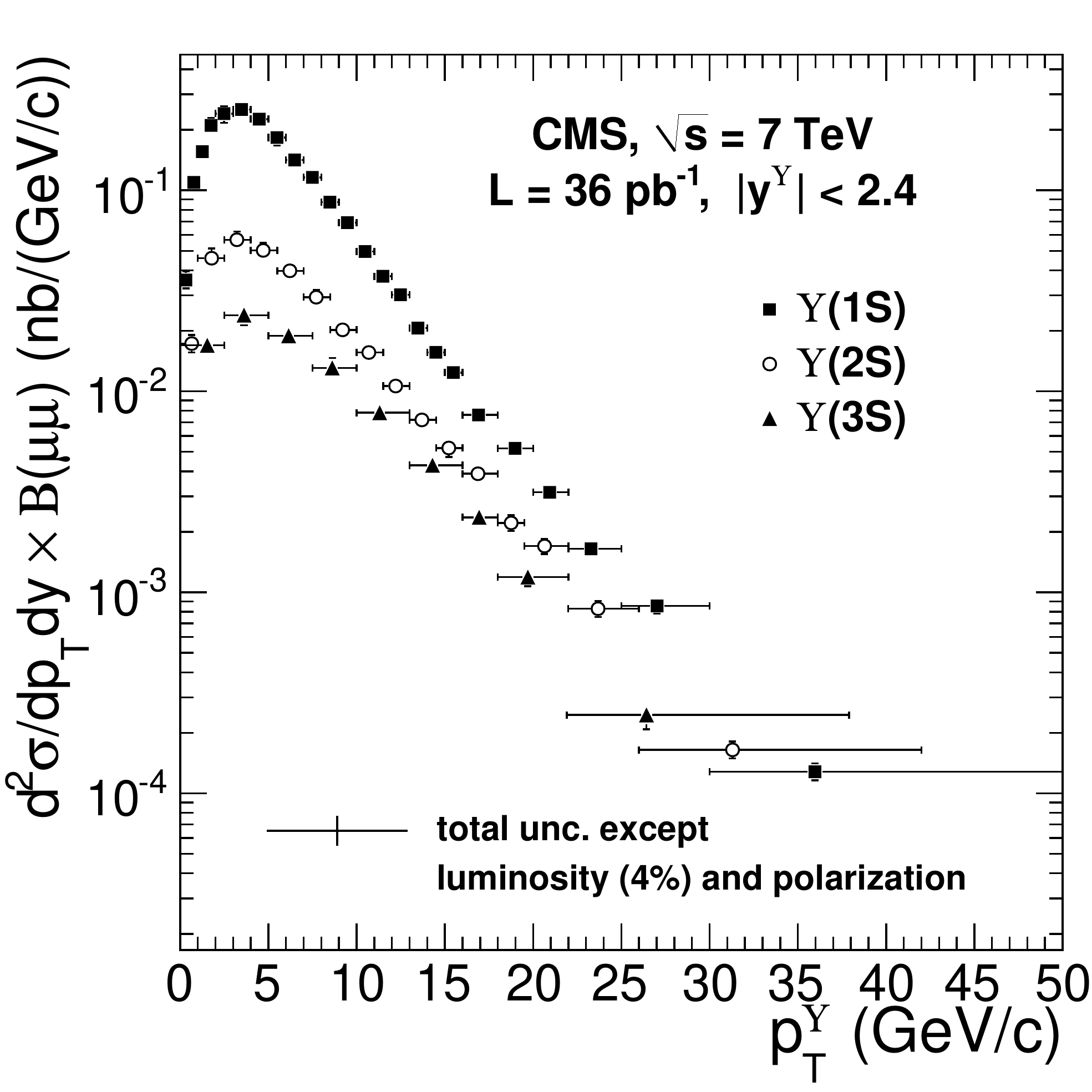}
  \caption{Acceptance-corrected differential cross sections as a function of $\pt^{\PgU}$ in the rapidity range $\abs{y^{\PgU}}<2.4$.}
  \label{fig:xsec_overlay}
\end{figure}

\begin{figure}[htb]
  \centering
  \includegraphics[width=0.45\textwidth]{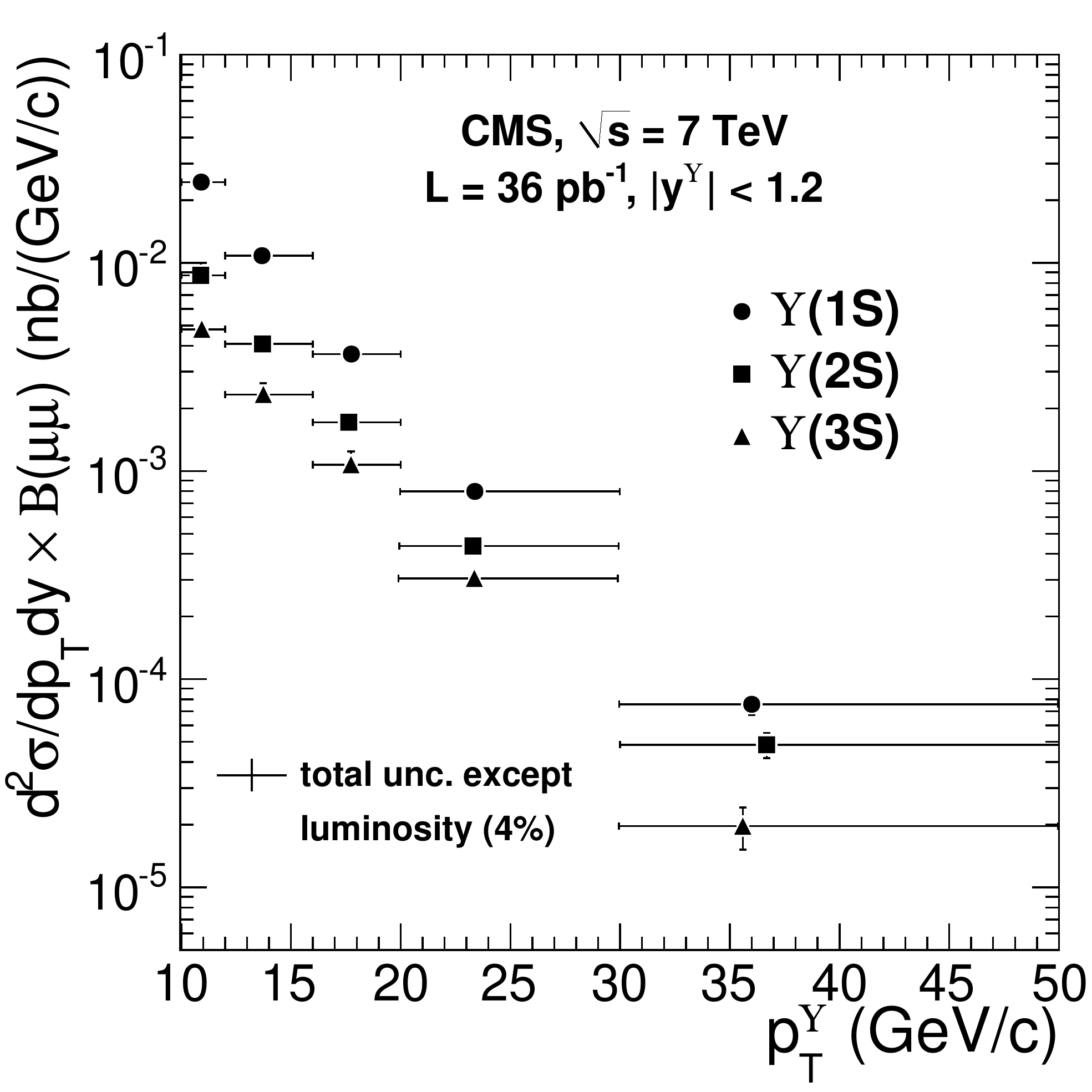}
  \caption{Acceptance-corrected differential cross sections as a function of $\pt^{\PgU}$ for $\abs{y^{\PgU}}<1.2$. The error bars represent the total uncertainties, including the systematic uncertainties from the measurement of the \upsn polarization~\cite{cms_ypol}, but not the uncertainty (4\%) in the integrated luminosity.}
  \label{fig:xsec_overlay_cmspol}
\end{figure}

\begin{figure}[htb]
  \centering
  \includegraphics[width=0.45\textwidth]{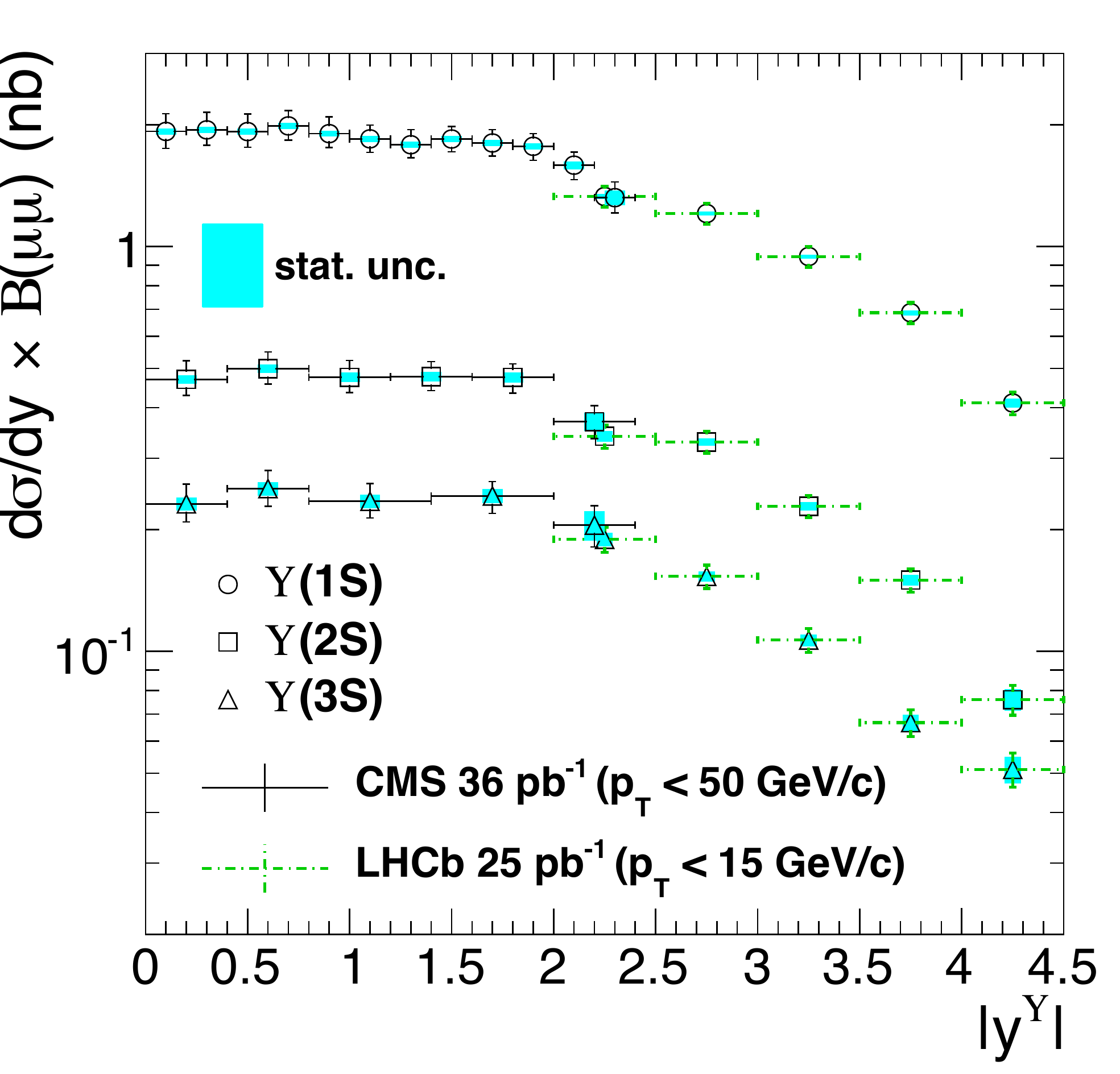}\label{fig:LHCb}
    \caption{Acceptance-corrected differential production cross sections as a function of rapidity, and comparison with LHCb results~\cite{bib-lhcb-paper}. The bands represent the statistical uncertainty and the error bars represent the total uncertainty, except for those from the \upsn polarization.}
  \label{fig:xsec_rapdiff_ns}
\end{figure}

\begin{figure*}[htbp]
  \centering
  \includegraphics[width=0.45\textwidth]{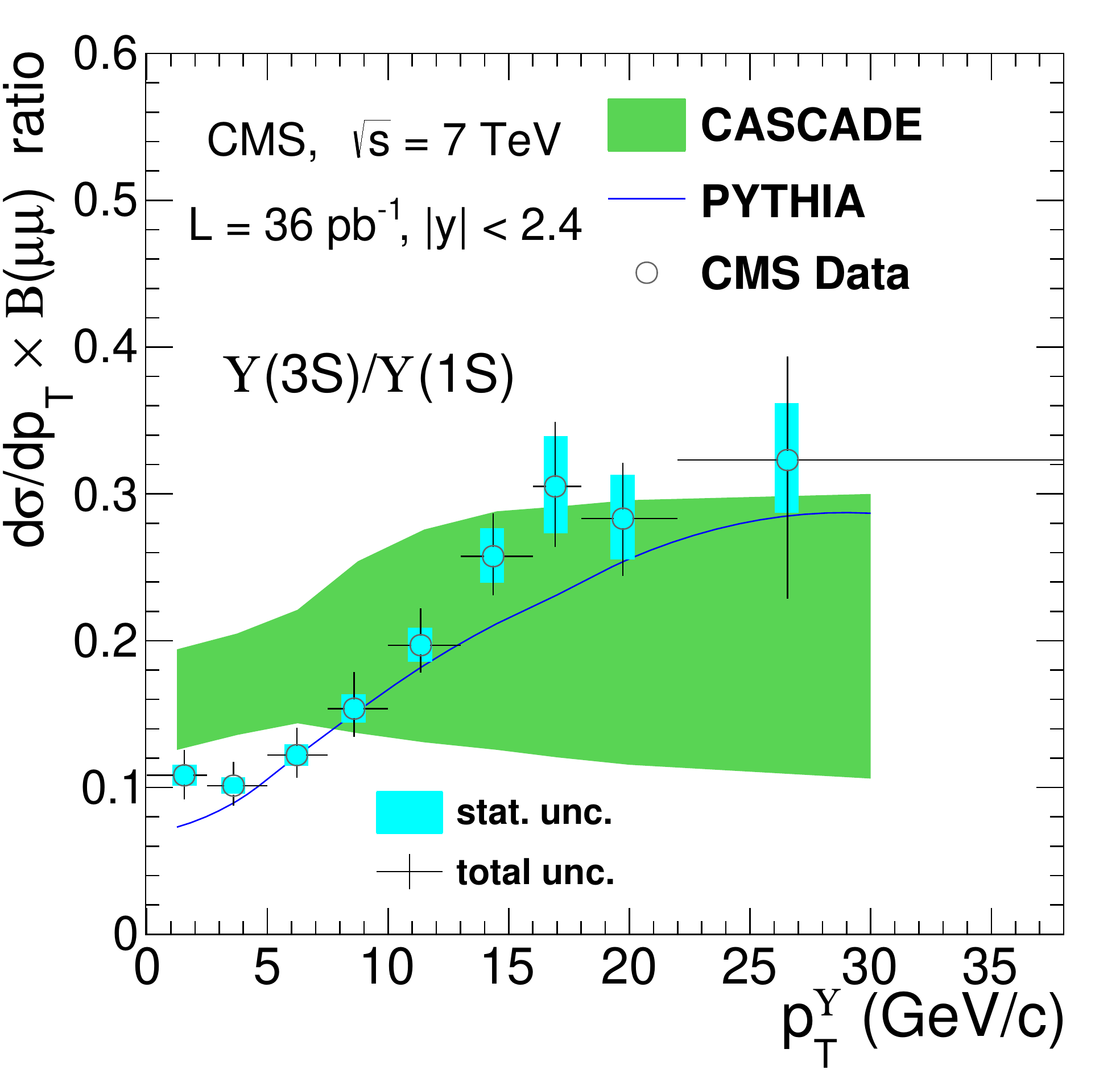}\label{fig:xsec_ratio_31}
  \includegraphics[width=0.45\textwidth]{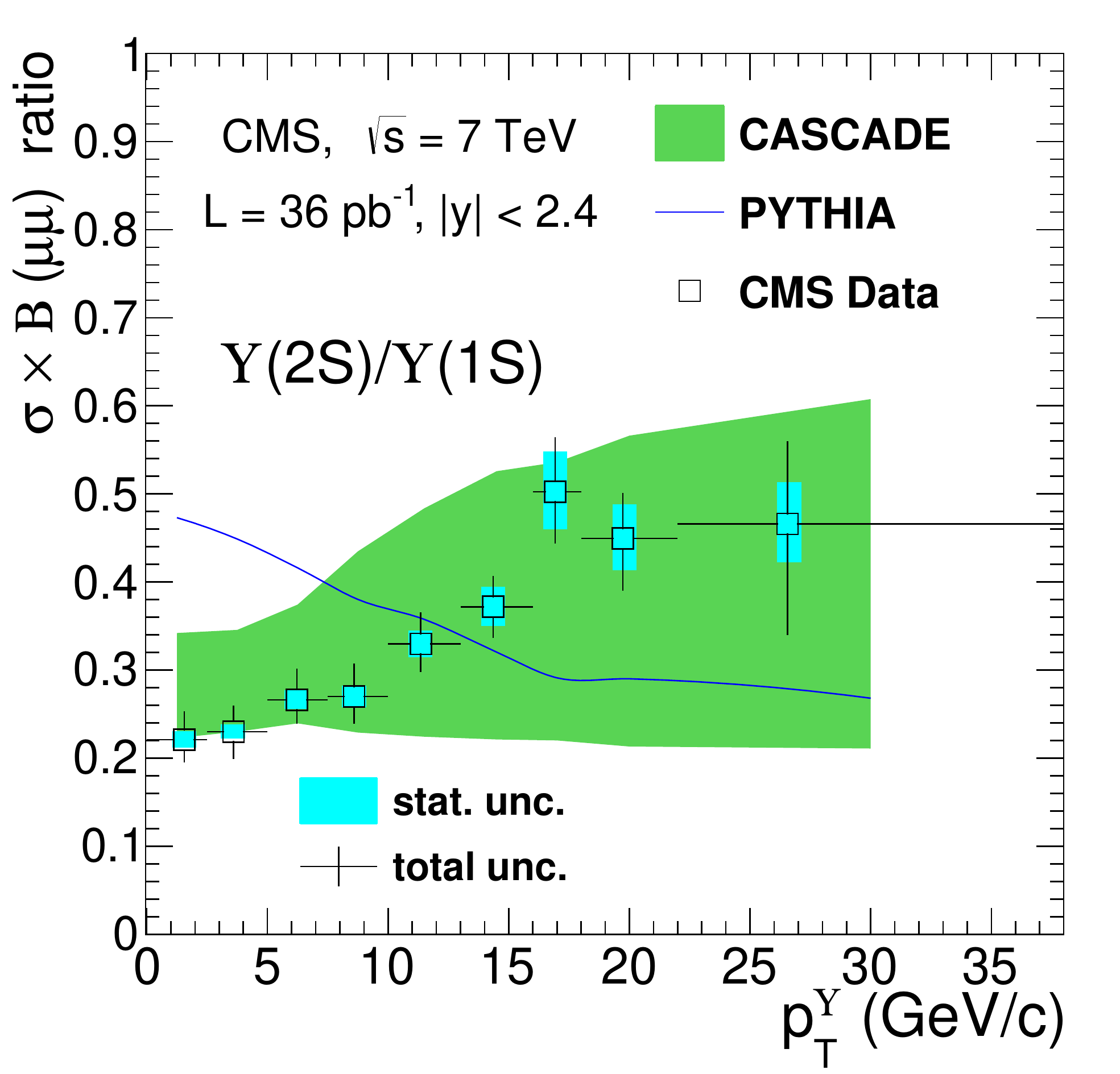}\label{fig:xsec_ratio_21}\\
  \includegraphics[width=0.45\textwidth]{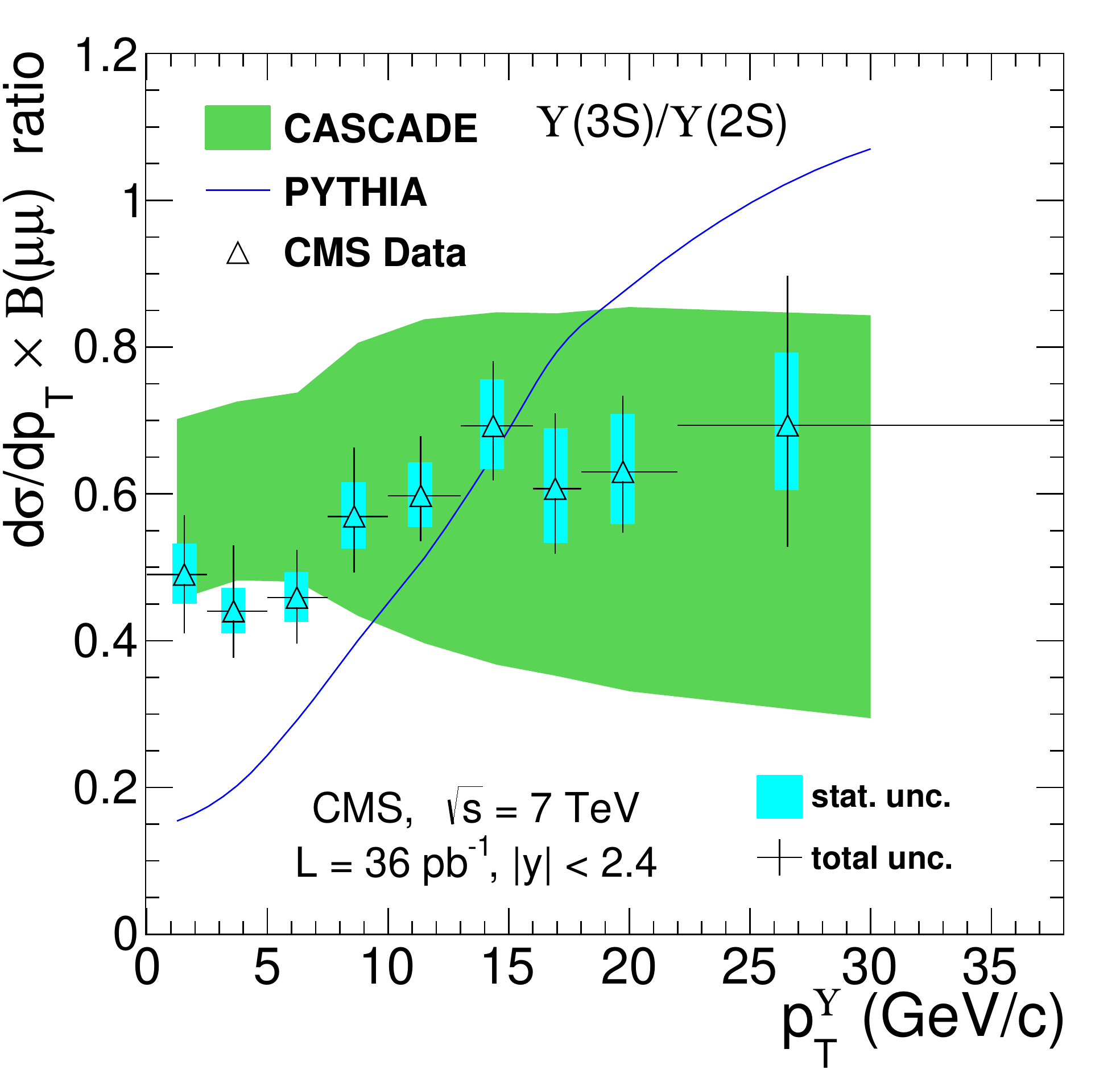}\label{fig:xsec_ratio_32}
  \caption{Ratios of acceptance-corrected differential cross sections as a function of $\pt^{\PgU}$ in the rapidity range $\abs{y^{\PgU}}<2.4$, along with predictions from \CASCADE (bands) and \PYTHIA (lines), for the $\PgUc/\PgUa$, $\PgUb/\PgUa$ and $\PgUc/\PgUb$. The width of a band indicates an estimate of the uncertainty in the prediction.}
  \label{fig:xsec_ratio}
\end{figure*}

The acceptance-corrected \upsn differential \pt cross sections for the rapidity range $\abs{y^{\PgU}}<2.4$ are plotted in Fig.~\ref{fig:xsec_overlay} and summarized in Table~\ref{table:cross section-y-ns}. Figure~\ref{fig:xsec_overlay_cmspol} shows the same for the ranges $10< \pt^{\PgU}<50\GeVc$, $\abs{y^{\PgU}}<1.2$ used in Ref.~\cite{cms_ypol} and includes the systematic uncertainties from the polarization measurement of Ref.~\cite{cms_ypol}, as explained in Section~\ref{systematics}. The \upsn differential \pt cross sections for six different rapidity bins are given in \suppMaterial. 
The $\pt^{\PgU}$ dependence of the cross section in the six exclusive rapidity intervals shows a similar trend within the uncertainties.
The \upsn \pt-integrated, differential rapidity cross section results are shown in Fig.~\ref{fig:xsec_rapdiff_ns}. Similar to the fiducial differential rapidity cross sections, the acceptance-corrected cross sections are approximately flat from $\abs{y^{\PgU}}=0$ to about 2.0, where they then begin to fall.
In Fig.~\ref{fig:xsec_rapdiff_ns}, a comparison with similar results from the LHCb Collaboration~\cite{bib-lhcb-paper} is also shown.
The two sets of measurements are complementary in their rapidity coverage and consistent within the uncertainties in the region of overlap.
The fiducial cross sections and the acceptance-corrected cross sections exhibit similar $\pt^{\PgU}$ and $\abs{y^{\PgU}}$ dependencies. However, the decrease in the cross section at large values of the rapidity is greater for the fiducial cross section than for the acceptance-corrected cross section because the acceptance also decreases with rapidity.
A comparison to the normalized differential \pt cross section results from CDF~\cite{bib-cdfups} and D0~\cite{bib-d0ups}, provided in \suppMaterial, indicates a harder spectrum at the LHC.  
Comparisons to results from ATLAS~\cite{yplb_ATLAS}, shown also in \suppMaterial, show good agreement. 
The ratios of the \upsn differential \pt cross sections are plotted in Fig.~\ref{fig:xsec_ratio}, along with comparisons to the \CASCADE and \PYTHIA predictions.
The ratios increase with $\pt^{\PgU}$, as they do for the fiducial cross sections.
The predictions for the ratios from \CASCADE have relatively large uncertainty bands;
this arises as a consequence of the asymmetric variation of the uncertainty of the predictions in Fig.~\ref{fig:CASCADE_fiducial} as a function of $\pt^{\PgU}$.
The \CASCADE prediction is consistent with the $\PgUb/\PgUa$ and $\PgUc/\PgUb$ measurements, while it disagrees with the $\PgUc/\PgUa$ results at low \pt. The \PYTHIA prediction agrees with the measured $\PgUc/\PgUa$ values, but is inconsistent with the $\PgUb/\PgUa$ and $\PgUc/\PgUb$ results.

\begin{figure*}[htbp]
  \centering
  \includegraphics[width=0.45\textwidth]{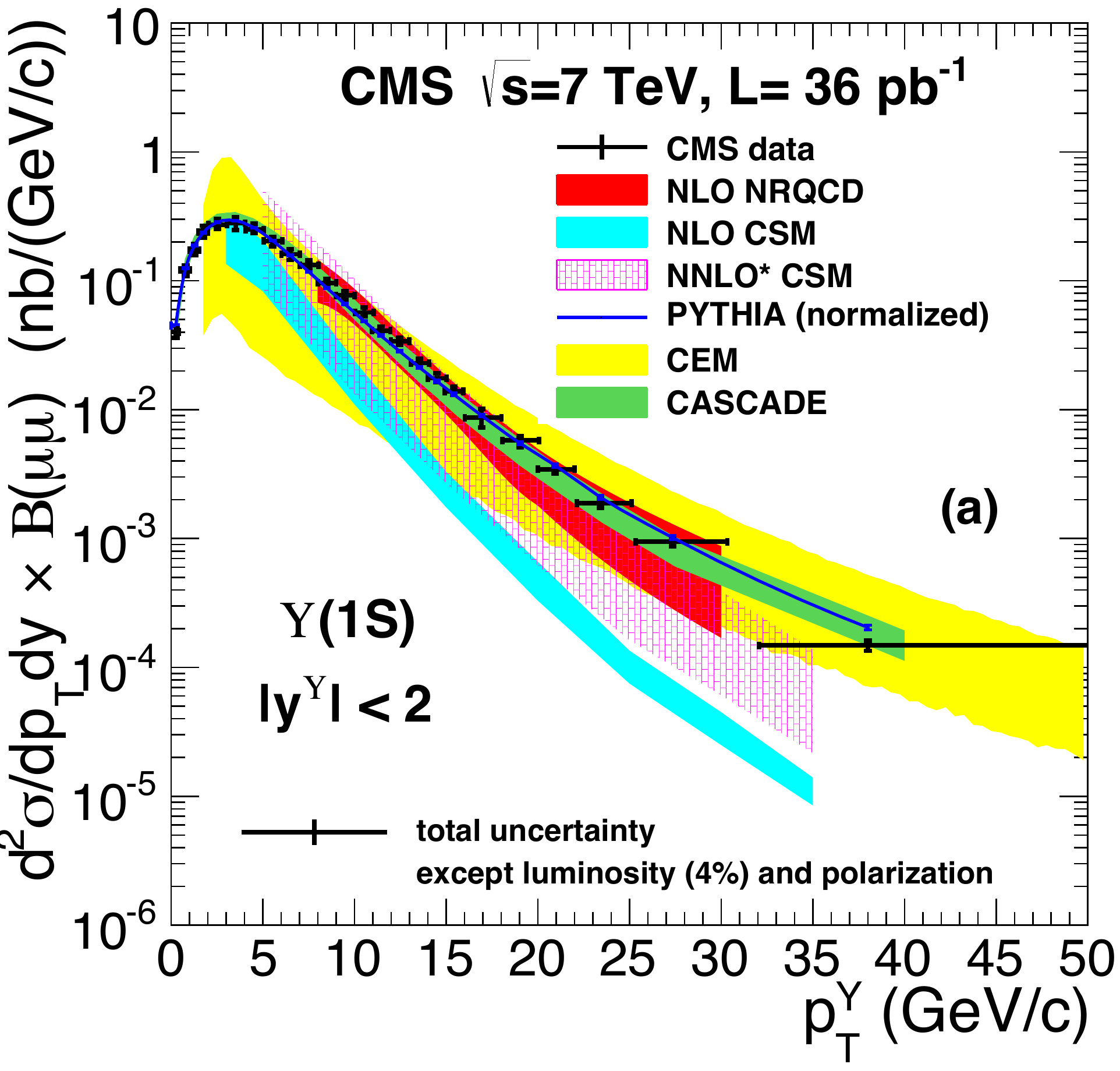}\label{fig:theoryPlot1Sscaled_36pb_y20}
  \includegraphics[width=0.45\textwidth]{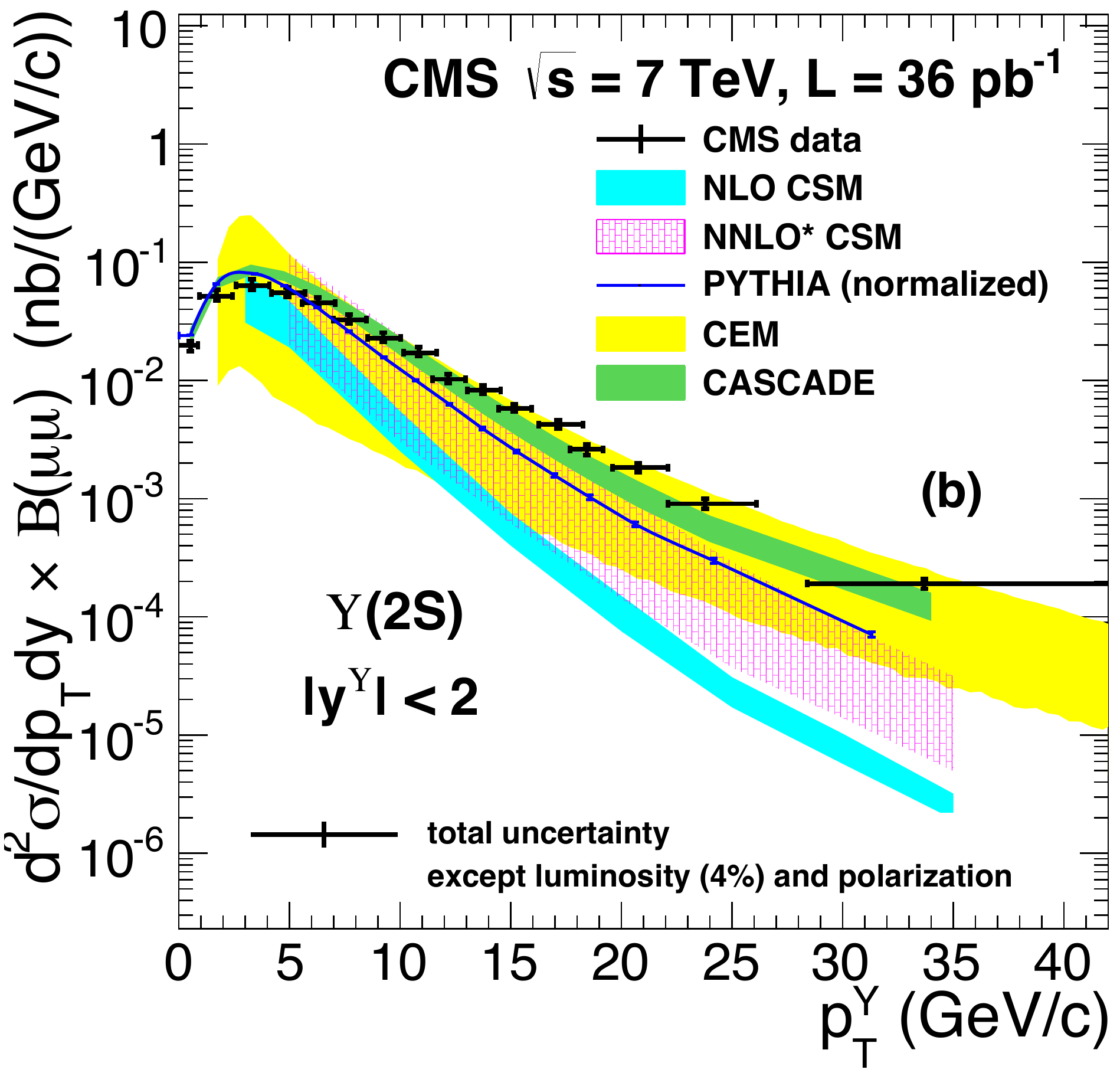}\label{fig:theoryPlot2Sscaled_36pb_y20}\\
  \includegraphics[width=0.45\textwidth]{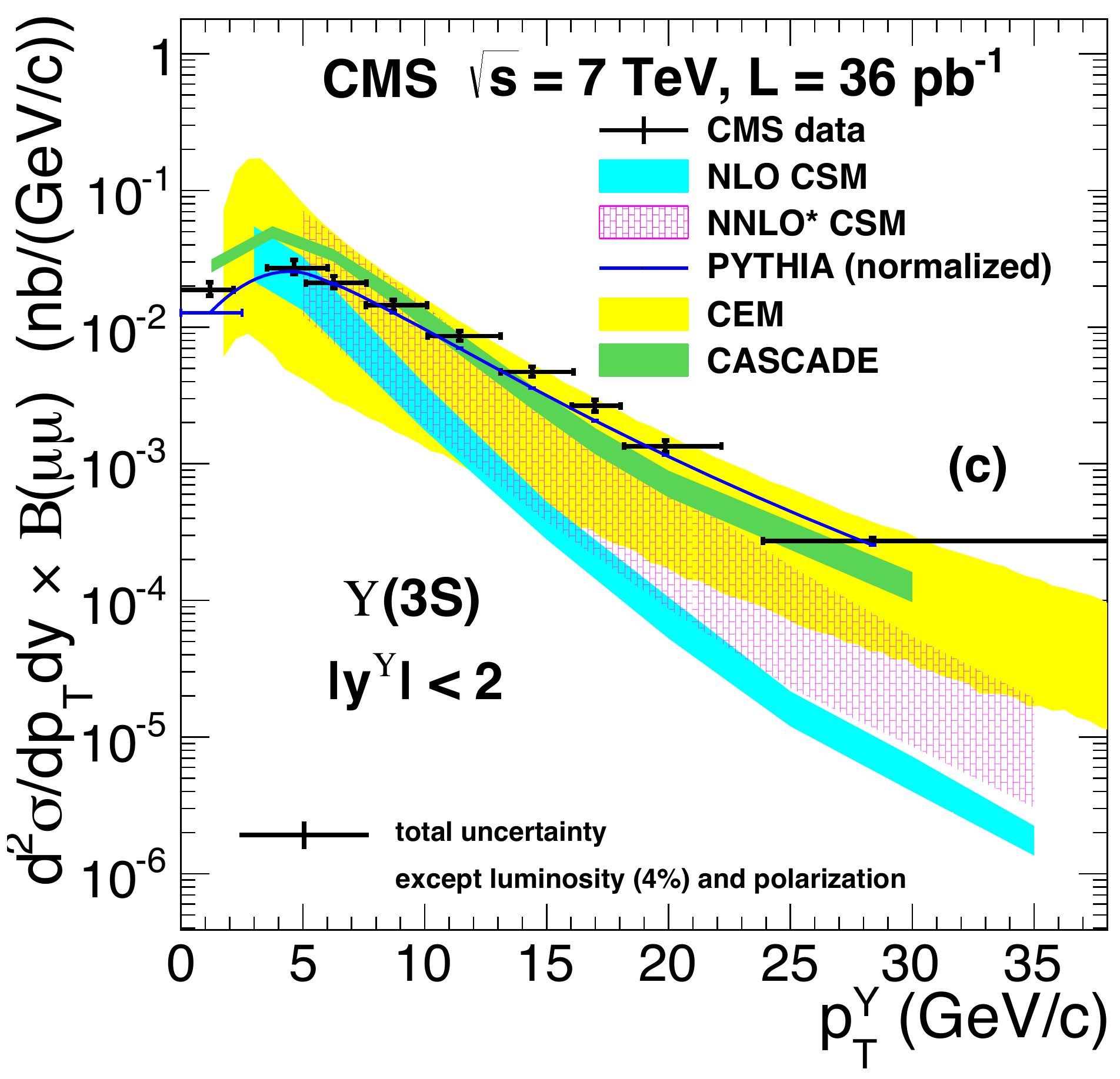}\label{fig:theoryPlot3Sscaled_36pb_y20}
  \includegraphics[width=0.45\textwidth]{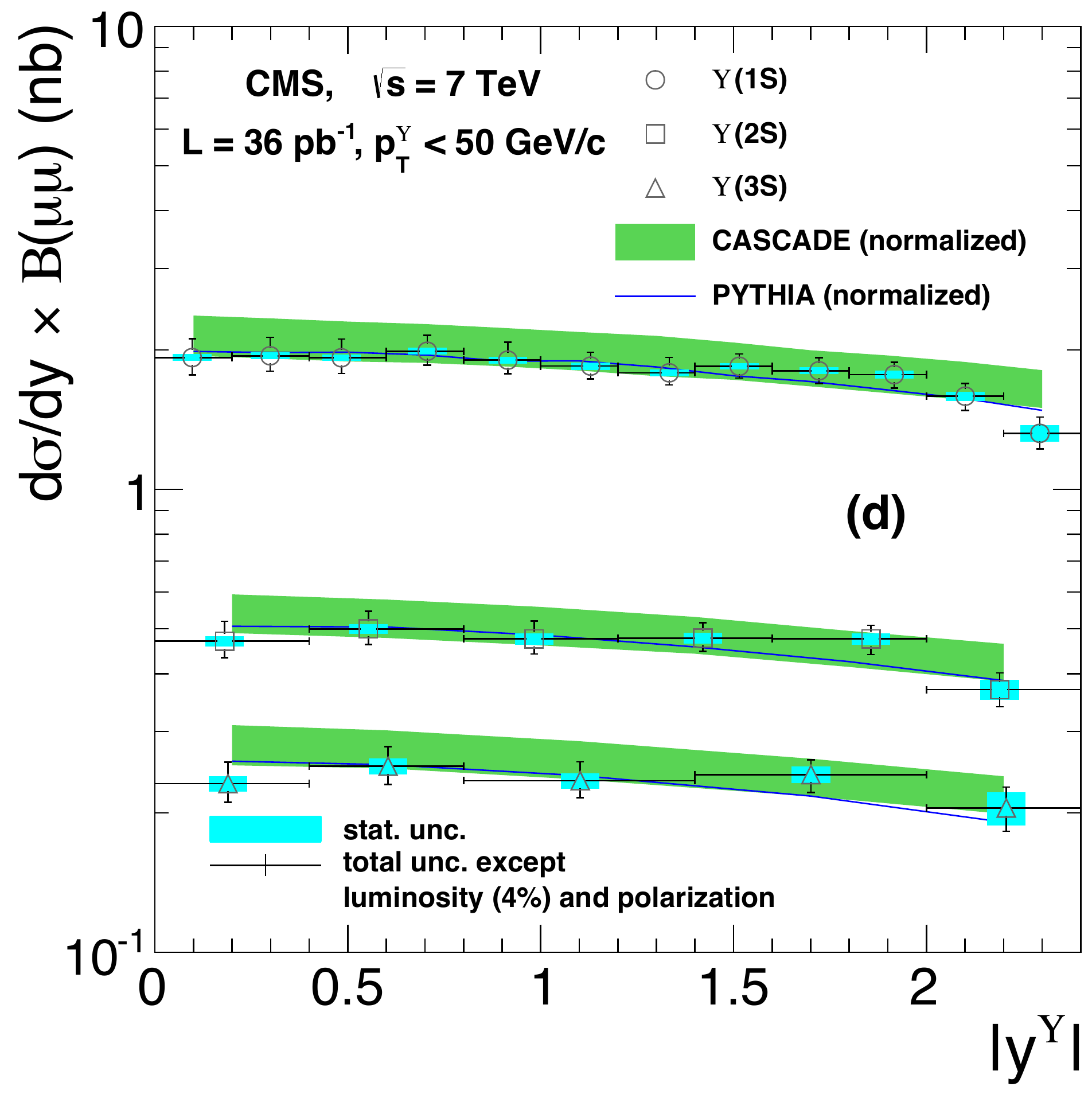}\label{fig:theory_y_full}
  \caption{Acceptance-corrected differential cross sections of (a) \PgUa, (b) \PgUb, and (c) \PgUc\ as a function of $\pt^{\PgU}$ in the rapidity range $\abs{y^{\PgU}}<2$, and comparison to various theoretical predictions. (d) Acceptance-corrected differential cross section of the \upsn as a function of rapidity and comparison to \CASCADE and \PYTHIA.
The \PYTHIA prediction is normalized to the measured total cross section, in order to facilitate the comparison of the shape of the dependences; for the rapidity differential results (d), the normalized \CASCADE prediction is also shown.
The width of a band indicates an estimate of the uncertainty in the prediction by the author of the prediction.}
  \label{fig:theory}
\end{figure*}

The acceptance-corrected differential \pt and rapidity \upsn cross sections and the theoretical predictions are shown in Fig.~\ref{fig:theory}. The measurements and predictions in Figs.~\ref{fig:theory}~(a,b,c) are for $\abs{y^{\PgU}}<2.0$ and assume unpolarized \upsn production.  Comparisons are made to the \CASCADE MC generator; the normalized \PYTHIA (as explained in Section~\ref{crosssection_noAcc_results});
the color-evaporation model (CEM)~\cite{Frawley:2008kk} with feed-down not included; nonrelativistic QCD (NRQCD) to next-to-leading order (NLO) including feed-down, as described in Ref.~\cite{bib-nrqcd3}; the color-singlet model (CSM) to NLO and NNLO*~\cite{bib-LansbergNNLO},
with feed-down accounted for by scaling the \PgUa\ and \PgUb\ direct-production cross sections by factors 2 and 1.43, respectively~\cite{bib-LansbergNNLO}, and no feed-down for the \PgUc.
The theoretical predictions are based on published models for 
\upsn production, and, except for NRQCD~\cite{bib-nrqcd3}, are made for lower $\sqrt{s}$~\cite{bib-CASCADE,Frawley:2008kk,bib-LansbergNNLO}. These models have been updated by their respective authors to $\sqrt{s} = 7\TeV$ when relevant. The updates are unpublished and are in the form of private communications.
Our measured \PgUa\ cross section is in good agreement with NRQCD, for the prediction provided for \pt in 8--30 \GeVc .
The CEM predictions for the three states are, within their uncertainties, also compatible with the data.
The data agree with \CASCADE for the \PgUa\ and \PgUb, but the agreement is not as satisfactory for the \PgUc\ when judged on the basis of the smaller uncertainties quoted by this prediction. 
The NLO CSM does not describe the data, while the NNLO* CSM shows improved agreement within the large uncertainties.
The total cross section predicted by \PYTHIA is higher than the measured cross section by about a factor 2; in Fig.~\ref{fig:theory}, the \PYTHIA predictions are for this reason normalized to the measured \upsn cross sections.
The \pt dependence of the cross section predicted by \PYTHIA agrees with the data for the \PgUa\ and \PgUc\, but not for the \PgUb. \CASCADE and \PYTHIA also describe the rapidity dependence over the range of the measurement, as shown in Fig.~\ref{fig:theory}~(d). Complete tables of results for the differential cross sections for the three \ups states are available in \suppMaterial, including variations for extreme polarization scenarios. 

\section{Summary}
\label{sec:summary}

Measurements of the \upsn differential and total production cross sections from proton-proton collisions at $\sqrt{s} = 7$\TeV with the CMS detector have been presented. The results have been shown in two ways: as acceptance-corrected cross sections, and fiducial cross sections in which both muons from the \upsn decay are within the detector acceptance. The latter cross sections are independent of the assumed \upsn polarizations. The differential cross sections have been given as a function of $\pt^{\PgU}$ and $\abs{y^{\PgU}}$, and compared to theoretical predictions.
The differential cross sections as a function of $\pt^{\PgU}$ and $y^{\PgU}$ for each \upsn state have also been measured and compared to theoretical predictions.
Finally, the \ups\ cross section ratios have been given. The dominant sources of systematic uncertainty in the cross section measurements
arise from the determination of the muon identification and trigger
efficiencies, and the integrated luminosity.

The measurements are consistent with previous CMS results based on less than 10\% of the integrated luminosity analyzed here. These earlier measurements have been extended in terms of both the precision attained and the kinematic reach.
In addition, this paper expands upon the previous result by the inclusion of fiducial cross section measurements and the polarization systematics, utilizing the recent \ups polarization results from CMS.
The results are compared to the ATLAS and LHCb Collaborations' measurements, and are found to be consistent in the regions of overlap.
Comparisons to measurements by the CDF, D0, and LHCb Collaborations also illustrate the achieved extension in kinematic coverage. 
The results presented here will allow for a more precise determination of the parameters of the various bottomonium production models.

\section*{Acknowledgements}
We congratulate our colleagues in the CERN accelerator departments for the excellent performance of the LHC and thank the technical and administrative staffs at CERN and at other CMS institutes for their contributions to the success of the CMS effort. In addition, we gratefully acknowledge the computing centres and personnel of the Worldwide LHC Computing Grid for delivering so effectively the computing infrastructure essential to our analyses. Finally, we acknowledge the enduring support for the construction and operation of the LHC and the CMS detector provided by the following funding agencies: BMWF and FWF (Austria); FNRS and FWO (Belgium); CNPq, CAPES, FAPERJ, and FAPESP (Brazil); MEYS (Bulgaria); CERN; CAS, MoST, and NSFC (China); COLCIENCIAS (Colombia); MSES (Croatia); RPF (Cyprus); MoER, SF0690030s09 and ERDF (Estonia); Academy of Finland, MEC, and HIP (Finland); CEA and CNRS/IN2P3 (France); BMBF, DFG, and HGF (Germany); GSRT (Greece); OTKA and NKTH (Hungary); DAE and DST (India); IPM (Iran); SFI (Ireland); INFN (Italy); NRF and WCU (Republic of Korea); LAS (Lithuania); CINVESTAV, CONACYT, SEP, and UASLP-FAI (Mexico); MSI (New Zealand); PAEC (Pakistan); MSHE and NSC (Poland); FCT (Portugal); JINR (Armenia, Belarus, Georgia, Ukraine, Uzbekistan); MON, RosAtom, RAS and RFBR (Russia); MSTD (Serbia); SEIDI and CPAN (Spain); Swiss Funding Agencies (Switzerland); NSC (Taipei); ThEPCenter, IPST and NSTDA (Thailand); TUBITAK and TAEK (Turkey); NASU (Ukraine); STFC (United Kingdom); DOE and NSF (USA). Individuals have received support from the Marie-Curie programme and the European Research Council and EPLANET (European Union); the Leventis Foundation; the A. P. Sloan Foundation; the Alexander von Humboldt Foundation; the Belgian Federal Science Policy Office; the Fonds pour la Formation \`a la Recherche dans l'Industrie et dans l'Agriculture (FRIA-Belgium); the Agentschap voor Innovatie door Wetenschap en Technologie (IWT-Belgium); the Ministry of Education, Youth and Sports (MEYS) of Czech Republic; the Council of Science and Industrial Research, India; the Compagnia di San Paolo (Torino); and the HOMING PLUS programme of Foundation for Polish Science, cofinanced from European Union, Regional Development Fund.

\bibliography{auto_generated}   

\providecommand{\href}[2]{#2}\begingroup\raggedright\begin{thebibliography}{10}%
\makeatletter
\providecommand{\hrefCMSnoop }[0]{\@secondoftwo}%
\makeatother
\providecommand{\doi}{\texttt{doi:}\begingroup \urlstyle{tt}\Url}

\bibitem{yellow}
\hrefCMSnoop {} {N.~Brambilla {et~al.}, ``{Heavy quarkonium: progress, puzzles,
  and opportunities}'',} \textit{ Eur. Phys. J. C} \textbf{ 71} (2011) 1534,
  \href{http://dx.doi.org/10.1140/epjc/s10052-010-1534-9}{\doi{10.1140/epjc/s10052-010-1534-9}},
\href{http://www.arXiv.org/abs/1010.5827}{\texttt{ arXiv:1010.5827}}.

\bibitem{bib-cdfups}
\hrefCMSnoop {} {{ CDF} Collaboration, ``{Upsilon production and polarization
  in $p\bar{p}$ collisions at $\sqrt{s}=$ 1.8-TeV}'',} \textit{ Phys. Rev.
  Lett.} \textbf{ 88} (2002) 161802,
\href{http://dx.doi.org/10.1103/PhysRevLett.88.161802}{\doi{10.1103/PhysRevLett.88.161802}}.

\bibitem{bib-d0ups}
\hrefCMSnoop {} {{ D0} Collaboration, ``{Measurement of inclusive differential
  cross sections for $\Upsilon$(1S) production in $p \bar{p}$ collisions at
  $\sqrt{s}$ = 1.96-TeV}'',} \textit{ Phys. Rev. Lett.} \textbf{ 94} (2005)
  232001,
  \href{http://dx.doi.org/10.1103/PhysRevLett.100.049902}{\doi{10.1103/PhysRevLett.100.049902}},
  \href{http://www.arXiv.org/abs/hep-ex/0502030}{\texttt{
  arXiv:hep-ex/0502030}}.
And \textit{Erratum, Phys. Rev. Lett.} {\textbf{100}} (2008) 049902,
  \doi{10.1103/PhysRevLett.100.049902}.

\bibitem{bib-lhcb-paper}
\hrefCMSnoop {} {{ LHCb} Collaboration, ``{Measurement of Upsilon production in
  $\Pp\Pp$ collisions at $\sqrt{s}=7\TeV$}'',} \textit{ Eur. Phys. J. C}
  \textbf{ 72} (2012) 2025,
  \href{http://dx.doi.org/10.1140/epjc/s10052-012-2025-y}{\doi{10.1140/epjc/s10052-012-2025-y}},
\href{http://www.arXiv.org/abs/1202.6579}{\texttt{ arXiv:1202.6579}}.

\bibitem{yplb_ATLAS}
\hrefCMSnoop {} {{ ATLAS} Collaboration, ``{Measurement of Upsilon production
  in 7 TeV pp collisions at ATLAS}'',} \textit{ Phys. Rev. D} \textbf{ 87}
  (2013) 052004,
  \href{http://dx.doi.org/10.1103/PhysRevD.87.052004}{\doi{10.1103/PhysRevD.87.052004}},
\href{http://www.arXiv.org/abs/1211.7255}{\texttt{ arXiv:1211.7255}}.

\bibitem{yprd10}
\hrefCMSnoop {} {{ CMS} Collaboration, ``{Upsilon production cross section in
  $\Pp\Pp$ collisions at $\sqrt{s}=7\TeV$}'',} \textit{ Phys. Rev. D} \textbf{
  83} (2011) 112004,
  \href{http://dx.doi.org/10.1103/PhysRevD.83.112004}{\doi{10.1103/PhysRevD.83.112004}},
\href{http://www.arXiv.org/abs/1012.5545}{\texttt{ arXiv:1012.5545}}.

\bibitem{cms_ypol}
\hrefCMSnoop {} {{ CMS} Collaboration, ``{Measurement of the $\PgUa$, $\PgUb$
  and $\PgUc$ polarizations in $\Pp\Pp$ collisions at $\sqrt{s}=7\TeV$}'',}
  \textit{ Phys. Rev. Lett.} \textbf{ 110} (2013) 081802,
  \href{http://dx.doi.org/10.1103/PhysRevLett.110.081802}{\doi{10.1103/PhysRevLett.110.081802}},
  \href{http://www.arXiv.org/abs/1209.2922}{\texttt{ arXiv:1209.2922}}.

\bibitem{JINST}
\hrefCMSnoop {} {{ CMS} Collaboration, ``{The CMS experiment at the CERN
  LHC}'',} \textit{ JINST} \textbf{ 3} (2008) S08004,
  \href{http://dx.doi.org/10.1088/1748-0221/3/08/S08004}{\doi{10.1088/1748-0221/3/08/S08004}}.

\bibitem{bib-PYTHIA}
\hrefCMSnoop {} {T.~Sj{\"o}strand, S.~Mrenna, and P.~Z. Skands, ``{$\PYTHIA$}
  6.4 physics and manual'',} \textit{ JHEP} \textbf{ 5} (2006) 026,
  \href{http://dx.doi.org/10.1088/1126-6708/2006/05/026}{\doi{10.1088/1126-6708/2006/05/026}},
  \href{http://www.arXiv.org/abs/hep-ph/0603175}{\texttt{
  arXiv:hep-ph/0603175}}.

\bibitem{bib-cdfPythia}
\hrefCMSnoop {} {M.~Kr{\"a}mer, ``Quarkonium production at high-energy
  colliders'',} \textit{ Prog. Part. Nucl. Phys.} \textbf{ 47} (2001) 141,
  \href{http://dx.doi.org/10.1016/S0146-6410(01)00154-5}{\doi{10.1016/S0146-6410(01)00154-5}},
  \href{http://www.arXiv.org/abs/hep-ph/0106120}{\texttt{
  arXiv:hep-ph/0106120}}.

\bibitem{bib-PYTHIA-tuning}
\href {http://cdsweb.cern.ch/record/1042611} {M.~Bargiotti and V.~Vagnoni,
  ``Heavy Quarkonia sector in $\PYTHIA$ 6.324 : tuning, validation and
  perspectives at LHC(b)'',} LHCB Note LHCb-2007-042, (2007).

\bibitem{bib-cteq6l1}
J.~Pumplin\hrefCMSnoop {} { {et~al.}, ``{New generation of parton distributions
  with uncertainties from global QCD analysis}'',} \textit{ JHEP} \textbf{ 07}
  (2002) 012,
  \href{http://dx.doi.org/10.1088/1126-6708/2002/07/012}{\doi{10.1088/1126-6708/2002/07/012}},
\href{http://www.arXiv.org/abs/hep-ph/0201195}{\texttt{ arXiv:hep-ph/0201195}}.

\bibitem{bib-photos1}
\hrefCMSnoop {} {E.~Barberio, B.~van Eijk, and Z.~W\c{a}s, ``{$\PHOTOS$---a
  universal Monte Carlo for QED radiative corrections in decays}'',} \textit{
  Comput. Phys. Commun.} \textbf{ 66} (1991) 115,
  \href{http://dx.doi.org/10.1016/0010-4655(91)90012-A}{\doi{10.1016/0010-4655(91)90012-A}}.

\bibitem{bib-photos2}
\hrefCMSnoop {} {E.~Barberio and Z.~W{\c a}s, ``{$\PHOTOS$---a universal Monte
  Carlo for QED radiative corrections: version 2.0}'',} \textit{ Comput. Phys.
  Commun.} \textbf{ 79} (1994) 291,
  \href{http://dx.doi.org/10.1016/0010-4655(94)90074-4}{\doi{10.1016/0010-4655(94)90074-4}}.

\bibitem{bib-GEANT4}
\hrefCMSnoop {} {S.~Agostinelli {et~al.}, ``GEANT4---a simulation toolkit'',}
  \textit{ Nucl. Instrum. Meth. A} \textbf{ 506} (2003) 250,
  \href{http://dx.doi.org/10.1016/S0168-9002(03)01368-8}{\doi{10.1016/S0168-9002(03)01368-8}}.

\bibitem{bib-crystalball}
\href {http://www.slac.stanford.edu/pubs/slacreports/slac-r-236.html} {M.~J.
  Oreglia, ``A study of the reactions $\psi^\prime \to \gamma \gamma \psi$''}.
\newblock PhD thesis, Stanford University, 1980.
\newblock {SLAC} Report {SLAC-R-236}, see Appendix {D}.

\bibitem{bib-pdg}
\hrefCMSnoop {} {{Particle Data Group}, J.~Beringer {et~al.}, ``{Review of
  Particle Physics}'',} \textit{ Phys. Rev. D} \textbf{ 86} (2012) 010001,
  \href{http://dx.doi.org/10.1103/PhysRevD.86.010001}{\doi{10.1103/PhysRevD.86.010001}}.

\bibitem{PbPb}
\hrefCMSnoop {} {{ CMS} Collaboration, ``{Observation of sequential \ups
  suppression in PbPb collisions}'',} \textit{ Phys. Rev. Lett.} \textbf{ 109}
  (2012) 222301,
  \href{http://dx.doi.org/10.1103/PhysRevLett.109.222301}{\doi{10.1103/PhysRevLett.109.222301}},
\href{http://www.arXiv.org/abs/1208.2826}{\texttt{ arXiv:1208.2826}}.

\bibitem{bib-evtgen}
\hrefCMSnoop {} {{D. J. Lange}, ``The {\EVTGEN} particle decay simulation
  package'',} \textit{ Nucl. Instrum. Meth. A} \textbf{ 462} (2001) 152,
  \href{http://dx.doi.org/10.1016/S0168-9002(01)00089-4}{\doi{10.1016/S0168-9002(01)00089-4}}.

\bibitem{bib:CS}
\hrefCMSnoop {} {J.~C. Collins and D.~E. Soper, ``{Angular Distribution of
  Dileptons in High-Energy Hadron Collisions}'',} \textit{ Phys. Rev. D}
  \textbf{ 16} (1977) 2219,
  \href{http://dx.doi.org/10.1103/PhysRevD.16.2219}{\doi{10.1103/PhysRevD.16.2219}}.

\bibitem{bib-trackingefficiency}
\href {http://cdsweb.cern.ch/record/1279139} {{ CMS} Collaboration,
  ``Measurement of Tracking Efficiency'',} CMS Physics Analysis Summary
  CMS-PAS-TRK-10-002, (2010).

\bibitem{bib-trackermomentum}
\href {http://cdsweb.cern.ch/record/1279137} {{ CMS} Collaboration,
  ``Measurement of Momentum Scale and Resolution using Low-mass Resonances and
  Cosmic Ray Muons'',} CMS Physics Analysis Summary CMS-PAS-TRK-10-004, (2010).

\bibitem{bib-lumi-pas}
\href {http://cdsweb.cern.ch/record/1279145} {{ CMS} Collaboration,
  ``Measurement of {CMS} Luminosity'',} CMS Physics Analysis Summary
  CMS-PAS-EWK-10-004, (2010).

\bibitem{bib-CASCADE}
H.~Jung\hrefCMSnoop {} { {et~al.}, ``{The CCFM Monte Carlo generator $\CASCADE$
  version 2.2.03}'',} \textit{ Eur. Phys. J. C} \textbf{ 70} (2010) 1237,
  \href{http://dx.doi.org/10.1140/epjc/s10052-010-1507-z}{\doi{10.1140/epjc/s10052-010-1507-z}},
\href{http://www.arXiv.org/abs/1008.0152}{\texttt{ arXiv:1008.0152}}.

\bibitem{PhysRevD.86.054015}
\hrefCMSnoop {} {S.~P. Baranov, ``Prompt \upsn production at the {LHC} in view
  of the $k_t$-factorization approach'',} \textit{ Phys. Rev. D} \textbf{ 86}
  (2012) 054015,
  \href{http://dx.doi.org/10.1103/PhysRevD.86.054015}{\doi{10.1103/PhysRevD.86.054015}}.

\bibitem{Frawley:2008kk}
\hrefCMSnoop {} {A.~D. Frawley, T.~Ullrich, and R.~Vogt, ``{Heavy flavor in
  heavy-ion collisions at RHIC and RHIC II}'',} \textit{ Phys. Rept.} \textbf{
  462} (2008) 125,
  \href{http://dx.doi.org/10.1016/j.physrep.2008.04.002}{\doi{10.1016/j.physrep.2008.04.002}},
  \href{http://www.arXiv.org/abs/0806.1013}{\texttt{ arXiv:0806.1013}}.
And private communication.

\bibitem{bib-nrqcd3}
\hrefCMSnoop {} {K.~Wang, Y.-Q. Ma, and K.-T. Chao, ``{$\PgUa$ prompt
  production at the Tevatron and LHC in nonrelativistic QCD}'',} \textit{ Phys.
  Rev. D} \textbf{ 85} (2012) 114003,
  \href{http://dx.doi.org/10.1103/PhysRevD.85.114003}{\doi{10.1103/PhysRevD.85.114003}},
  \href{http://www.arXiv.org/abs/1202.6012}{\texttt{ arXiv:1202.6012}}. And
  private communication.

\bibitem{bib-LansbergNNLO}
\hrefCMSnoop {} {J.~P. Lansberg, ``{$\JPsi$} production at $\sqrt{s}=1.96$ and
  7 {TeV}: Color-Singlet Model, {NNLO*} and polarisation'',} \textit{ J. Phys.
  G} \textbf{ 38} (2011) 124110,
  \href{http://dx.doi.org/10.1088/0954-3899/38/12/124110}{\doi{10.1088/0954-3899/38/12/124110}},
  \href{http://www.arXiv.org/abs/1107.0292}{\texttt{ arXiv:1107.0292}}. And
  private communication.

\end{thebibliography}\endgroup

\ifthenelse{\boolean{cms@external}}{}{
\clearpage
\appendix
\section{Supplementary Material\label{app:suppMat}}
\setcounter{table}{0}
\setcounter{figure}{0}
\renewcommand{\thetable}{\thesection.\arabic{table}}
\renewcommand{\thefigure}{\thesection.\arabic{figure}}

\begin{figure*}[!ht]
  \centering
  \includegraphics[width=1.0\textwidth]{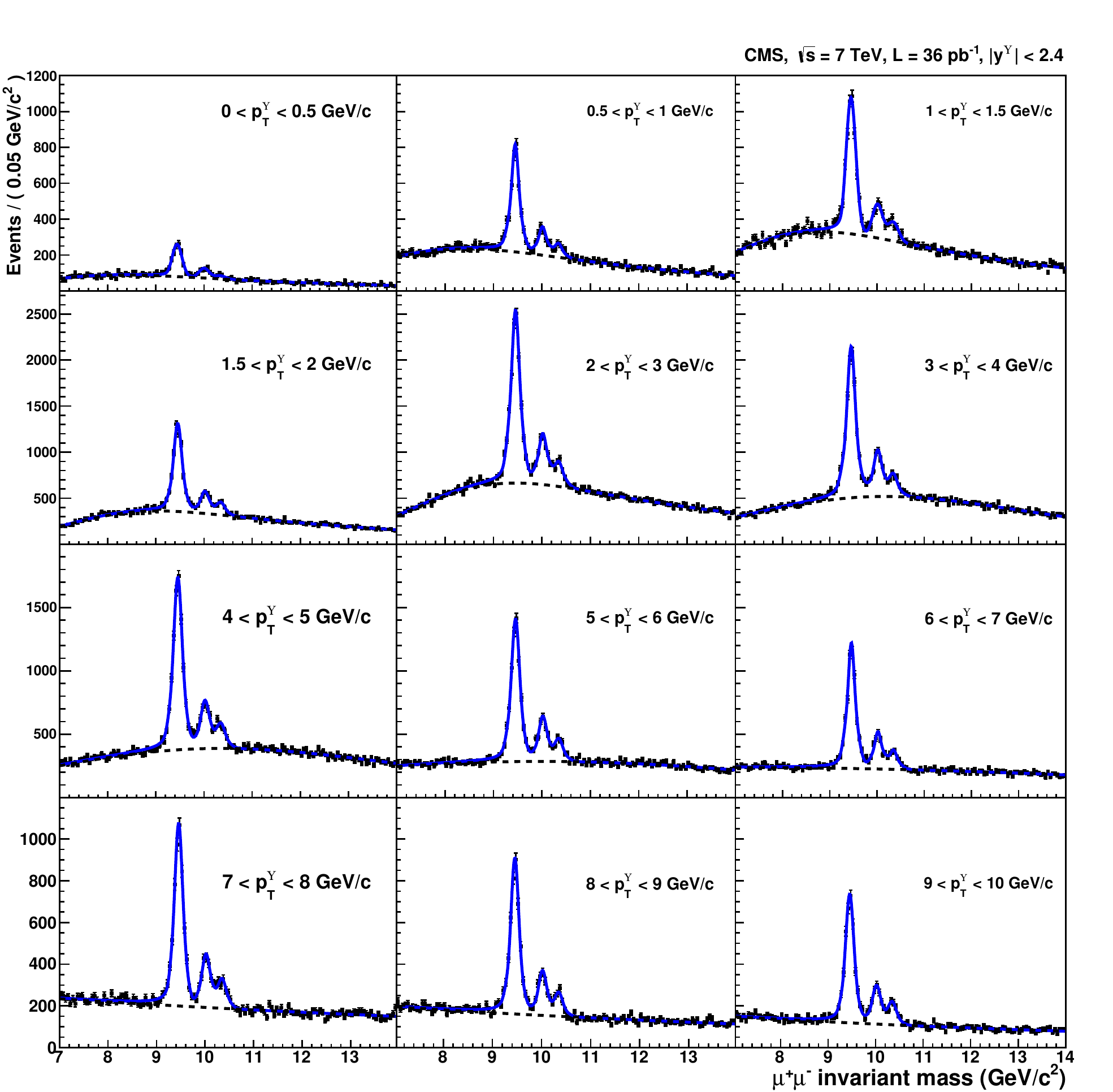}\label{fig:massfits_lowpt}
  \caption{Fit to the dimuon invariant-mass distribution in the specified \pt regions for $|y| < 2.4$, before accounting for acceptance and efficiency. The solid line shows the result of the fit described in the text, with the dashed line representing the background component.}
 \label{fig:massFits_low}
\end{figure*}

\clearpage

\begin{figure*}[!ht]
  \centering
  \includegraphics[width=1.0\textwidth]{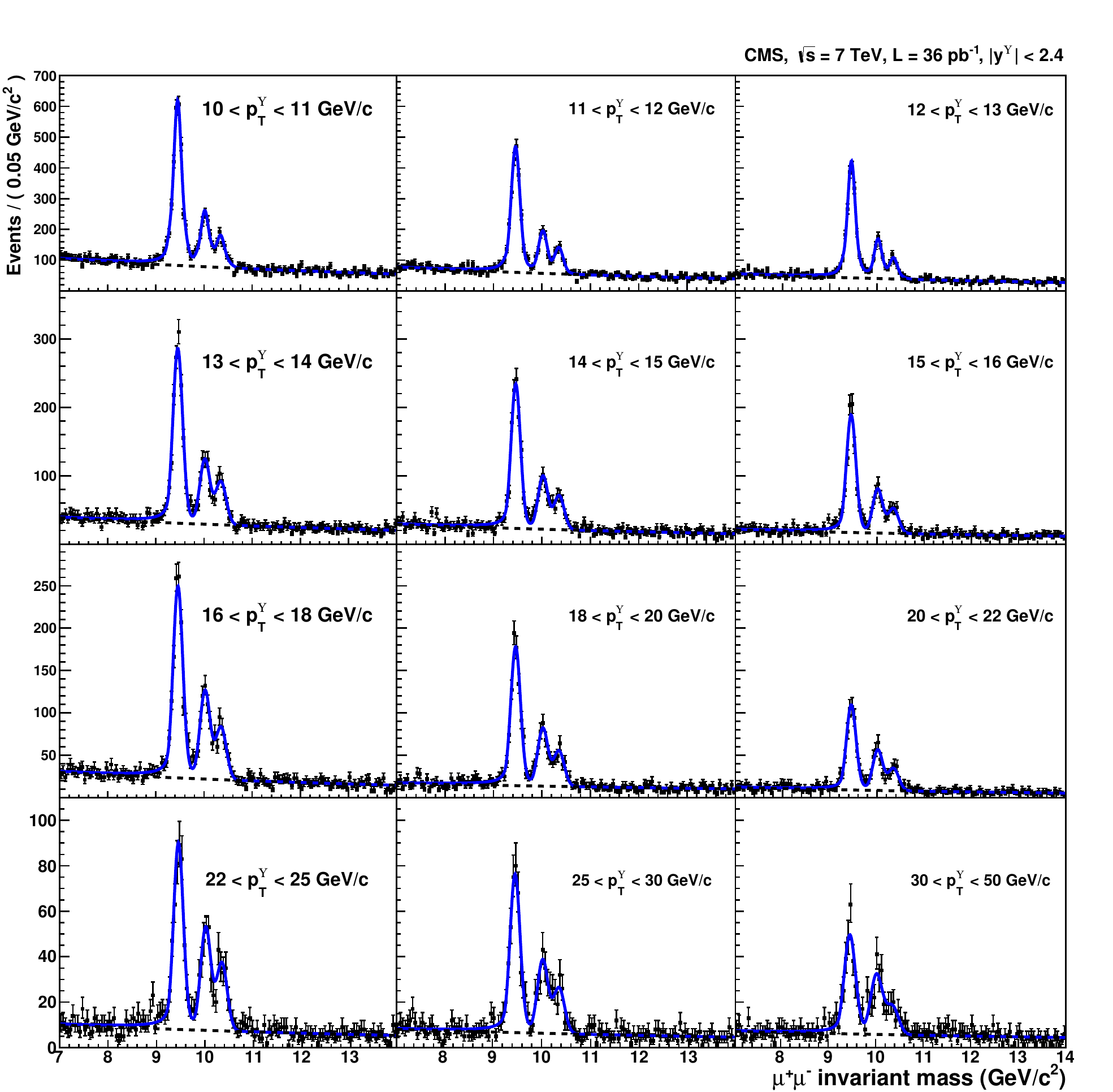}\label{fig:massfits_highpt}
  \caption{Fit to the dimuon invariant-mass distribution in the specified \pt regions for $|y| < 2.4$, before accounting for acceptance and efficiency. The solid line shows the result of the fit described in the text, with the dashed line representing the background component.}
 \label{fig:massFits_high}
\end{figure*}

\clearpage

\begin{figure}[htb]
  \centering
  \includegraphics[width=0.45\textwidth]{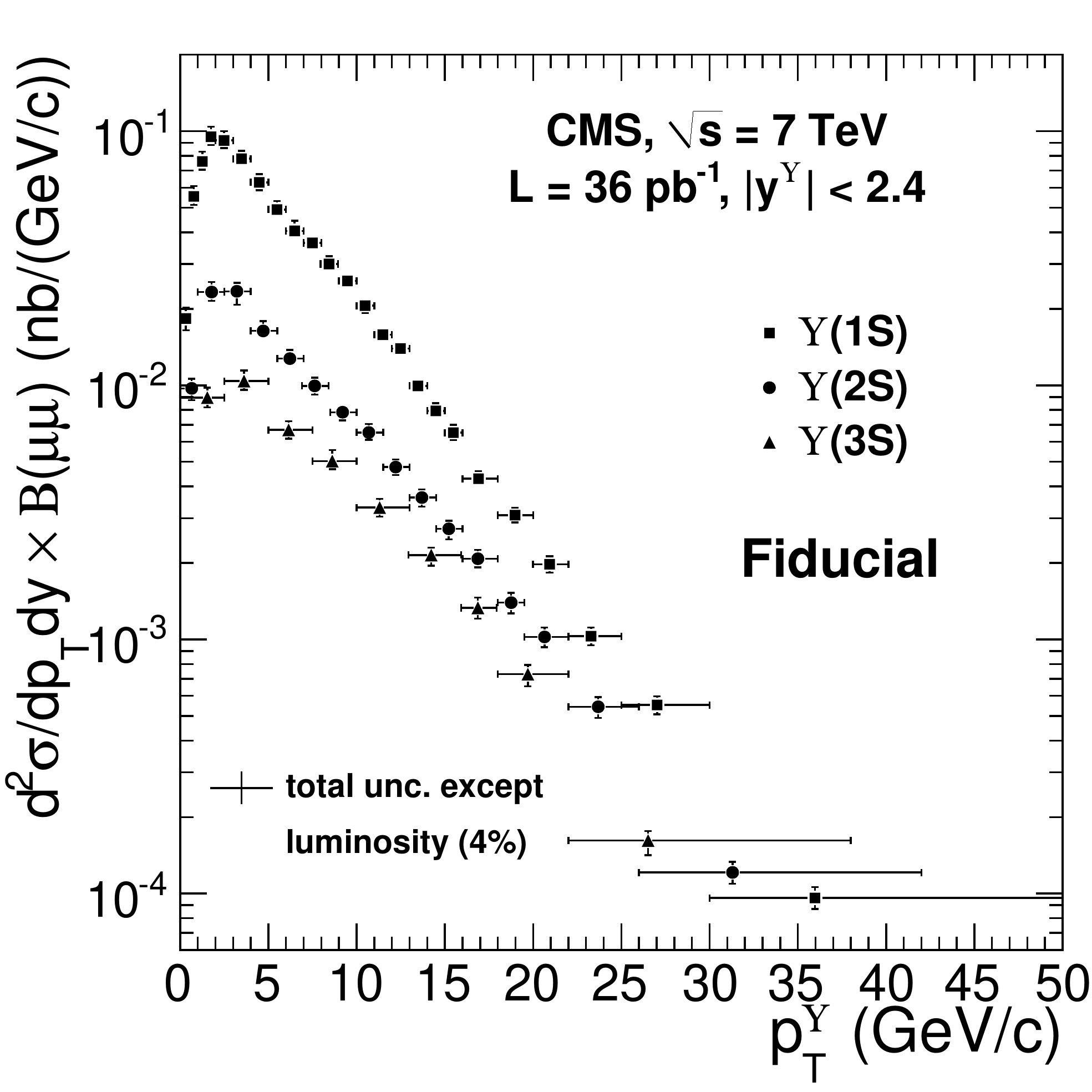}
  \caption{Measured \upsn differential fiducial cross sections as a function of $\pt^{\PgU}$ in the rapidity range $\abs{y^{\PgU}}<2.4$. The error bars indicate the total uncertainties, except for the 4\% uncertainty in the integrated luminosity.}
  \label{fig:xsec_overlay_noAcc}
\end{figure}

\begin{figure*}[htbp]
  \centering
  \includegraphics[width=0.45\textwidth]{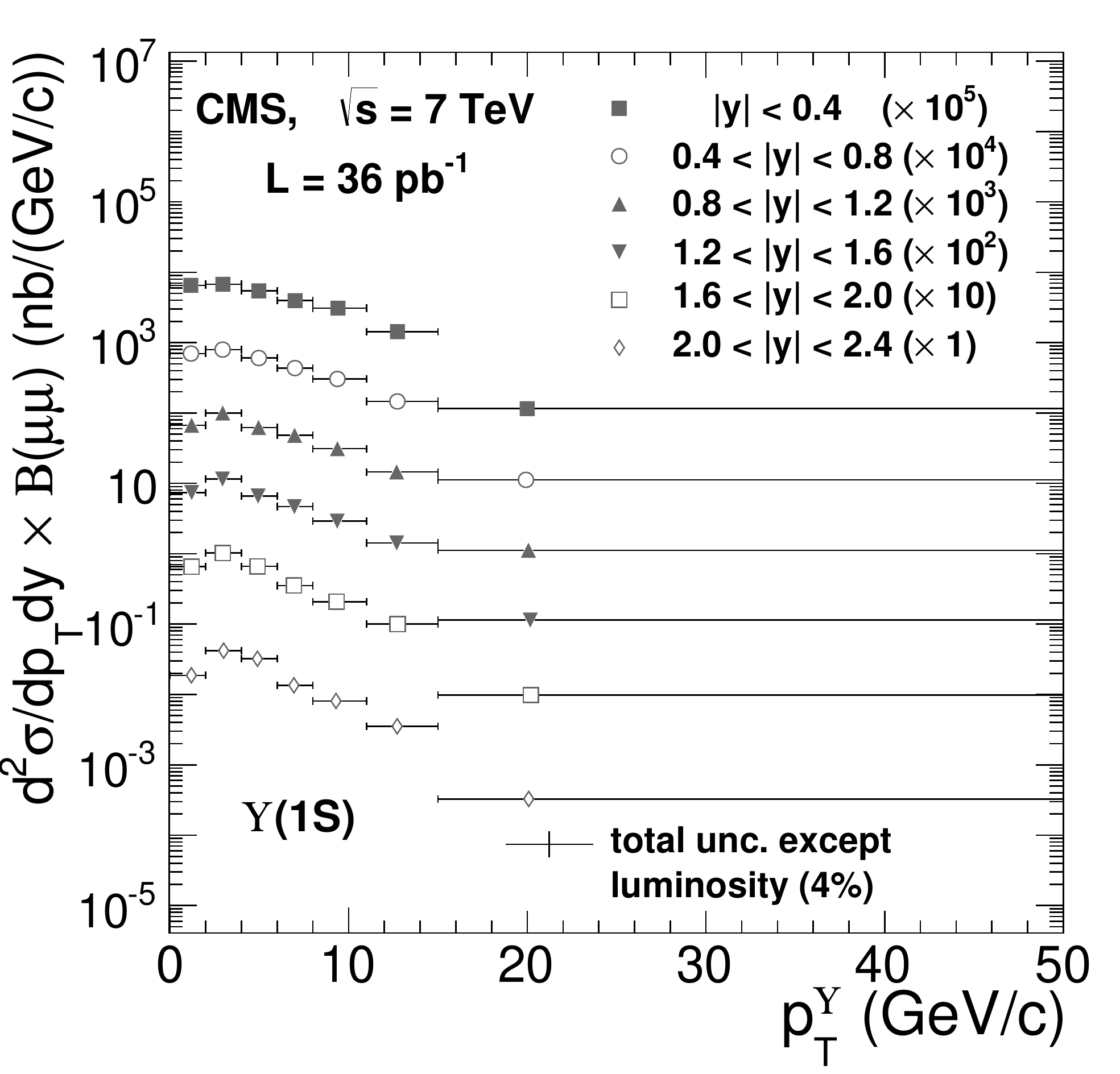}\label{fig:xsec_1s_6ybin_noAcc}\\
  \includegraphics[width=0.45\textwidth]{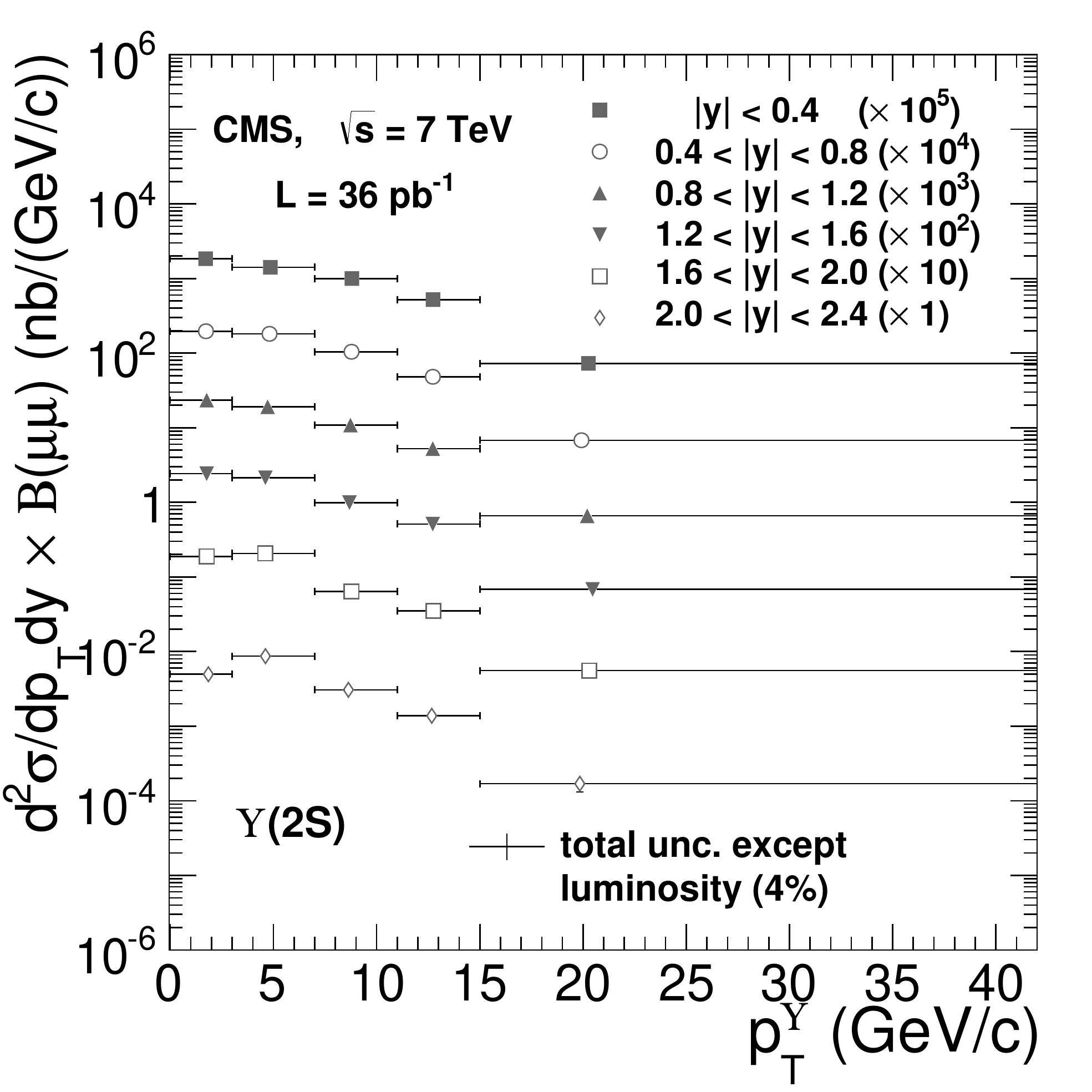}\label{fig:xsec_2s_6ybin_noAcc}
  \includegraphics[width=0.45\textwidth]{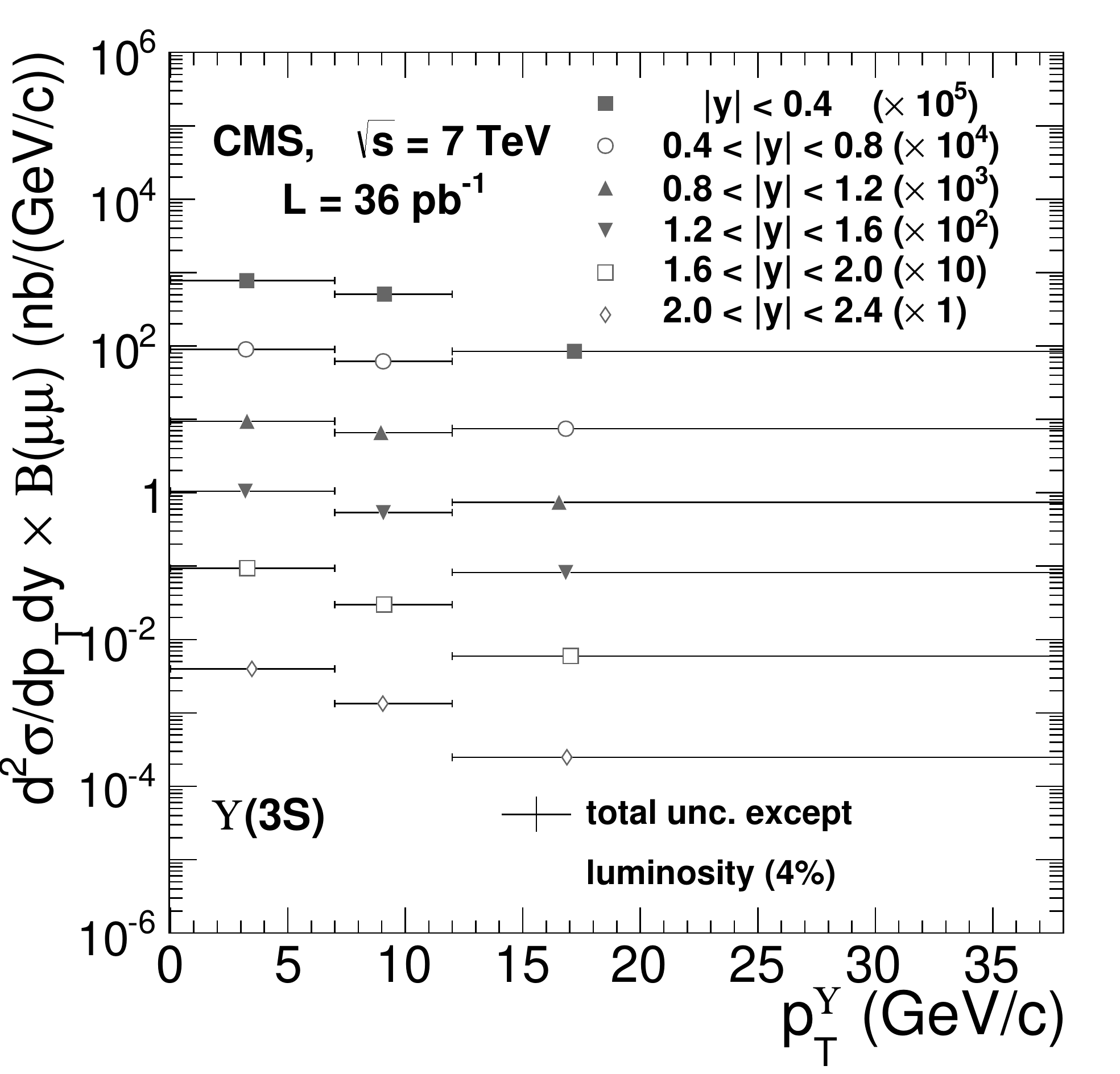}\label{fig:xsec_3s_6ybin_noAcc}
  \caption{The \PgU(nS) differential fiducial cross sections as a function of $\pt^{\PgU}$ in six rapidity ranges, scaled for clarity by the factors shown in the figures.}
  \label{fig:xsec_6ybin_noAcc}
\end{figure*}

\begin{figure}[htbp]
  \centering
  \includegraphics[width=0.45\textwidth]{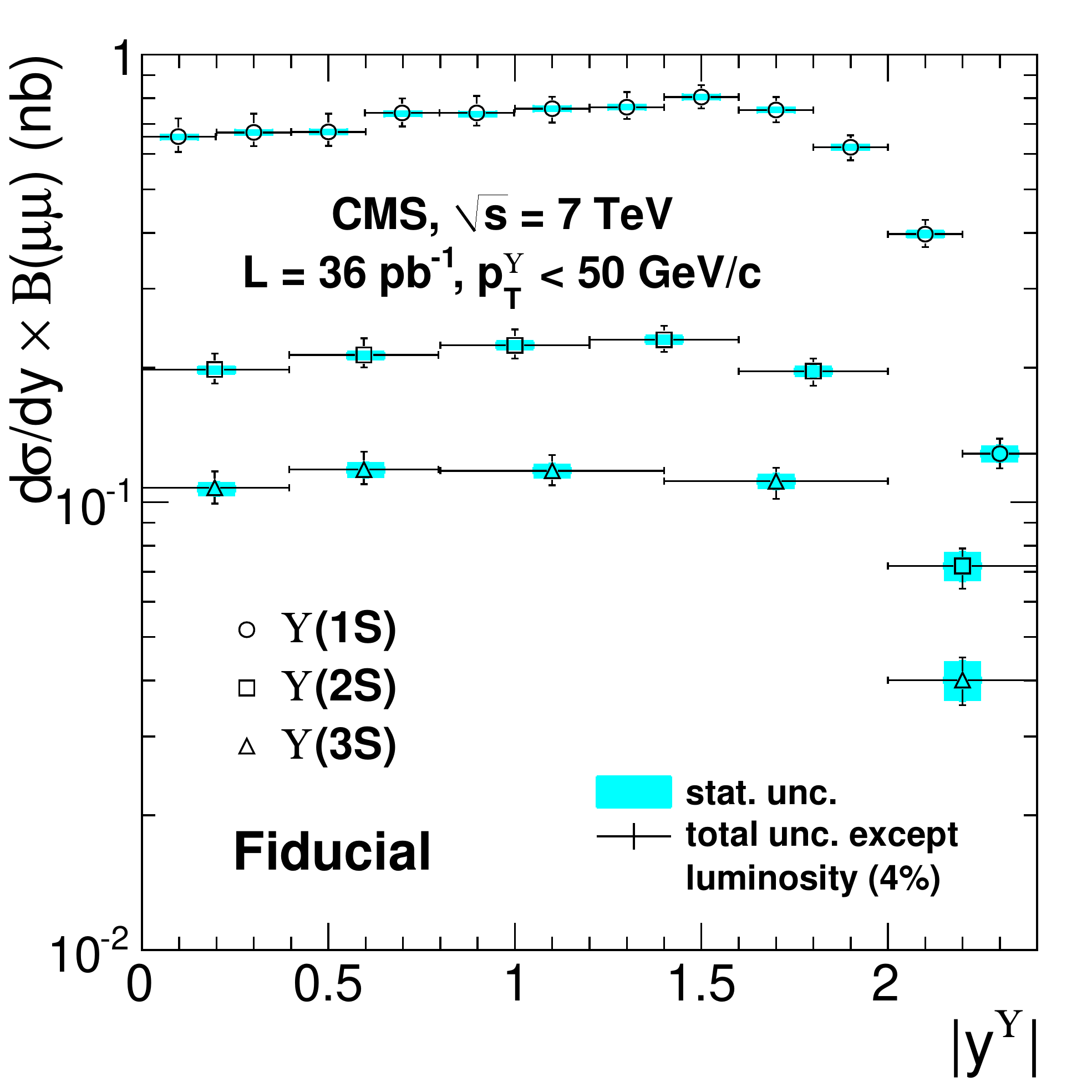}
    \caption{Differential fiducial cross sections of the \upsn as a function of rapidity. The regions show the statistical uncertainties and the error bars show the total uncertainties, except for the 4\% uncertainty in the integrated luminosity.}
  \label{fig:xsec_rapdiff_ns_noAcc}
\end{figure}

\begin{figure}[htb]
  \centering
  \includegraphics[width=0.45\textwidth]{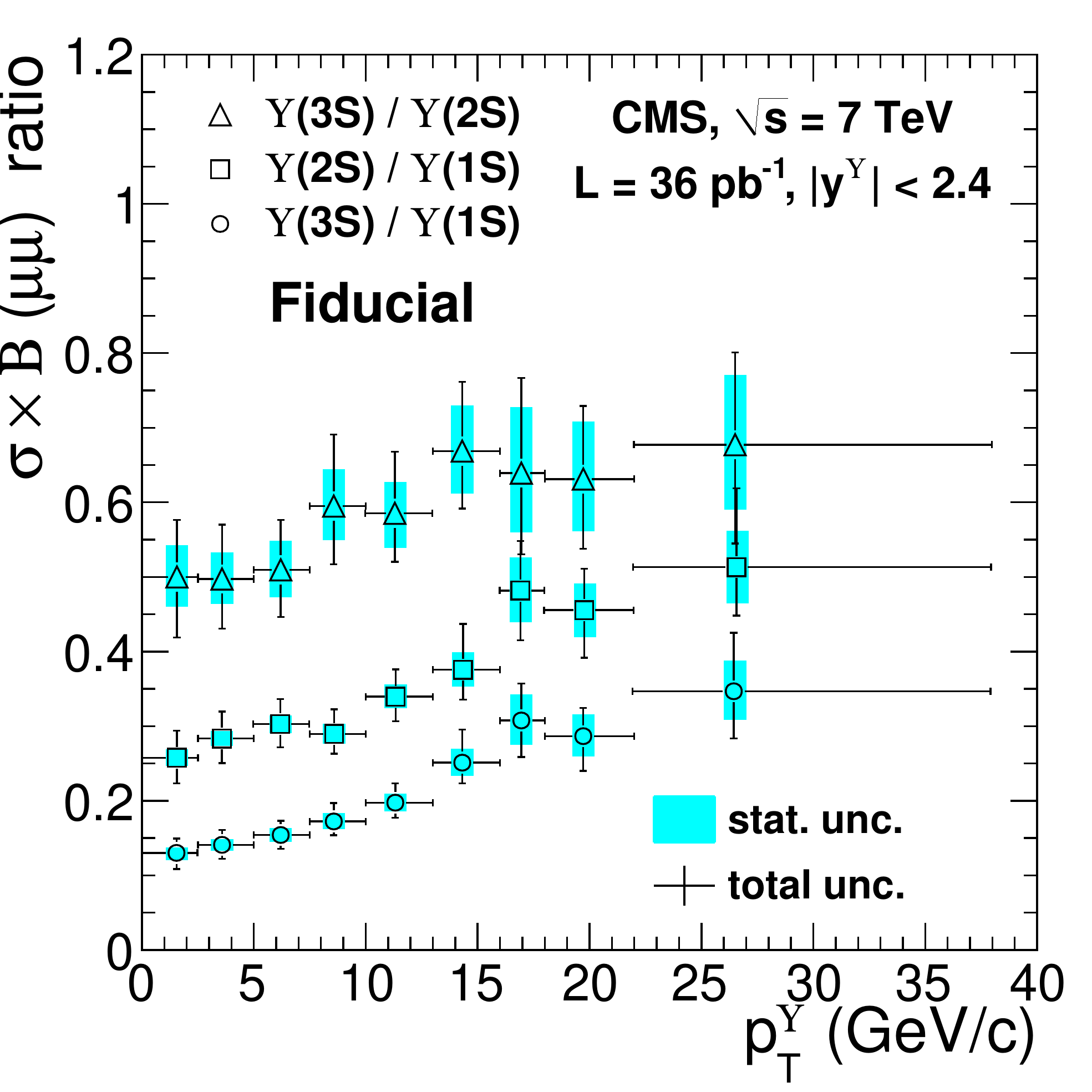}
  \caption{Ratios of the differential fiducial cross sections for the \upsn as a function of $\pt^{\PgU}$ in the rapidity range $\abs{y^{\PgU}}<2.4$.}
  \label{fig:xsec_ratio_noAcc}
\end{figure}

\begin{figure*}[htb]
  \centering
  \includegraphics[width=0.45\textwidth]{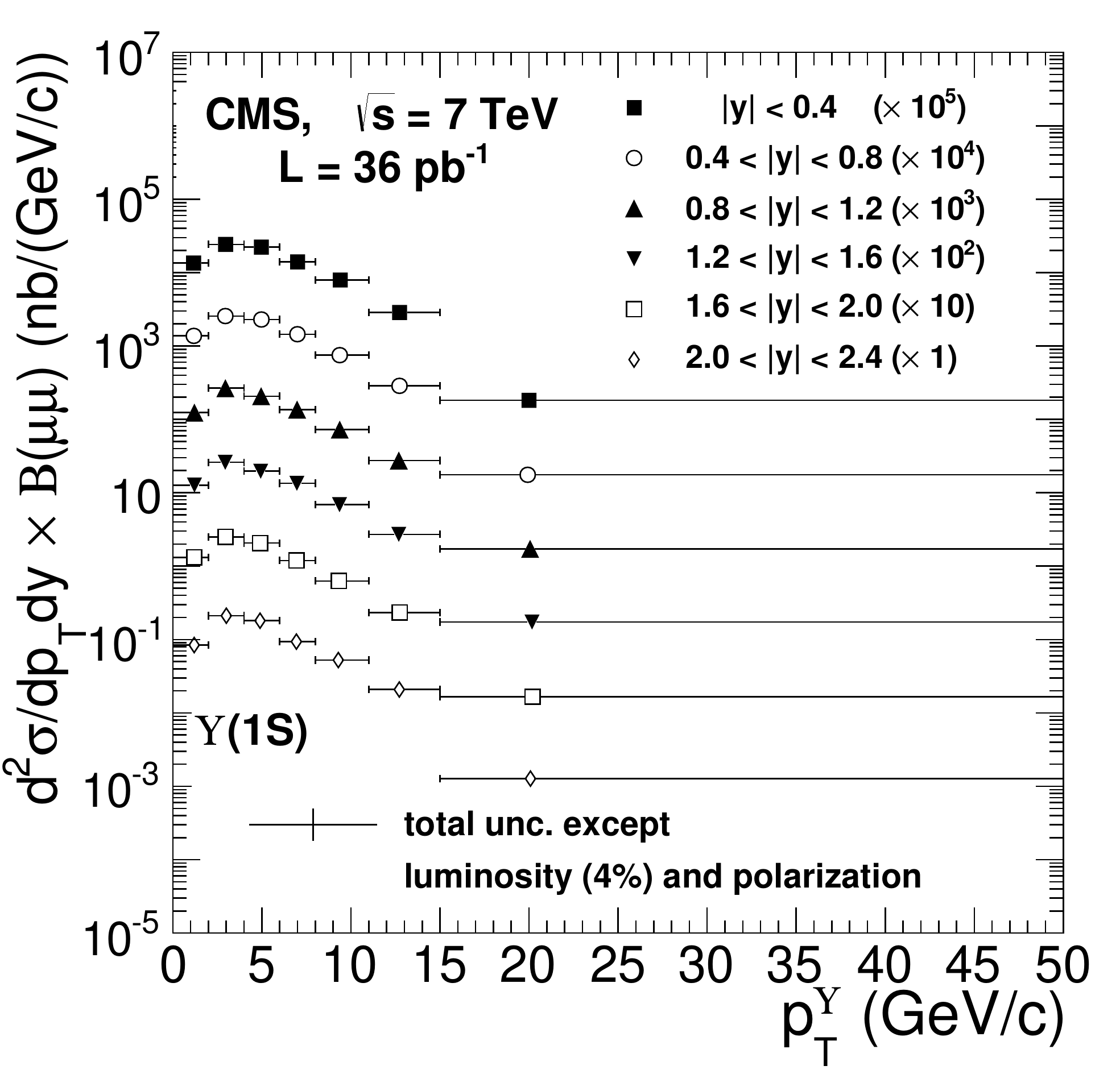}\label{fig:xsec_1s_5ybin}\\
  \includegraphics[width=0.45\textwidth]{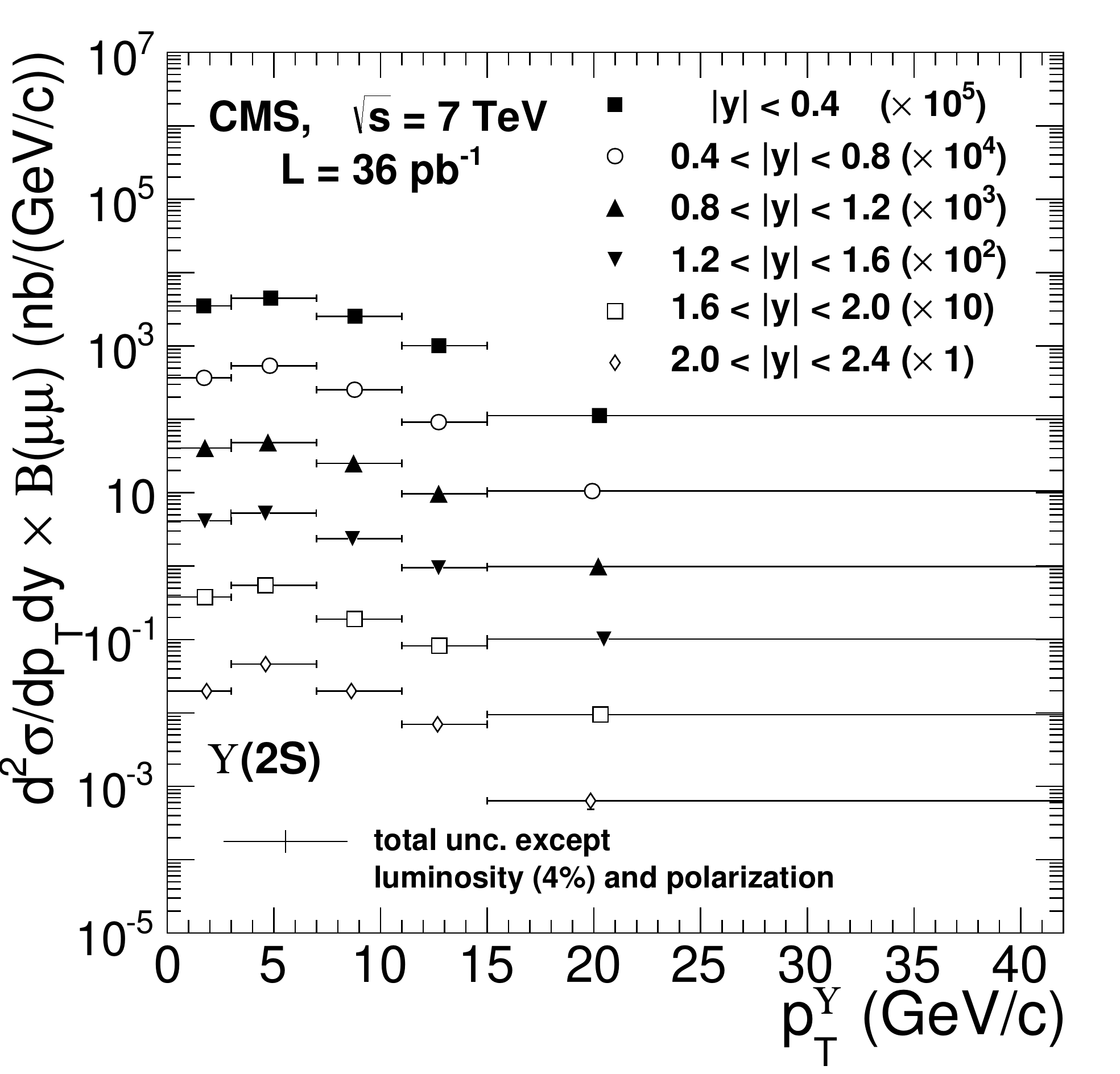}\label{fig:xsec_2s_5ybin}
  \includegraphics[width=0.45\textwidth]{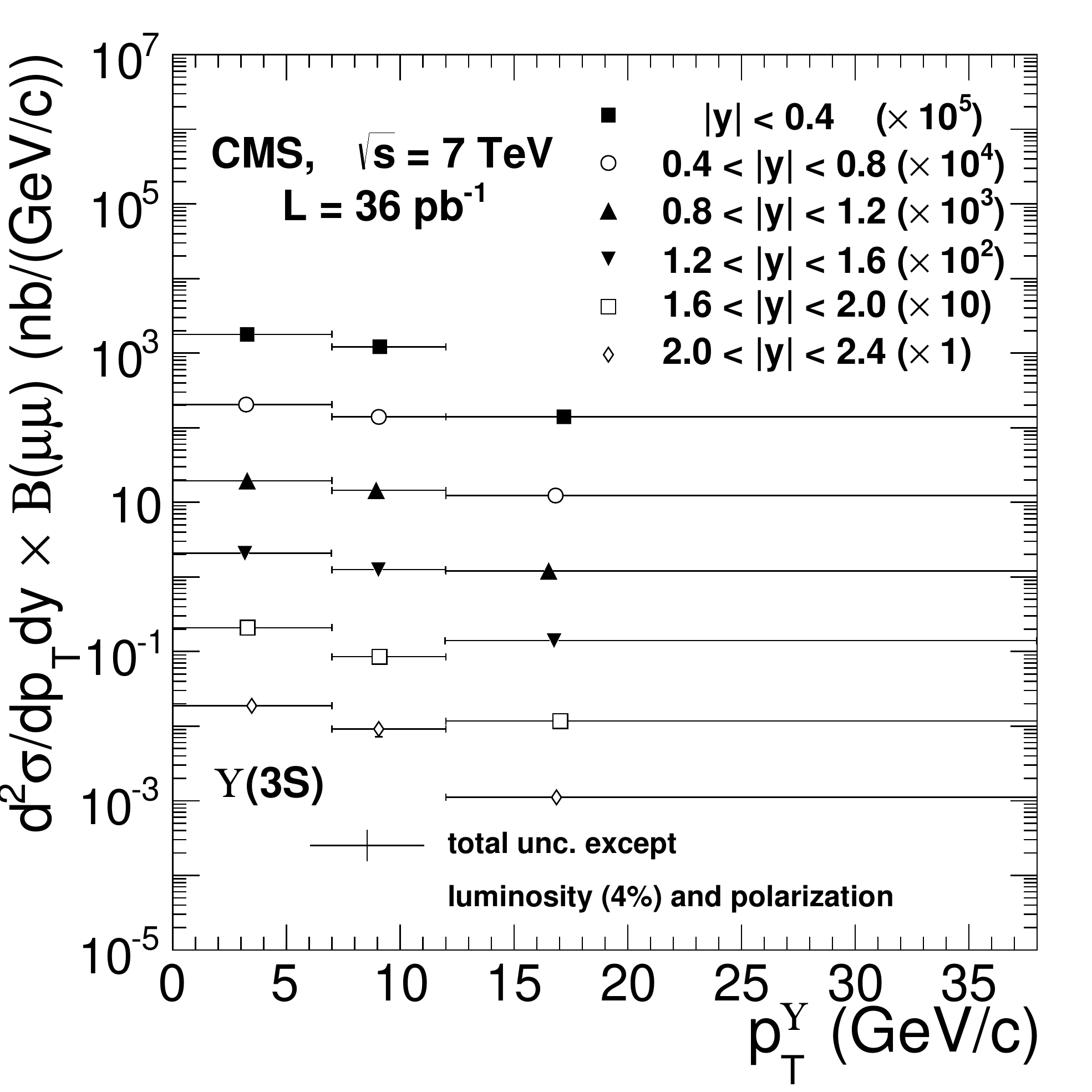}\label{fig:xsec_3s_5ybin}
  \caption{Differential production cross sections of the \PgU(nS) as a function of $\pt^{\PgU}$ in six rapidity regions. The measurements have been multiplied by the factors shown in the figure for clarity.}
  \label{fig:xsec_6ybin}
\end{figure*}

\begin{figure}[htb]
  \centering
  \includegraphics[width=0.45\textwidth]{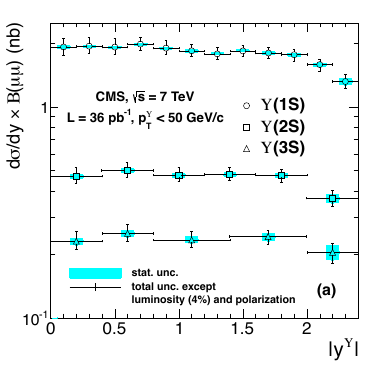}\label{fig:xsec_rapdiff}
    \caption{Acceptance-corrected differential production cross sections as a function of rapidity. The bands represent the statistical uncertainty and the error bars represent the total uncertainty, except for those from the \upsn polarization and the integrated luminosity.}
  \label{fig:xsec_rapdiff_ns_A}
\end{figure}

\begin{figure}[htb]
  \centering
  \includegraphics[width=0.45\textwidth]{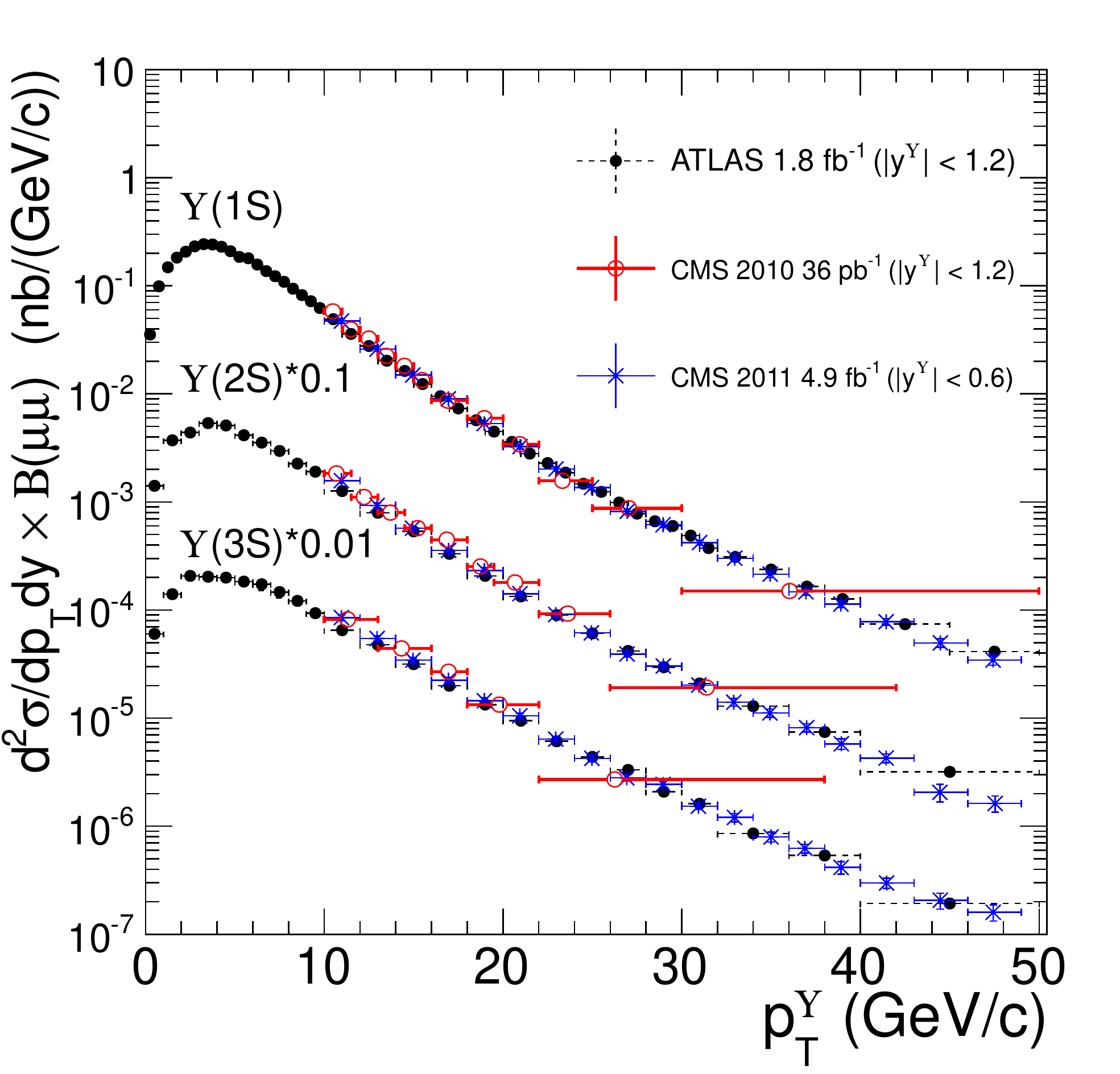}
  \includegraphics[width=0.45\textwidth]{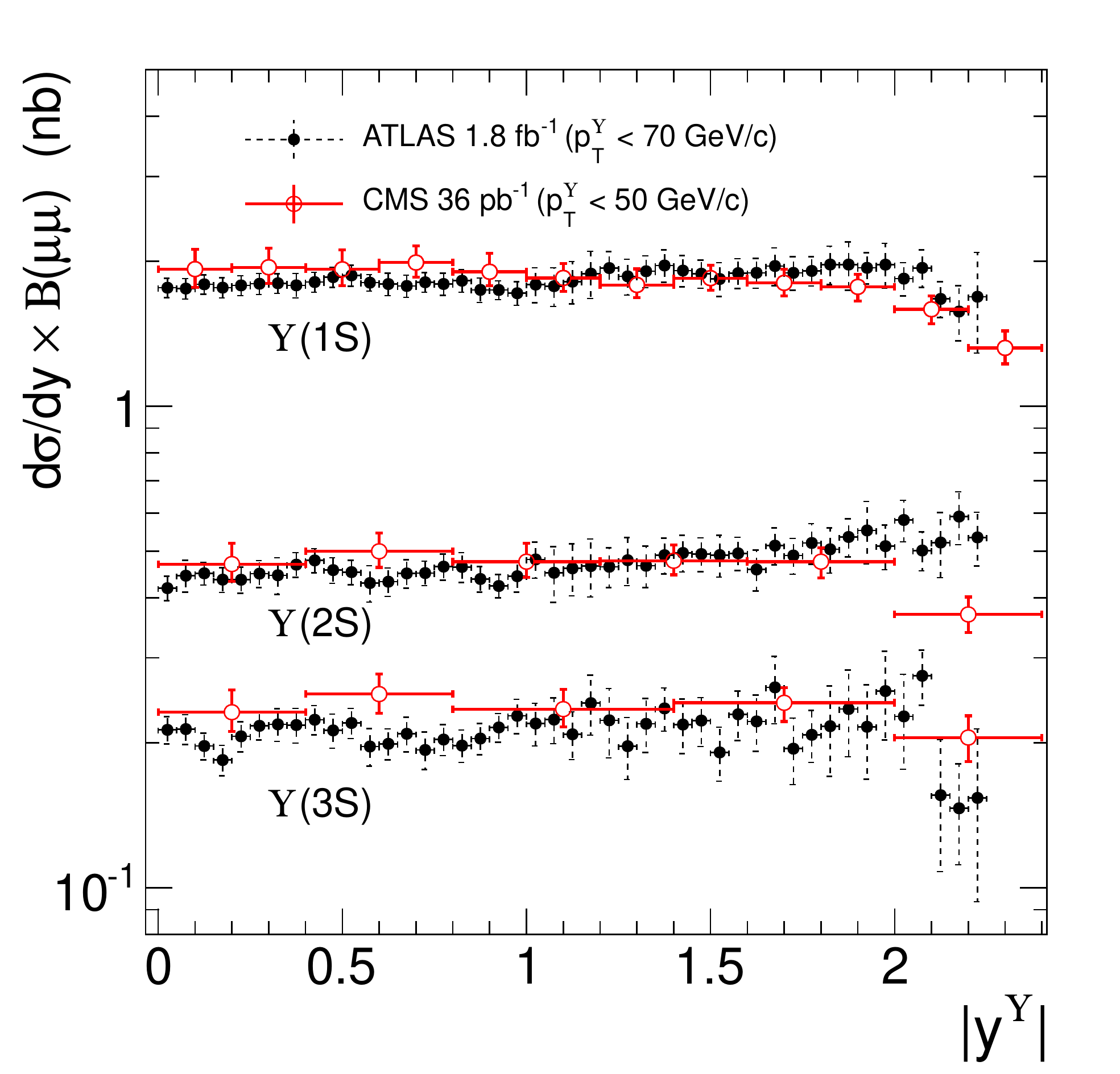}
    \caption{Comparison of the $\upsn$ acceptance-corrected differential cross section results to the ATLAS results, as a function of $\pt^\PgU$ (left) and $y^\PgU$ (right).} 
  \label{fig:compare_atlas}
\end{figure}

\begin{figure}[htb]
  \centering
  \includegraphics[width=0.45\textwidth]{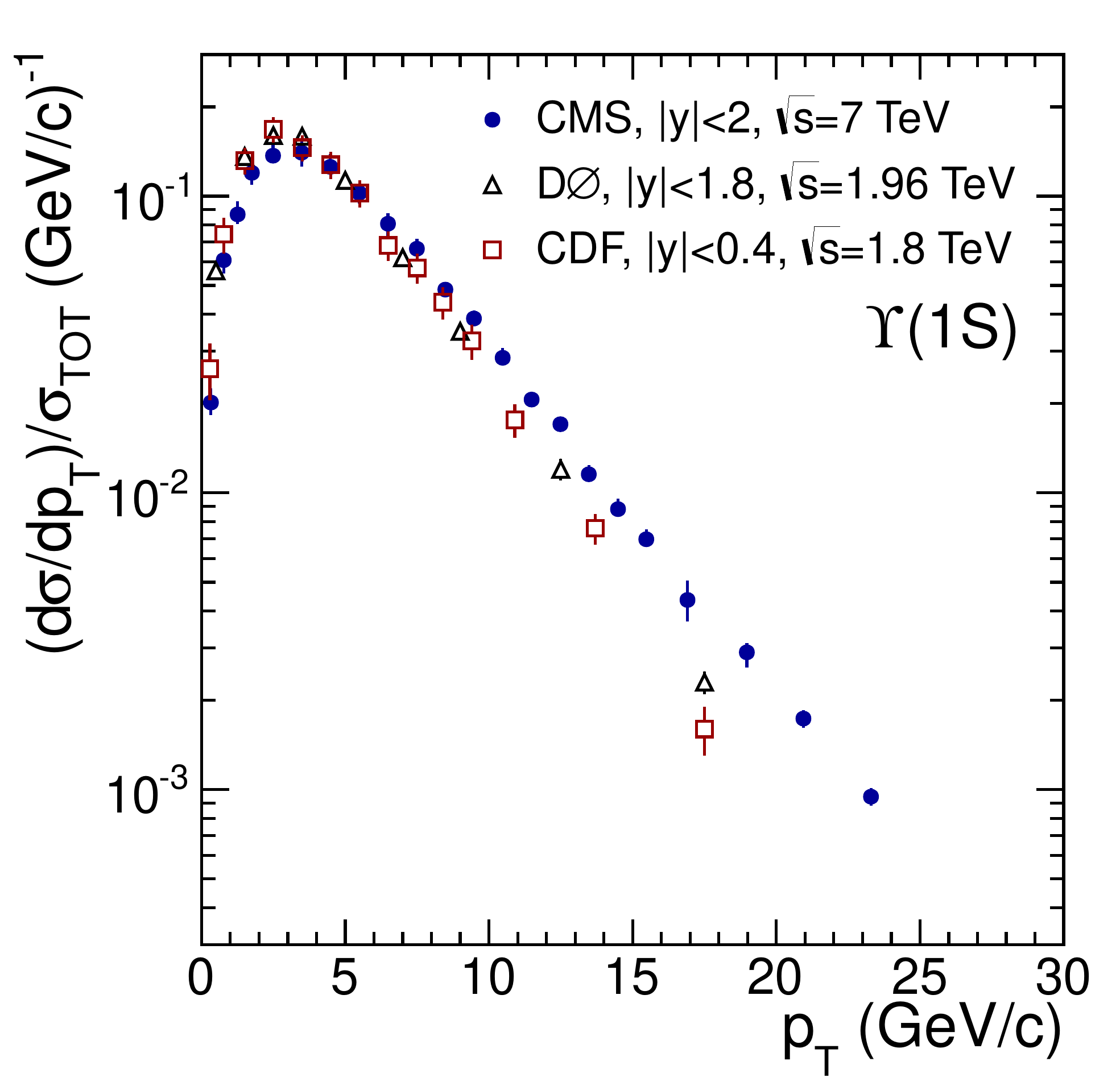}\\
  \includegraphics[width=0.45\textwidth]{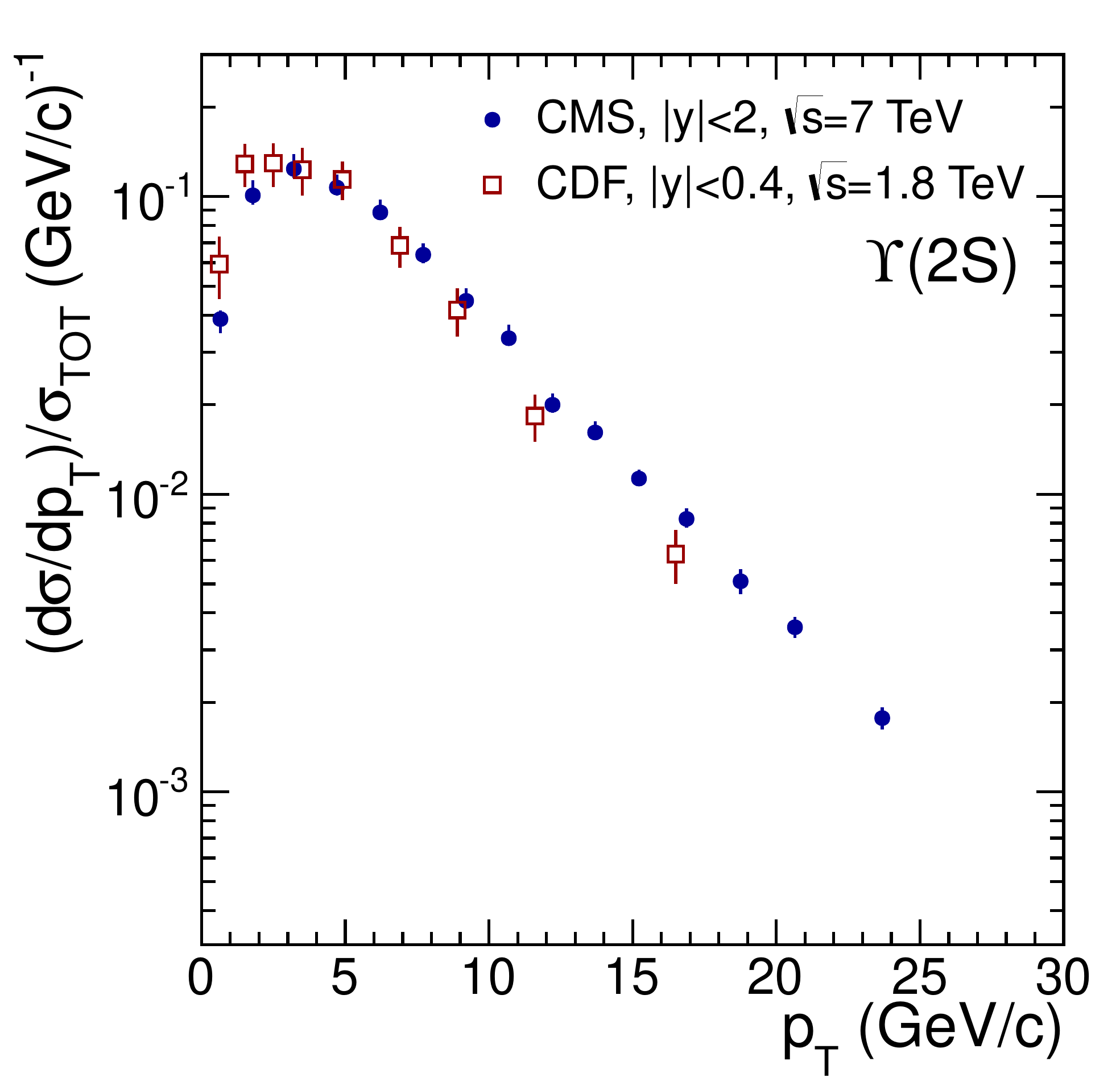}
  \includegraphics[width=0.45\textwidth]{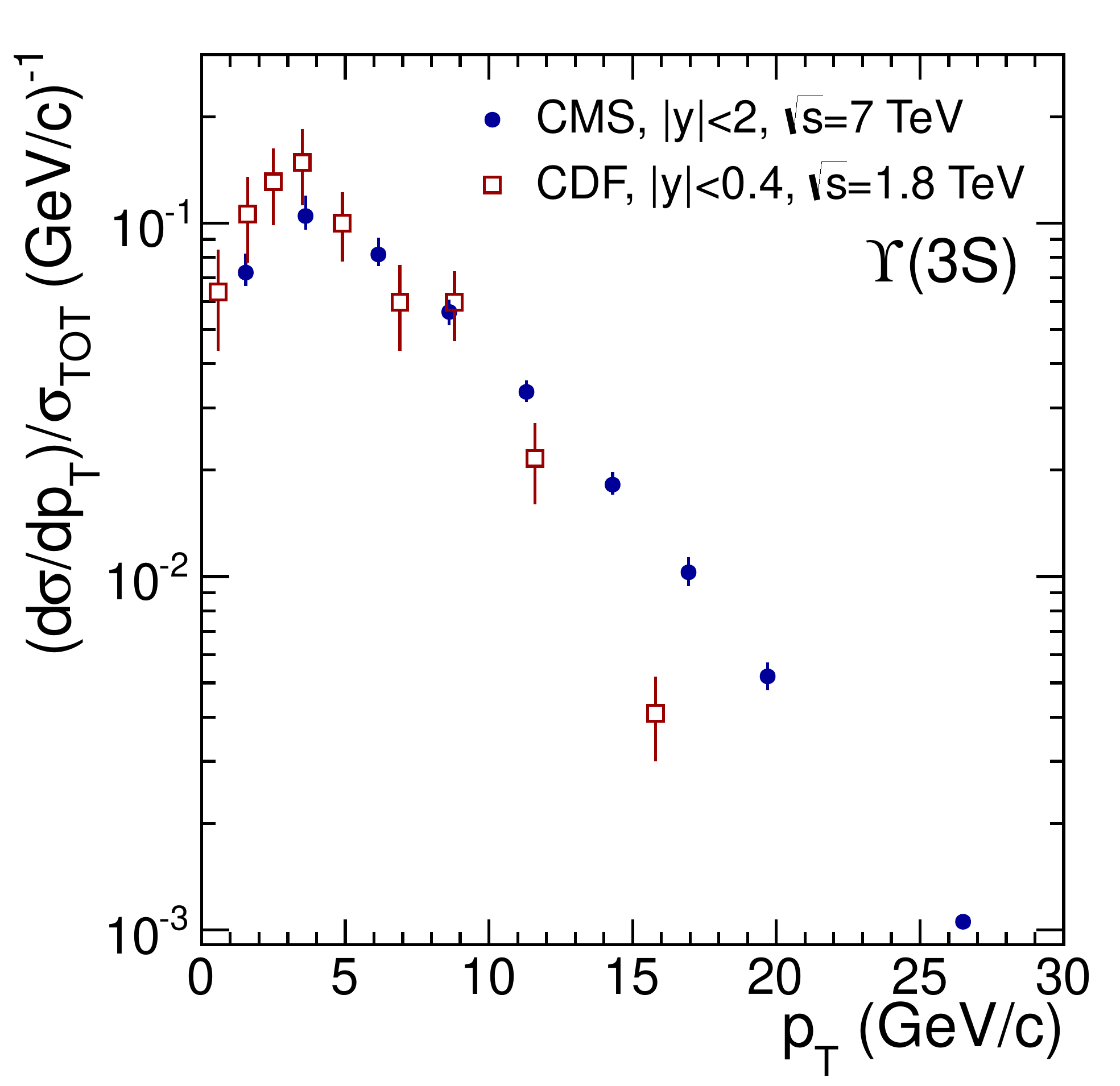}
    \caption{Comparison of acceptance-corrected differential cross section results, normalized by $\sigma_\text{TOT}=\sum(d\sigma/d\pt)\Delta\pt$, to the CDF and D0 results, as a function of $\pt$, for \PgUa\ (top), \PgUb\ (left), \PgUc\ (right).} 
  \label{fig:compare_tevatron}
\end{figure}

\clearpage

\begin{table*}[!ht]
\topcaption{The product of the fiducial $\PgUa$ production cross sections, $\sigma$, integrated and differential in $\pt^{\PgU}$, and the dimuon branching fraction, $\mathcal{B}$, integrated over the rapidity range $|y^{\PgU}| < 2.4$. The statistical uncertainty (stat.), the sum of the systematic uncertainties in quadrature $(\Sigma_\text{syst.})$, and the total uncertainty ($\Delta \sigma$; including stat., $\sum_\text{syst.}$, and the uncertainty in the integrated luminosity) are in percent.  The numbers in parentheses are negative variations.}
  \centering
  $|y^\PgU|<2.4$\\
   \PgUa\\\begin{tabular}{c c c c c } \hline
\pt (\GeVc)  & $\sigma  \cdot {\cal B}~(\rm nb)$ & 	$\frac{\rm stat.} {\sigma}$ &$\frac{\sum_{\rm{syst.}}}{\sigma}$ & $\frac{\Delta\sigma}{\sigma}$
\\ \hline 
0--0.5	&$0.0440$&		${ 5.4}$ &$   8\,(   8)$ & $  11\,(  11)$\\ 
0.5--1	&$0.133$&		${ 3.1}$ &$   8\,(   8)$ & $  10\,(  10)$\\ 
1--1.5	&$0.182$&		${ 2.5}$ &$   8\,(   8)$ & $   9\,(   9)$\\ 
1.5--2	&$0.228$&		${ 2.4}$ &$   8\,(   8)$ & $  10\,(   9)$\\ 
2--3	&$0.442$&		${ 1.6}$ &$   8\,(   7)$ & $   9\,(   8)$\\ 
3--4	&$0.374$&		${ 1.8}$ &$   6\,(   6)$ & $   8\,(   7)$\\ 
4--5	&$0.302$&		${ 1.8}$ &$   7\,(   7)$ & $   8\,(   8)$\\ 
5--6	&$0.236$&		${ 2.0}$ &$   7\,(   6)$ & $   8\,(   7)$\\ 
6--7	&$0.195$&		${ 2.0}$ &$   8\,(   7)$ & $   9\,(   9)$\\ 
7--8	&$0.174$&		${ 2.1}$ &$   5\,(   5)$ & $   7\,(   6)$\\ 
8--9	&$0.144$&		${ 2.3}$ &$   6\,(   5)$ & $   7\,(   7)$\\ 
9--10	&$0.1235$&		${ 2.4}$ &$   5\,(   4)$ & $   7\,(   6)$\\ 
10--11	&$0.0988$&		${ 2.5}$ &$   6\,(   5)$ & $   8\,(   7)$\\ 
11--12	&$0.0759$&		${ 2.8}$ &$   4\,(   4)$ & $   7\,(   6)$\\ 
12--13	&$0.0670$&		${ 2.9}$ &$   4\,(   4)$ & $   7\,(   6)$\\ 
13--14	&$0.0477$&		${ 3.3}$ &$   5\,(   4)$ & $   7\,(   7)$\\ 
14--15	&$0.0381$&		${ 3.6}$ &$   5\,(   5)$ & $   7\,(   7)$\\ 
15--16	&$0.0312$&		${ 4.0}$ &$   5\,(   4)$ & $   8\,(   7)$\\ 
16--18	&$0.0412$&		${ 3.5}$ &$   5\,(   5)$ & $   7\,(   7)$\\ 
18--20	&$0.0296$&		${ 4.0}$ &$   5\,(   4)$ & $   7\,(   7)$\\ 
20--22	&$0.0187$&		${ 5.1}$ &$   4\,(   4)$ & $   8\,(   8)$\\ 
22--25	&$0.0148$&		${ 5.8}$ &$   4\,(   4)$ & $   8\,(   8)$\\ 
25--30	&$0.0133$&		${ 6.1}$ &$   4\,(   4)$ & $   8\,(   8)$\\ 
30--50	&$0.00923$&		${ 7.8}$ &$   6\,(   6)$ & $  11\,(  10)$\\ 
0--50	&$3.06$&		${ 0.6}$ &$   6\,(   6)$ & $   8\,(   7)$\\\hline
\end{tabular}

  \label{table:cross-section-y-1s-noAcc}
\end{table*}

\clearpage

\begin{table*}[!ht]
\topcaption{The product of the fiducial $\PgUb$ production cross sections, $\sigma$, integrated and differential in $\pt^{\PgU}$, and the dimuon branching fraction, $\mathcal{B}$, integrated over the rapidity range $|y^{\PgU}| < 2.4$. The statistical uncertainty (stat.), the sum of the systematic uncertainties in quadrature $(\Sigma_\text{syst.})$, and the total uncertainty ($\Delta \sigma$; including stat., $\sum_\text{syst.}$, and the uncertainty in the integrated luminosity) are in percent. The numbers in parentheses are negative variations. }
  \centering
   $|y^\PgU|<2.4$\\
   \PgUb\\ \begin{tabular}{c c c c c } \hline
\pt (\GeVc)  & $\sigma  \cdot {\cal B}~(\rm nb)$ & 	$\frac{\rm stat.} {\sigma}$ &$\frac{\sum_{\rm{syst.}}}{\sigma}$ & $\frac{\Delta\sigma}{\sigma}$
\\ \hline
0--1	&$0.0467$&		${ 6.3}$ &$   7\,(   8)$ & $  10\,(  11)$\\ 
1--2.5	&$0.168$&		${ 3.4}$ &$   8\,(   8)$ & $  10\,(  10)$\\ 
2.5--4	&$0.169$&		${ 3.1}$ &$   8\,(  11)$ & $   9\,(  12)$\\ 
4--5.5	&$0.118$&		${ 3.3}$ &$   8\,(   7)$ & $  10\,(   9)$\\ 
5.5--7	&$0.0917$&		${ 3.6}$ &$   6\,(   5)$ & $   8\,(   8)$\\ 
7--8.5	&$0.0716$&		${ 3.4}$ &$   7\,(   7)$ & $   9\,(   9)$\\ 
8.5--10	&$0.0564$&		${ 4.0}$ &$   5\,(   5)$ & $   8\,(   8)$\\ 
10--11.5	&$0.0470$&		${ 4.1}$ &$   6\,(   5)$ & $   8\,(   8)$\\ 
11.5--13	&$0.0343$&		${ 4.6}$ &$   4\,(   4)$ & $   7\,(   8)$\\ 
13--14.5	&$0.0260$&		${ 5.2}$ &$   5\,(   5)$ & $   8\,(   8)$\\ 
14.5--16	&$0.0196$&		${ 5.7}$ &$   4\,(   6)$ & $   8\,(   9)$\\ 
16--18	&$0.0198$&		${ 5.5}$ &$   6\,(   5)$ & $   9\,(   8)$\\ 
18--19.5	&$0.01005$&		${ 7.5}$ &$   4\,(   5)$ & $   9\,(  10)$\\ 
19.5--22	&$0.0123$&		${ 6.8}$ &$   5\,(   5)$ & $   9\,(   9)$\\ 
22--26	&$0.0104$&		${ 7.4}$ &$   4\,(   5)$ & $   9\,(  10)$\\ 
26--42	&$0.00930$&		${ 8.0}$ &$   5\,(   5)$ & $  10\,(  10)$\\ 
0--42	&$0.910$&		${ 1.2}$ &$   6\,(   5)$ & $   7\,(   7)$\\\hline
\end{tabular}
\\
  \label{table:cross-section-y-2s-noAcc}
\end{table*}

\begin{table*}[!ht]
\topcaption{The product of the fiducial $\PgUc$ production cross sections, $\sigma$, integrated and differential in $\pt^{\PgU}$, and the dimuon branching fraction, $\mathcal{B}$, integrated over the rapidity range $|y^{\PgU}| < 2.4$. The statistical uncertainty (stat.), the sum of the systematic uncertainties in quadrature $(\Sigma_\text{syst.})$, and the total uncertainty ($\Delta \sigma$; including stat., $\sum_\text{syst.}$, and the uncertainty in the integrated luminosity) are in percent.  The numbers in parentheses are negative variations.}
\centering
$|y^\PgU|<2.4$\\
 \PgUc\\\begin{tabular}{c c c c c } \hline
\pt (\GeVc)  & $\sigma  \cdot {\cal B}~(\rm nb)$ & 	$\frac{\rm stat.} {\sigma}$ &$\frac{\sum_{\rm{syst.}}}{\sigma}$ & $\frac{\Delta\sigma}{\sigma}$
\\ \hline
0--2.5	&$0.107$&		${ 5.3}$ &$   7\,(   7)$ & $  10\,(  10)$\\ 
2.5--5	&$0.125$&		${ 4.5}$ &$   8\,(   8)$ & $  10\,(  10)$\\ 
5--7.5	&$0.0801$&		${ 4.7}$ &$   6\,(   6)$ & $   9\,(   8)$\\ 
7.5--10	&$0.0604$&		${ 4.8}$ &$   9\,(   8)$ & $  11\,(  10)$\\ 
10--13	&$0.0476$&		${ 4.5}$ &$   6\,(   7)$ & $   8\,(   9)$\\ 
13--16	&$0.0308$&		${ 5.1}$ &$   5\,(   6)$ & $   8\,(   9)$\\ 
16--18	&$0.0127$&		${ 7.5}$ &$   6\,(   5)$ & $  10\,(  10)$\\ 
18--22	&$0.0140$&		${ 6.9}$ &$   7\,(   7)$ & $  11\,(  11)$\\ 
22--38	&$0.0124$&		${ 7.4}$ &$   9\,(   9)$ & $  12\,(  12)$\\ 
0--38	&$0.490$&		${ 2.0}$ &$   6\,(   6)$ & $   8\,(   7)$\\ 
\hline
\end{tabular}
\\
\label{table:cross-section-y-3s-noAcc}
\end{table*}

\begin{table*}[!ht]
\topcaption{The product of the fiducial $\PgU$(nS) production cross sections, $\sigma$, integrated and differential in $\pt^{\PgU}$, and the respective dimuon branching fraction, $\mathcal{B}$, integrated over the rapidity range $|y^{\PgU}| < 0.4$. The statistical uncertainty (stat.), the sum of the systematic uncertainties in quadrature $(\Sigma_\text{syst.})$, and the total uncertainty ($\Delta \sigma$; including stat., $\sum_\text{syst.}$, and the uncertainty in the integrated luminosity) are in percent.  The numbers in parentheses are negative variations.}
\centering
$|y^\PgU|<0.4$\\
\begin{tabular}{c| c c c c c } \hline
\multirow{2}{*}{ } & \pt (\rm \GeVc)  & $\sigma  \cdot {\cal B}~(\rm nb)$ &     $\frac{\rm stat.} {\sigma}$ &$\frac{\sum_{\rm{syst.}}}{\sigma}$ & $\frac{\Delta\sigma}{\sigma}$
\\ \hline
\multirow{7}{*}{$\PgUa$} 
&0--2	&$0.104$&	      ${ 3.5}$ &$  14\,(  13)$ & $  15\,(  14)$\\ 
&2--4	&$0.108$&	      ${ 3.1}$ &$  10\,(  11)$ & $  11\,(  12)$\\ 
&4--6	&$0.0870$&	      ${ 2.9}$ &$  10\,(   9)$ & $  11\,(  11)$\\ 
&6--8	&$0.0632$&	      ${ 3.3}$ &$  14\,(  14)$ & $  15\,(  15)$\\ 
&8--11	&$0.0741$&	      ${ 2.9}$ &$   7\,(   7)$ & $   8\,(   8)$\\ 
&11--15	&$0.0458$&	      ${ 3.3}$ &$   5\,(   5)$ & $   7\,(   7)$\\ 
&15--50	&$0.0323$&	      ${ 3.7}$ &$   4\,(   5)$ & $   7\,(   7)$\\ 
\hline
\multirow{5}{*}{$\PgUb$}
&0--3	&$0.0442$&	      ${ 7.0}$ &$  12\,(  11)$ & $  15\,(  14)$\\
&3--7	&$0.0453$&	      ${ 4.9}$ &$  11\,(  11)$ & $  12\,(  12)$\\
&7--11	&$0.0320$&	      ${ 5.3}$ &$   8\,(   7)$ & $  10\,(  10)$\\
&11--15	&$0.0167$&	      ${ 6.1}$ &$   5\,(   5)$ & $   9\,(   9)$\\
&15--42	&$0.0157$&	      ${ 5.7}$ &$   4\,(   5)$ & $   8\,(   9)$\\
\hline
\multirow{3}{*}{$\PgUc$}
&0--7	&$0.0434$&	     ${ 6.5}$ &$  14\,(  12)$ & $  16\,(  15)$\\
&7--12	&$0.0203$&	     ${ 7.5}$ &$  10\,(   9)$ & $  13\,(  13)$\\
&12--38	&$0.0176$&	     ${ 5.9}$ &$   5\,(   5)$ & $   9\,(   9)$\\
\hline
\end{tabular}

\label{table:cross-section-y-ranges-1-noAcc}
\end{table*}

\begin{table*}[!ht]
\topcaption{The product of the fiducial $\PgU$(nS) production cross sections, $\sigma$, integrated and differential in $\pt^{\PgU}$, and the respective dimuon branching fraction, $\mathcal{B}$, integrated over the rapidity range $0.4<|y^\PgU|<0.8$. The statistical uncertainty (stat.), the sum of the systematic uncertainties in quadrature $(\Sigma_\text{syst.})$, and the total uncertainty ($\Delta \sigma$; including stat., $\sum_\text{syst.}$, and the uncertainty in the integrated luminosity) are in percent.  The numbers in parentheses are negative variations.}
\centering
$0.4<|y^\PgU|<0.8$\\
\begin{tabular}{c| c c c c c } \hline
\multirow{2}{*}{ } & \pt (\rm\GeVc)  & $\sigma \cdot {\cal B}~(\rm nb)$ &     $\frac{\rm stat.} {\sigma}$ &$\frac{\sum_{\rm{syst.}}}{\sigma}$ & $\frac{\Delta\sigma}{\sigma}$
\\ \hline
\multirow{7}{*}{$\PgUa$}
&0--2	&$0.112$&	      ${ 3.3}$ &$  12\,(  12)$ & $  13\,(  13)$\\ 
&2--4	&$0.127$&	      ${ 3.0}$ &$   9\,(   9)$ & $  10\,(  11)$\\ 
&4--6	&$0.0967$&	      ${ 2.8}$ &$   8\,(   8)$ & $   9\,(   9)$\\ 
&6--8	&$0.0695$&	      ${ 3.3}$ &$  12\,(  12)$ & $  13\,(  13)$\\ 
&8--11	&$0.0731$&	      ${ 2.9}$ &$   6\,(   5)$ & $   8\,(   7)$\\ 
&11--15	&$0.0467$&	      ${ 3.3}$ &$   6\,(   6)$ & $   8\,(   8)$\\ 
&15--50	&$0.0314$&	      ${ 0.8}$ &$   4\,(   4)$ & $   6\,(   6)$\\ 
\hline
\multirow{5}{*}{$\PgUb$}
&0--3	&$0.0472$&	      ${ 7.2}$ &$  12\,(  11)$ & $  14\,(  14)$\\
&3--7	&$0.0580$&	      ${ 4.5}$ &$   8\,(   8)$ & $  10\,(  10)$\\
&7--11	&$0.0334$&	      ${ 5.2}$ &$   7\,(   5)$ & $  10\,(   8)$\\
&11--15	&$0.0154$&	      ${ 6.6}$ &$   5\,(   6)$ & $   9\,(  10)$\\
&15--42	&$0.0147$&	      ${ 5.9}$ &$   4\,(   5)$ & $   8\,(   8)$\\
\hline
\multirow{3}{*}{$\PgUc$}
&0--7	&$0.0504$&	      ${ 6.8}$ &$  13\,(  11)$ & $  15\,(  13)$\\
&7--12	&$0.0247$&	      ${ 6.8}$ &$   7\,(   7)$ & $  11\,(  10)$\\
&12--38	&$0.0155$&	      ${ 6.5}$ &$   5\,(   5)$ & $   9\,(   9)$\\
\hline
\end{tabular}

\label{table:cross-section-y-ranges-2-noAcc}
\end{table*}

\begin{table*}[!ht]
\topcaption{The product of the fiducial $\PgU$(nS) production cross sections, $\sigma$, integrated and differential in $\pt^{\PgU}$, and the respective dimuon branching fraction, $\mathcal{B}$, integrated over the rapidity range $0.8<|y^\PgU|<1.2$. The statistical uncertainty (stat.), the sum of the systematic uncertainties in quadrature $(\Sigma_\text{syst.})$, and the total uncertainty ($\Delta \sigma$; including stat., $\sum_\text{syst.}$, and the uncertainty in the integrated luminosity) are in percent.  The numbers in parentheses are negative variations.}
\centering
$0.8<|y^\PgU|<1.2$\\
\begin{tabular}{c| c c c c c } \hline
\multirow{2}{*}{ } & \pt (\rm\GeVc)  & $\sigma \cdot {\cal B}~(\rm nb)$ &     $\frac{\rm stat.} {\sigma}$ &$\frac{\sum_{\rm{syst.}}}{\sigma}$ & $\frac{\Delta\sigma}{\sigma}$
\\ \hline
\multirow{7}{*}{$\Upsilon(\rm 1S)$}
&0--2	&$0.107$&	       ${ 3.7}$ &$   9\,(   9)$ & $  11\,(  10)$\\ 
&2--4	&$0.160$&	       ${ 2.7}$ &$   9\,(   9)$ & $  10\,(  10)$\\ 
&4--6	&$0.0995$&	       ${ 2.9}$ &$   7\,(   6)$ & $   8\,(   8)$\\ 
&6--8	&$0.0770$&	       ${ 3.1}$ &$  12\,(  12)$ & $  13\,(  13)$\\ 
&8--11	&$0.0746$&	       ${ 2.9}$ &$   5\,(   8)$ & $   7\,(   9)$\\ 
&11--15	&$0.0464$&	       ${ 3.4}$ &$   6\,(   6)$ & $   8\,(   8)$\\ 
&15--50	&$0.0313$&	       ${ 3.8}$ &$   5\,(   6)$ & $   7\,(   8)$\\ 
\hline
\multirow{5}{*}{$\Upsilon(\rm 2S)$}
&0--3	&$0.0561$&	     ${ 6.1}$  &$  11\,(  10)$ & $  13\,(  12)$\\
&3--7	&$0.0610$&	     ${ 5.0}$  &$   8\,(   8)$ & $  10\,(  10)$\\
&7--11	&$0.0347$&	     ${ 5.3}$  &$  14\,(  14)$ & $  15\,(  15)$\\
&11--15	&$0.0168$&	     ${ 6.2}$  &$   6\,(   7)$ & $   9\,(  10)$\\
&15--42	&$0.0142$&	     ${ 6.2}$  &$   6\,(   7)$ & $  10\,(  10)$\\
\hline
\multirow{3}{*}{$\Upsilon(\rm 3S)$}
&0--7	&$0.0526$&	     ${ 7.7}$ &$  17\,(  17)$ & $  19\,(  19)$\\
&7--12	&$0.0264$&	     ${ 6.9}$ &$  11\,(  14)$ & $  14\,(  17)$\\
&12--38	&$0.0154$&	     ${ 6.8}$ &$  10\,(  10)$ & $  13\,(  13)$\\
\hline
\end{tabular}

\label{table:cross-section-y-ranges-3-noAcc}
\end{table*}

\begin{table*}[!ht]
\topcaption{The product of the fiducial $\PgU$(nS) production cross sections, $\sigma$, integrated and differential in $\pt^{\PgU}$, and the respective dimuon branching fraction, $\mathcal{B}$, integrated over the rapidity range $1.2<|y^\PgU|<1.6$. The statistical uncertainty (stat.), the sum of the systematic uncertainties in quadrature $(\Sigma_\text{syst.})$, and the total uncertainty ($\Delta \sigma$; including stat., $\sum_\text{syst.}$, and the uncertainty in the integrated luminosity) are in percent.  The numbers in parentheses are negative variations.}
\centering
$1.2<|y^\PgU|<1.6$\\
\begin{tabular}{c|c c c c c } \hline
\multirow{2}{*}{ } & \pt (\rm\GeVc)  & $\sigma \cdot {\cal B}~(\rm nb)$ &     $\frac{\rm stat.} {\sigma}$ &$\frac{\sum_{\rm{syst.}}}{\sigma}$ & $\frac{\Delta\sigma}{\sigma}$
\\ \hline
\multirow{7}{*}{$\Upsilon(\rm 1S)$}
&0--2	&$0.119$&	     ${ 3.5}$ &$   6\,(  12)$ & $   8\,(  13)$\\ 
&2--4	&$0.184$&	     ${ 2.4}$ &$   7\,(   6)$ & $   8\,(   8)$\\ 
&4--6	&$0.1059$&	     ${ 2.8}$ &$   6\,(   5)$ & $   8\,(   7)$\\ 
&6--8	&$0.0747$&	     ${ 3.0}$ &$   5\,(   6)$ & $   7\,(   7)$\\ 
&8--11	&$0.0697$&	     ${ 3.0}$ &$   5\,(   5)$ & $   7\,(   7)$\\ 
&11--15	&$0.0456$&	     ${ 3.3}$ &$   6\,(   6)$ & $   8\,(   8)$\\ 
&15--50	&$0.0319$&	     ${ 3.8}$ &$   7\,(   8)$ & $   9\,(  10)$\\ 
\hline
\multirow{5}{*}{$\Upsilon(\rm 2S)$}
&0--3	&$0.0582$&	    ${ 6.2}$ &$   6\,(   6)$ & $  10\,(   9)$\\
&3--7	&$0.0684$&	    ${ 4.5}$ &$   7\,(   6)$ & $   9\,(   8)$\\
&7--11	&$0.0317$&	    ${ 5.5}$ &$   5\,(   6)$ & $   9\,(   9)$\\
&11--15	&$0.0164$&	    ${ 6.5}$ &$   5\,(   5)$ & $   9\,(   9)$\\
&15--42	&$0.0147$&	    ${ 6.4}$ &$   8\,(   9)$ & $  11\,(  12)$\\
\hline
\multirow{3}{*}{$\Upsilon(\rm 3S)$}
&0--7	&$0.0589$&	       ${ 6.9}$ &$   6\,(   5)$ & $  10\,(  10)$\\
&7--12	&$0.0214$&	       ${ 7.7}$ &$   9\,(   9)$ & $  13\,(  12)$\\
&12--38	&$0.0171$&	       ${ 6.7}$ &$   8\,(   8)$ & $  11\,(  11)$\\
\hline
\end{tabular}

\label{table:cross-section-y-ranges-4-noAcc}
\end{table*}

\begin{table*}[!ht]
\topcaption{The product of the fiducial $\PgU$(nS) production cross sections, $\sigma$, integrated and differential in $\pt^{\PgU}$, and the respective dimuon branching fraction, $\mathcal{B}$, integrated over the rapidity range $1.6<|y^\PgU|<2.0$. The statistical uncertainty (stat.), the sum of the systematic uncertainties in quadrature $(\Sigma_\text{syst.})$, and the total uncertainty ($\Delta \sigma$; including stat., $\sum_\text{syst.}$, and the uncertainty in the integrated luminosity) are in percent.  The numbers in parentheses are negative variations.}
\centering
$1.6<|y^\PgU|<2.0$\\
\begin{tabular}{c|c c c c c } \hline
\multirow{2}{*}{ }& \pt (\rm\GeVc)  & $\sigma \cdot {\cal B}~(\rm nb)$ &     $\frac{\rm stat.} {\sigma}$ &$\frac{\sum_{\rm{syst.}}}{\sigma}$ & $\frac{\Delta\sigma}{\sigma}$
\\ \hline
\multirow{7}{*}{$\PgUa$}
&0--2	&$0.105$&	    ${ 3.1}$ &$   9\,(   8)$ & $  10\,(   9)$\\ 
&2--4	&$0.164$&	    ${ 2.5}$ &$   6\,(   5)$ & $   8\,(   7)$\\ 
&4--6	&$0.1058$&	    ${ 2.7}$ &$   5\,(   4)$ & $   7\,(   6)$\\ 
&6--8	&$0.0564$&	    ${ 3.2}$ &$   4\,(   4)$ & $   6\,(   6)$\\ 
&8--11	&$0.0500$&	    ${ 3.4}$ &$   6\,(   6)$ & $   8\,(   8)$\\ 
&11--15	&$0.0321$&	    ${ 4.1}$ &$   5\,(   5)$ & $   8\,(   8)$\\ 
&15--50	&$0.0274$&	    ${ 4.4}$ &$   6\,(   6)$ & $   8\,(   8)$\\ 
\hline
\multirow{5}{*}{$\PgUb$}
&0--3	&$0.0452$&	     ${ 6.2}$ &$   8\,(   8)$ & $  11\,(  11)$\\
&3--7	&$0.0661$&	     ${ 4.3}$ &$   7\,(   7)$ & $   9\,(   9)$\\
&7--11	&$0.0205$&	     ${ 7.0}$ &$  10\,(  10)$ & $  13\,(  13)$\\
&11--15	&$0.0112$&	     ${ 8.3}$ &$   5\,(   5)$ & $  11\,(  11)$\\
&15--42	&$0.0120$&	     ${ 8.1}$ &$   6\,(   5)$ & $  11\,(  11)$\\
\hline
\multirow{3}{*}{$\PgUc$}
&0--7	&$0.0525$&	    ${ 7.2}$ &$   9\,(   9)$ & $  12\,(  12)$\\
&7--12	&$0.0120$&	    ${10.6}$ &$  15\,(  15)$ & $  19\,(  19)$\\
&12--38	&$0.0124$&	    ${ 8.8}$ &$   6\,(   6)$ & $  11\,(  12)$\\
\hline
\end{tabular}

\label{table:cross-section-y-ranges-5-noAcc}
\end{table*}

\begin{table*}[!ht]
\topcaption{The product of the fiducial $\PgU$(nS) production cross sections, $\sigma$, integrated and differential in $\pt^{\PgU}$, and the respective dimuon branching fraction, $\mathcal{B}$, integrated over the rapidity range $2.0<|y^\PgU|<2.4$. The statistical uncertainty (stat.), the sum of the systematic uncertainties in quadrature $(\Sigma_\text{syst.})$, and the total uncertainty ($\Delta \sigma$; including stat., $\sum_\text{syst.}$, and the uncertainty in the integrated luminosity) are in percent.  The numbers in parentheses are negative variations.}
\centering
$2.0<|y^\PgU|<2.4$\\
\begin{tabular}{c|c c c c c } \hline
\multirow{2}{*}{ } & \pt (\rm\GeVc) & $\sigma \cdot {\cal B}~(\rm nb)$ &     $\frac{\rm stat.} {\sigma}$ &$\frac{\sum_{\rm{syst.}}}{\sigma}$ & $\frac{\Delta\sigma}{\sigma}$
\\ \hline
\multirow{7}{*}{$\PgUa$}
&0--2	&$0.0299$&	     ${ 5.3}$ &$   9\,(   9)$ & $  11\,(  11)$\\ 
&2--4	&$0.0671$&	     ${ 2.8}$ &$   7\,(   5)$ & $   8\,(   7)$\\ 
&4--6	&$0.0515$&	     ${ 3.7}$ &$  12\,(  12)$ & $  13\,(  13)$\\ 
&6--8	&$0.0216$&	     ${ 5.6}$ &$   6\,(   7)$ & $   9\,(  10)$\\ 
&8--11	&$0.0194$&	     ${ 5.8}$ &$   6\,(   6)$ & $   9\,(   9)$\\ 
&11--15	&$0.0113$&	     ${ 7.2}$ &$   9\,(   8)$ & $  12\,(  12)$\\ 
&15--50	&$0.0092$&	     ${ 9.1}$ &$  13\,(  12)$ & $  16\,(  16)$\\ 
\hline
\multirow{5}{*}{$\PgUb$}
&0--3	&$0.0119$&	    ${13.2}$ &$  11\,(  10)$ & $  18\,(  17)$\\
&3--7	&$0.0278$&	    ${ 7.6}$ &$   9\,(   8)$ & $  12\,(  12)$\\
&7--11	&$0.0098$&	    ${10.9}$ &$   9\,(   8)$ & $  14\,(  14)$\\
&11--15	&$0.00440$&	    ${14.7}$ &$   8\,(   7)$ & $  17\,(  17)$\\
&15--42	&$0.00365$&	    ${20.4}$ &$  10\,(   9)$ & $  23\,(  23)$\\
\hline
\multirow{3}{*}{$\PgUc$}
&0--7	&$0.0223$&	     ${11.2}$ &$  10\,(   9)$ & $  16\,(  15)$\\
&7--12	&$0.0054$&	     ${19.3}$ &$   7\,(   7)$ & $  21\,(  21)$\\
&12--38	&$0.00518$&	     ${15.8}$ &$   8\,(   8)$ & $  18\,(  18)$\\
\hline
\end{tabular}

\label{table:cross-section-y-ranges-6-noAcc}
\end{table*}

\begin{table*}[!ht]
\topcaption{The product of the fiducial $\PgU$(nS) production cross sections, $\sigma$, integrated and differential in $|y^{\PgU}|$, and the respective dimuon branching fraction, $\mathcal{B}$, integrated over the \pt range $\pt^\PgU<50\GeVc$. The statistical uncertainty (stat.), the sum of the systematic uncertainties in quadrature $(\Sigma_\text{syst.})$, and the total uncertainty ($\Delta \sigma$; including stat., $\sum_\text{syst.}$, and the uncertainty in the integrated luminosity) are in percent.  The numbers in parentheses are negative variations.}
\centering
$\pt^\PgU<50\GeVc$\\
\begin{tabular}{c|c c c c c } \hline
&$|y|$ & $\sigma \cdot {\cal B}~(\rm nb)$ &     $\frac{\rm stat.} {\sigma}$ &$\frac{\sum_{\rm{syst.}}}{\sigma}$ & $\frac{\Delta\sigma}{\sigma}$
\\ \hline
\multirow{7}{*}{$\PgUa$}
&0--0.2	&$0.262$&	     ${ 1.7}$ &$   9\,(   9)$ & $  10\,(  10)$\\ 
&0.2--0.4	&$0.268$&	     ${ 1.7}$ &$   9\,(   9)$ & $  10\,(  10)$\\ 
&0.4--0.6	&$0.269$&	     ${ 1.7}$ &$   9\,(   9)$ & $  10\,(  10)$\\ 
&0.6--0.8	&$0.295$&	     ${ 1.7}$ &$   7\,(   6)$ & $   8\,(   8)$\\ 
&0.8--1.0	&$0.295$&	     ${ 1.7}$ &$   8\,(   8)$ & $   9\,(   9)$\\ 
&1.0--1.2	&$0.303$&	     ${ 1.8}$ &$   6\,(   6)$ & $   7\,(   7)$\\ 
&1.2--1.4	&$0.305$&	     ${ 1.7}$ &$   7\,(   7)$ & $   8\,(   8)$\\ 
&1.4--1.6	&$0.322$&	     ${ 1.7}$ &$   5\,(   4)$ & $   7\,(   6)$\\ 
&1.6--1.8	&$0.301$&	     ${ 1.7}$ &$   6\,(   5)$ & $   7\,(   7)$\\ 
&1.8--2	&$0.248$&	     ${ 1.9}$ &$   6\,(   5)$ & $   7\,(   7)$\\ 
&2--2.2	&$0.159$&	     ${ 2.3}$ &$   7\,(   5)$ & $   8\,(   7)$\\ 
&2.2--2.4	&$0.0514$&	     ${ 4.3}$ &$   6\,(   5)$ & $   9\,(   7)$\\ 
&0--2.4	&$3.08$&	     ${ 0.5}$ &$   6\,(   6)$ & $   8\,(   7)$\\ 
\hline
\multirow{5}{*}{$\PgUb$}
&0--0.4	&$0.158$&	    ${ 2.4}$ &$  8\,(   7)$ & $   9\,(   9)$\\
&0.4--0.8	&$0.170$&	    ${ 2.6}$ &$  8\,(   7)$ & $   9\,(   9)$\\
&0.8--1.2	&$0.179$&	    ${ 2.8}$ &$  7\,(   7)$ & $   9\,(   9)$\\
&1.2--1.6	&$0.185$&	    ${ 2.7}$ &$  6\,(   5)$ & $   8\,(   7)$\\
&1.6--2	&$0.157$&	    ${ 2.9}$ &$  6\,(   6)$ & $   8\,(   7)$\\
&2--2.4	&$0.0577$&	    ${ 7.6}$ &$  7\,(   6)$ & $  11\,(  11)$\\
&0--2.4	&$0.907$&	    ${ 1.2}$ &$  6\,(   5)$ & $   7\,(   7)$\\
\hline
\multirow{3}{*}{$\PgUc$}
&0--0.4	&$0.0858$&	     ${ 3.7}$ &$  8\,(   8)$ & $  10\,(  10)$\\
&0.4--0.8	&$0.0946$&	     ${ 4.0}$ &$  8\,(   7)$ & $  10\,(   9)$\\
&0.8--1.4	&$0.141$&	     ${ 3.8}$ &$  7\,(   7)$ & $   9\,(   9)$\\
&1.4--2	&$0.134$&	     ${ 4.0}$ &$  7\,(   7)$ & $   9\,(   9)$\\
&2--2.4	&$0.0321$&	     ${10.3}$ &$  7\,(   5)$ & $  13\,(  12)$\\
&0--2.4	&$0.487$&	     ${ 2.0}$ &$  6\,(   6)$ & $   8\,(   7)$\\
\hline
\end{tabular}

\label{table:cross-section-rapdiff-noAcc}
\end{table*}

\begin{table*}
  \centering
  \topcaption{The ratio of $\PgUc/\PgUa$ fiducial cross sections, and its \pt dependence, integrated over the rapidity range $|y^\PgU|<2.4$, with statistical and systematic uncertainties combined in quadrature. The numbers in parentheses are negative variations.}
  \begin{tabular}{cccc} \hline 
\pt (\GeVc) & \PgUc/\PgUa  \\ \hline
0--2 &$0.129\pm0.009\,(0.008)\pm0.017\,(0.020)$\\ 
2--5 &$0.141\pm0.008\,(0.008)\pm0.018\,(0.016)$\\ 
5--8 &$0.154\pm0.009\,(0.009)\pm0.016\,(0.016)$\\ 
8--10 &$0.172\pm0.011\,(0.011)\pm0.022\,(0.014)$\\ 
10--13 &$0.197\pm0.012\,(0.012)\pm0.022\,(0.016)$\\ 
13--16 &$0.251\pm0.019\,(0.018)\pm0.037\,(0.021)$\\ 
16--18 &$0.308\pm0.035\,(0.033)\pm0.031\,(0.037)$\\ 
18--22 &$0.287\pm0.029\,(0.027)\pm0.023\,(0.037)$\\ 
22--38 &$0.345\pm0.042\,(0.038)\pm0.055\,(0.051)$\\
0--38 &$0.158\pm0.004\,(0.004)\pm0.016\,(0.016)$\\  
\hline 
 \end{tabular}

  \label{tab:xsec_ratio-noAcc_1S}
\end{table*}

\begin{table*}
  \centering
  \topcaption{The ratio of $\PgUb/\PgUa$ fiducial cross sections, and its \pt dependence, integrated over the rapidity range $|y^\PgU|<2.4$, with statistical and systematic uncertainties combined in quadrature. The numbers in parentheses are negative variations.}
  \begin{tabular}{cccc} \hline 
\pt (\GeVc) & \PgUb/\PgUa \\ \hline
0--2 &$0.258\pm0.011\,(0.011)\pm0.034\,(0.032)$\\ 
2--5 &$0.283\pm0.010\,(0.010)\pm0.035\,(0.031)$\\ 
5--8 &$0.302\pm0.012\,(0.012)\pm0.031\,(0.028)$\\ 
8--10 &$0.289\pm0.014\,(0.013)\pm0.030\,(0.022)$\\ 
10--13 &$0.339\pm0.016\,(0.016)\pm0.032\,(0.029)$\\ 
13--16 &$0.376\pm0.023\,(0.022)\pm0.053\,(0.034)$\\ 
16--18 &$0.481\pm0.045\,(0.042)\pm0.046\,(0.053)$\\ 
18--22 &$0.454\pm0.037\,(0.035)\pm0.040\,(0.053)$\\ 
22--38 &$0.511\pm0.051\,(0.047)\pm0.083\,(0.046)$\\ 
0--38 &$0.296\pm0.005\,(0.005)\pm0.031\,(0.029)$\\ 
\hline 
 \end{tabular}

  \label{tab:xsec_ratio-noAcc_2S}
\end{table*}

\begin{table*}
  \centering
  \topcaption{The ratio of $\PgUc/\PgUb$ fiducial cross sections, and its \pt dependence, integrated over the rapidity range $|y^\PgU|<2.4$, with statistical and systematic uncertainties combined in quadrature. The numbers in parentheses are negative variations.}
  \begin{tabular}{cccc} \hline 
\pt (\GeVc) & \PgUc/\PgUb \\ \hline
0--2 &$0.500\pm0.043\,(0.040)\pm0.062\,(0.072)$\\ 
2--5 &$0.497\pm0.036\,(0.034)\pm0.063\,(0.057)$\\ 
5--8 &$0.510\pm0.039\,(0.037)\pm0.053\,(0.051)$\\ 
8--10 &$0.595\pm0.049\,(0.046)\pm0.079\,(0.062)$\\ 
10--13 &$0.581\pm0.046\,(0.043)\pm0.067\,(0.049)$\\ 
13--16 &$0.669\pm0.061\,(0.057)\pm0.066\,(0.054)$\\ 
16--18 &$0.639\pm0.089\,(0.079)\pm0.083\,(0.077)$\\ 
18--22 &$0.631\pm0.077\,(0.070)\pm0.057\,(0.065)$\\ 
22--38 &$0.675\pm0.096\,(0.085)\pm0.068\,(0.101)$\\ 
0--38 &$0.534\pm0.017\,(0.016)\pm0.054\,(0.051)$\\ 
\hline 
 \end{tabular}

  \label{tab:xsec_ratio-noAcc_3S}
\end{table*}

\begin{table*}[!ht]
\topcaption{The product of the $\PgUa$ acceptance-corrected production cross sections, $\sigma$, integrated and differential in $\pt^{\PgU}$, and the dimuon branching fraction, $\mathcal{B}$, measured for four polarization scenarios, in the helicity frame (HX) and Colins--Soper (CS) frame, each for $\lambda_{\theta}=1$ and $\lambda_{\theta}=-1$,  integrated over the rapidity range $|y^\PgU|<2.4$. The statistical uncertainty (stat.), the sum of the systematic uncertainties in quadrature $(\Sigma_\text{syst.})$, and the total uncertainty ($\Delta \sigma$; including stat., $\sum_\text{syst.}$, and the uncertainty in the integrated luminosity) are in percent. For the four polarization scenarios the fractional change to the central value of the cross section relative to the unpolarized value is given in percent. The numbers in parentheses are negative variations.}
  \centering
  $|y^\PgU|<2.4$\\
   {\PgUa\  \\\begin{tabular}{c c c c c |  c c| c c} \hline
\multicolumn{5}{c|}{} & \multicolumn{2}{c|}{$\rm HX$} & \multicolumn{2}{c}{$\rm CS$} \\ 
\pt (\rm \GeVc)  & $\sigma  \cdot {\cal B}~(\rm nb)$ & 	$\frac{\rm stat.} {\sigma}$ &$\frac{\sum_{\rm{syst.}}}{\sigma}$ & $\frac{\Delta\sigma}{\sigma}$
& $\lambda_{\theta} = 1$ & $\lambda_{\theta} = -1$ & $\lambda_{\theta} = 1$ & $\lambda_{\theta} = -1$ \\ \hline
0--0.5	&$0.0859$&		${ 5.4}$ &$   8\,(   7)$ & $  11\,(  10)$ & $  +19$ & $  -24$ & $  +21$ & $  -26$\\ 
0.5--1	&$0.263$&		${ 3.3}$ &$   8\,(   7)$ & $   9\,(   9)$ & $  +22$ & $  -24$ & $  +23$ & $  -25$\\ 
1--1.5	&$0.374$&		${ 2.6}$ &$   8\,(   8)$ & $   9\,(   9)$ & $  +23$ & $  -22$ & $  +26$ & $  -24$\\ 
1.5--2	&$0.505$&		${ 2.4}$ &$   9\,(   8)$ & $  10\,(   9)$ & $  +16$ & $  -27$ & $  +19$ & $  -29$\\ 
2--3	&$1.16$&		${ 1.6}$ &$   8\,(  10)$ & $   9\,(  11)$ & $  +20$ & $  -25$ & $  +26$ & $  -27$\\ 
3--4	&$1.21$&		${ 2.1}$ &$   7\,(   6)$ & $   9\,(   8)$ & $  +20$ & $  -26$ & $  +21$ & $  -26$\\ 
4--5	&$1.084$&		${ 2.1}$ &$   7\,(   6)$ & $   8\,(   8)$ & $  +21$ & $  -25$ & $  +21$ & $  -24$\\ 
5--6	&$0.879$&		${ 1.9}$ &$   7\,(   9)$ & $   8\,(  10)$ & $  +19$ & $  -25$ & $  +16$ & $  -23$\\ 
6--7	&$0.680$&		${ 2.6}$ &$   6\,(   6)$ & $   8\,(   7)$ & $  +18$ & $  -24$ & $  +13$ & $  -23$\\ 
7--8	&$0.556$&		${ 2.0}$ &$   6\,(   5)$ & $   7\,(   7)$ & $  +20$ & $  -22$ & $  +12$ & $  -13$\\ 
8--9	&$0.419$&		${ 2.2}$ &$   5\,(   5)$ & $   7\,(   7)$ & $  +19$ & $  -22$ & $   +9$ & $   -9$\\ 
9--10	&$0.331$&		${ 2.3}$ &$   5\,(   4)$ & $   7\,(   6)$ & $  +19$ & $  -21$ & $   +7$ & $   -4$\\ 
10--11	&$0.238$&		${ 2.5}$ &$   5\,(   4)$ & $   7\,(   6)$ & $  +17$ & $  -20$ & $   +4$ & $   -1$\\ 
11--12	&$0.179$&		${ 2.9}$ &$   5\,(   4)$ & $   7\,(   6)$ & $  +18$ & $  -19$ & $   +6$ & $   +0$\\ 
12--13	&$0.145$&		${ 2.9}$ &$   5\,(   4)$ & $   7\,(   7)$ & $  +16$ & $  -20$ & $   +2$ & $   +2$\\ 
13--14	&$0.0990$&		${ 3.2}$ &$   4\,(   5)$ & $   7\,(   7)$ & $  +14$ & $  -21$ & $   -0$ & $   +3$\\ 
14--15	&$0.0750$&		${ 3.6}$ &$   5\,(   5)$ & $   8\,(   7)$ & $  +15$ & $  -20$ & $   -1$ & $   +5$\\ 
15--16	&$0.0595$&		${ 3.8}$ &$   5\,(   5)$ & $   7\,(   7)$ & $  +13$ & $  -19$ & $   -1$ & $   +5$\\ 
16--18	&$0.0732$&		${ 3.4}$ &$   5\,(   5)$ & $   7\,(   7)$ & $  +12$ & $  -20$ & $   -3$ & $   +6$\\ 
18--20	&$0.0500$&		${ 3.8}$ &$   5\,(   4)$ & $   7\,(   7)$ & $  +13$ & $  -18$ & $   -2$ & $   +7$\\ 
20--22	&$0.0302$&		${ 5.1}$ &$   5\,(   4)$ & $   8\,(   8)$ & $  +11$ & $  -18$ & $   -3$ & $   +6$\\ 
22--25	&$0.0237$&		${ 5.6}$ &$   5\,(   4)$ & $   8\,(   8)$ & $   +9$ & $  -18$ & $   -3$ & $   +4$\\ 
25--30	&$0.0205$&		${ 6.0}$ &$   5\,(   4)$ & $   9\,(   8)$ & $  +10$ & $  -16$ & $   -1$ & $   +4$\\ 
30--50	&$0.0123$&		${ 7.4}$ &$   6\,(   6)$ & $  10\,(  10)$ & $   +5$ & $  -17$ & $   -7$ & $   +3$\\ 
0--50	&$8.55$&		${ 0.6}$ &$   7\,(   6)$ & $   8\,(   7)$ & $  +19$ & $  -24$ & $  +16$ & $  -19$\\ 
\hline
\end{tabular}
}\\
  \label{table:cross-section-y-1s}
\end{table*}

\begin{table*}[!ht]
\topcaption{The product of the $\PgUb$ acceptance-corrected production cross sections, $\sigma$, integrated and differential in $\pt^{\PgU}$, and the dimuon branching fraction, $\mathcal{B}$, measured for four polarization scenarios, in the helicity frame (HX) and Colins--Soper (CS) frame, each for $\lambda_{\theta}=1$ and $\lambda_{\theta}=-1$,  integrated over the rapidity range $|y^\PgU|<2.4$. The statistical uncertainty (stat.), the sum of the systematic uncertainties in quadrature $(\Sigma_\text{syst.})$, and the total uncertainty ($\Delta \sigma$; including stat., $\sum_\text{syst.}$, and the uncertainty in the integrated luminosity) are in percent. For the four polarization scenarios the fractional change to the central value of the cross section relative to the unpolarized value is given in percent. The numbers in parentheses are negative variations.}
  \centering
   $|y^\PgU|<2.4$\\
   {\PgUb\  \\\begin{tabular}{c c c c c |  c c| c c} \hline
\multicolumn{5}{c|}{} & \multicolumn{2}{c|}{$\rm HX$} & \multicolumn{2}{c}{$\rm CS$} \\ 
\pt (\rm \GeVc)  & $\sigma \cdot {\cal B}~(\rm nb)$ & 	$\frac{\rm stat.} {\sigma}$ &$\frac{\sum_{\rm{syst.}}}{\sigma}$ & $\frac{\Delta\sigma}{\sigma}$
& $\lambda_{\theta} = 1$ & $\lambda_{\theta} = -1$ & $\lambda_{\theta} = 1$ & $\lambda_{\theta} = -1$ \\ \hline
0--1	&$0.0829$&		${ 5.9}$ &$   9\,(   8)$ & $  11\,(  11)$ & $  +18$ & $  -23$ & $  +19$ & $  -24$\\ 
1--2.5	&$0.331$&		${ 3.3}$ &$  11\,(  10)$ & $  12\,(  11)$ & $  +14$ & $  -23$ & $  +18$ & $  -25$\\ 
2.5--4	&$0.409$&		${ 3.1}$ &$   9\,(   8)$ & $  10\,(   9)$ & $  +22$ & $  -24$ & $  +21$ & $  -26$\\ 
4--5.5	&$0.362$&		${ 3.3}$ &$   8\,(   7)$ & $   9\,(   9)$ & $  +18$ & $  -22$ & $  +16$ & $  -24$\\ 
5.5--7	&$0.286$&		${ 3.6}$ &$   7\,(   6)$ & $   9\,(   8)$ & $  +17$ & $  -23$ & $  +15$ & $  -20$\\ 
7--8.5	&$0.212$&		${ 3.9}$ &$   7\,(   7)$ & $   9\,(   9)$ & $  +21$ & $  -20$ & $  +15$ & $  -12$\\ 
8.5--10	&$0.146$&		${ 4.0}$ &$   6\,(   6)$ & $   9\,(   8)$ & $  +20$ & $  -18$ & $  +11$ & $   -5$\\ 
10--11.5	&$0.1123$&		${ 4.1}$ &$   6\,(   6)$ & $   9\,(   8)$ & $  +19$ & $  -18$ & $   +8$ & $   -0$\\ 
11.5--13	&$0.0765$&		${ 4.6}$ &$   5\,(   5)$ & $   8\,(   8)$ & $  +17$ & $  -18$ & $   +6$ & $   +1$\\ 
13--14.5	&$0.0519$&		${ 5.1}$ &$   5\,(   5)$ & $   8\,(   8)$ & $  +14$ & $  -20$ & $   +0$ & $   +3$\\ 
14.5--16	&$0.0376$&		${ 5.7}$ &$   5\,(   7)$ & $   9\,(  10)$ & $  +14$ & $  -19$ & $   +1$ & $   +3$\\ 
16--18	&$0.0373$&		${ 5.3}$ &$   6\,(   5)$ & $   9\,(   8)$ & $  +13$ & $  -18$ & $   +1$ & $   +3$\\ 
18--19.5	&$0.0159$&		${ 7.4}$ &$   5\,(   4)$ & $  10\,(   9)$ & $  +13$ & $  -18$ & $   -3$ & $   +9$\\ 
19.5--22	&$0.0204$&		${ 6.6}$ &$   5\,(   5)$ & $   9\,(   9)$ & $  +11$ & $  -16$ & $   -1$ & $   +5$\\ 
22--26	&$0.0158$&		${ 7.2}$ &$   5\,(   4)$ & $  10\,(   9)$ & $  +11$ & $  -16$ & $   -3$ & $   +7$\\ 
26--42	&$0.0126$&		${ 7.7}$ &$   6\,(   5)$ & $  10\,(  10)$ & $  +10$ & $  -15$ & $   -4$ & $   +9$\\ 
0--42	&$2.21$&		${ 1.2}$ &$   7\,(   6)$ & $   8\,(   7)$ & $  +14$ & $  -24$ & $  +13$ & $  -20$\\ 
\hline
\end{tabular}
}\\
  \label{table:cross-section-y-2s}
\end{table*}

\begin{table*}[!ht]
\topcaption{The product of the $\PgUc$ acceptance-corrected production cross sections, $\sigma$, integrated and differential in $\pt^{\PgU}$, and the dimuon branching fraction, $\mathcal{B}$, measured for four polarization scenarios, in the helicity frame (HX) and Colins--Soper (CS) frame, each for $\lambda_{\theta}=1$ and $\lambda_{\theta}=-1$,  integrated over the rapidity range $|y^\PgU|<2.4$. The statistical uncertainty (stat.), the sum of the systematic uncertainties in quadrature $(\Sigma_\text{syst.})$, and the total uncertainty ($\Delta \sigma$; including stat., $\sum_\text{syst.}$, and the uncertainty in the integrated luminosity) are in percent. For the four polarization scenarios the fractional change to the central value of the cross section relative to the unpolarized value is given in percent. The numbers in parentheses are negative variations.}
\centering
$|y^\PgU|<2.4$\\
 {\PgUc\  \\\begin{tabular}{c c c c c |  c c| c c} \hline
\multicolumn{5}{c|}{} & \multicolumn{2}{c|}{$\rm HX$} & \multicolumn{2}{c}{$\rm CS$} \\ 
\pt (\rm \GeVc)  & $\sigma \cdot {\cal B}~(\rm nb)$ & 	$\frac{\rm stat.} {\sigma}$ &$\frac{\sum_{\rm{syst.}}}{\sigma}$ & $\frac{\Delta\sigma}{\sigma}$
& $\lambda_{\theta} = 1$ & $\lambda_{\theta} = -1$ & $\lambda_{\theta} = 1$ & $\lambda_{\theta} = -1$ \\ \hline
0--2.5	&$0.203$&		${ 5.3}$ &$   8\,(   8)$ & $  11\,(  10)$ & $  +17$ & $  -21$ & $  +20$ & $  -25$\\ 
2.5--5	&$0.287$&		${ 4.5}$ &$  10\,(  11)$ & $  12\,(  12)$ & $  +17$ & $  -22$ & $  +20$ & $  -25$\\ 
5--7.5	&$0.227$&		${ 4.6}$ &$   9\,(   8)$ & $  11\,(  10)$ & $  +16$ & $  -22$ & $  +20$ & $  -22$\\ 
7.5--10	&$0.157$&		${ 4.8}$ &$  11\,(  10)$ & $  12\,(  12)$ & $  +23$ & $  -16$ & $  +16$ & $   -5$\\ 
10--13	&$0.113$&		${ 4.3}$ &$   7\,(   5)$ & $   9\,(   8)$ & $  +20$ & $  -15$ & $  +12$ & $   -1$\\ 
13--16	&$0.0617$&		${ 5.0}$ &$   5\,(   5)$ & $   8\,(   8)$ & $  +14$ & $  -17$ & $   +4$ & $   +1$\\ 
16--18	&$0.0227$&		${ 7.4}$ &$   6\,(   5)$ & $  10\,(  10)$ & $  +12$ & $  -18$ & $   -1$ & $   +4$\\ 
18--22	&$0.0229$&		${ 7.0}$ &$   7\,(   6)$ & $  10\,(  10)$ & $  +12$ & $  -17$ & $   -1$ & $   +6$\\ 
22--38	&$0.0185$&		${ 7.6}$ &$  13\,(  13)$ & $  15\,(  15)$ & $  +10$ & $  -15$ & $   -0$ & $   +6$\\ 
0--38	&$1.11$&		${ 2.0}$ &$   9\,(   8)$ & $  10\,(   9)$ & $  +16$ & $  -21$ & $  +14$ & $  -17$\\ 
\hline
\end{tabular}
}\\
\label{table:cross-section-y-3s}
\end{table*}

\begin{table*}[!ht]
\topcaption{The product of the $\PgU$(nS) acceptance-corrected production cross sections, $\sigma$, integrated and differential in $\pt^{\PgU}$, and the respective dimuon branching fraction, $\mathcal{B}$, measured for four polarization scenarios, in the helicity frame (HX) and Colins--Soper (CS) frame, each for $\lambda_{\theta}=1$ and $\lambda_{\theta}=-1$,  integrated over the rapidity range $|y^\PgU|<0.4$. The statistical uncertainty (stat.), the sum of the systematic uncertainties in quadrature $(\Sigma_\text{syst.})$, and the total uncertainty ($\Delta \sigma$; including stat., $\sum_\text{syst.}$, and the uncertainty in the integrated luminosity) are in percent. For the four polarization scenarios the fractional change to the central value of the cross section relative to the unpolarized value is given in percent. The numbers in parentheses are negative variations.}
\centering
$|y^\PgU|<0.4$\\
\begin{tabular}{c|c c c c c |  c c| c c} \hline
&\multicolumn{5}{c|}{} & \multicolumn{2}{c|}{$\rm HX$} & \multicolumn{2}{c}{$\rm CS$} \\ 
&$\pt (\rm \GeVc)$  & $\sigma  \cdot {\cal B}~(\rm nb)$ & 	$\frac{\rm stat.} {\sigma}$ &$\frac{\sum_{\rm{syst.}}}{\sigma}$ & $\frac{\Delta\sigma}{\sigma}$
& $\lambda_{\theta} = 1$ & $\lambda_{\theta} = -1$ & $\lambda_{\theta} = 1$ & $\lambda_{\theta} = -1$ \\ \hline
\multirow{8}{*}{$\PgUa$}
&0--2	&$0.216$&		${ 3.4}$ &$  14\,(  13)$ & $  15\,(  14)$ & $  +10$ & $  -16$ & $  +21$ & $  -27$\\ 
&2--4	&$0.387$&		${ 3.2}$ &$  11\,(  10)$ & $  12\,(  12)$ & $  +15$ & $  -22$ & $  +20$ & $  -26$\\ 
&4--6	&$0.355$&		${ 3.0}$ &$  10\,(  10)$ & $  11\,(  11)$ & $  +21$ & $  -27$ & $  +16$ & $  -22$\\ 
&6--8	&$0.224$&		${ 3.2}$ &$   9\,(   9)$ & $  10\,(  10)$ & $  +17$ & $  -30$ & $   +3$ & $  -17$\\ 
&8--11	&$0.190$&		${ 2.9}$ &$   7\,(   7)$ & $   8\,(   8)$ & $  +22$ & $  -27$ & $   -0$ & $   +1$\\ 
&11--15	&$0.0914$&		${ 3.3}$ &$   5\,(   5)$ & $   7\,(   7)$ & $  +20$ & $  -25$ & $   -4$ & $  +10$\\ 
&15--50	&$0.0509$&		${ 3.7}$ &$   4\,(   4)$ & $   7\,(   7)$ & $  +17$ & $  -22$ & $   -5$ & $  +13$\\ 
\hline
\multirow{6}{*}{$\PgUb$}
&0--3	& $0.084$ &		${ 6.6}$ &$  16\,(  15)$ & $  18\,(  17)$ & $  +22$ & $  -13$ & $  +20$ & $  -25$\\ 
&3--7	& $0.143$ &		${ 4.9}$ &$  13\,(  16)$ & $  15\,(  17)$ & $  +27$ & $  -16$ & $  +15$ & $  -19$\\
&7--11	& $0.0813$ &		${ 5.3}$ &$   9\,(   8)$ & $  11\,(  11)$ &   $+21 $&   $-26 $&    $+2 $&    $-3$\\
&11--15	& $0.0323$ &		${ 6.1}$ &$   6\,(   5)$ & $   9\,(   9)$ &   $+20 $&   $-25 $&    $-4 $&    $+9$\\
&15--42	& $0.0243$ &		${ 5.7}$ &$   5\,(   4)$ & $   8\,(   8)$ &   $+16 $&   $-21 $&    $-5 $&   $+12$\\
\hline
\multirow{4}{*}{$\PgUc$}
&0--7	&$0.100$&		${ 6.5}$ &$  15\,(  14)$ & $  17\,(  16)$ & $  +12$ & $  -16$ & $  +16$ & $  -22$\\ 
&7--12	&$0.0487$&		${ 7.5}$ &$  11\,(  11)$ & $  14\,(  14)$ & $  +20$ & $  -25$ & $   +2$ & $   -3$\\ 
&12--38	&$0.0291$&		${ 5.9}$ &$   5\,(   5)$ & $   9\,(   9)$ & $  +17$ & $  -23$ & $   -5$ & $  +11$\\ 
\hline
\end{tabular}

\label{table:cross-section-y-ranges-1}
\end{table*}

\begin{table*}[!ht]
\topcaption{The product of the $\PgU$(nS) acceptance-corrected production cross sections, $\sigma$, integrated and differential in $\pt^{\PgU}$, and the respective dimuon branching fraction, $\mathcal{B}$, measured for four polarization scenarios, in the helicity frame (HX) and Colins--Soper (CS) frame, each for $\lambda_{\theta}=1$ and $\lambda_{\theta}=-1$,  integrated over the rapidity range $0.4<|y^\PgU|<0.8$. The statistical uncertainty (stat.), the sum of the systematic uncertainties in quadrature $(\Sigma_\text{syst.})$, and the total uncertainty ($\Delta \sigma$; including stat., $\sum_\text{syst.}$, and the uncertainty in the integrated luminosity) are in percent. For the four polarization scenarios the fractional change to the central value of the cross section relative to the unpolarized value is given in percent. The numbers in parentheses are negative variations.}
\centering
$0.4<|y^\PgU|<0.8$\\
\begin{tabular}{c|c c c c c |  c c| c c} \hline
&\multicolumn{5}{c|}{} & \multicolumn{2}{c|}{$\rm HX$} & \multicolumn{2}{c}{$\rm CS$} \\
&$\pt (\rm \GeVc)$  & $\sigma \cdot {\cal B}~(\rm nb)$ &       $\frac{\rm stat.} {\sigma}$ &$\frac{\sum_{\rm{syst.}}}{\sigma}$ & $\frac{\Delta\sigma}{\sigma}$
& $\lambda_{\theta} = 1$ & $\lambda_{\theta} = -1$ & $\lambda_{\theta} = 1$ & $\lambda_{\theta} = -1$ \\ \hline
\multirow{8}{*}{$\PgUa$}
&0--2	&$0.220$&		${ 4.5}$ &$  13\,(  12)$ & $  14\,(  13)$ & $  +20$ & $  -25$ & $  +21$ & $  -26$\\ 
&2--4	&$0.409$&		${ 3.0}$ &$  10\,(   9)$ & $  11\,(  10)$ & $  +21$ & $  -25$ & $  +22$ & $  -25$\\ 
&4--6	&$0.367$&		${ 2.9}$ &$   8\,(   8)$ & $   9\,(   9)$ & $  +20$ & $  -25$ & $  +16$ & $  -21$\\ 
&6--8	&$0.231$&		${ 3.3}$ &$   7\,(   6)$ & $   9\,(   8)$ & $  +13$ & $  -27$ & $   +2$ & $  -15$\\ 
&8--11	&$0.180$&		${ 2.9}$ &$   6\,(   5)$ & $   8\,(   7)$ & $  +19$ & $  -23$ & $   -0$ & $   +2$\\ 
&11--15	&$0.0915$&		${ 3.3}$ &$   6\,(   6)$ & $   8\,(   8)$ & $  +18$ & $  -23$ & $   -4$ & $  +10$\\ 
&15--50	&$0.0492$&		${ 3.8}$ &$   4\,(   4)$ & $   7\,(   7)$ & $  +15$ & $  -20$ & $   -5$ & $  +13$\\ 
\hline
\multirow{6}{*}{$\PgUb$}
&0--3	& $0.088$ &		${ 7.3}$ &$  14\,(  13)$ & $  16\,(  15)$ & $  +18$ & $  -23$ & $  +19$ & $  -24$\\
&3--7	& $0.172$ &		${ 4.5}$ &$  10\,(   9)$ & $  11\,(  11)$ & $  +17$ & $  -23$ & $  +15$ & $  -20$\\
&7--11	& $0.0811$ &		${ 5.2}$ & $   6\,(   6)$ & $   9\,(   9)$&   $+17 $&   $-23 $&    $+1 $&    $-2$\\
&11--15	& $0.0293$ &		${ 6.6}$ & $   5\,(   5)$ & $   9\,(   9)$&   $+17 $&   $-22 $&    $-4 $&    $+9$\\
&15--42	& $0.0227$ &		${ 5.9}$ & $   4\,(   4)$ & $   8\,(   8)$&   $+15 $&   $-20 $&    $-5 $&   $+12$\\
\hline
\multirow{4}{*}{$\PgUc$}
&0--7	&$0.115$&		${ 6.7}$ &$  14\,(  13)$ & $  16\,(  15)$ & $  +16$ & $  -22$ & $  +16$ & $  -21$\\ 
&7--12	&$0.0558$&		${ 6.8}$ &$   7\,(   6)$ & $  11\,(  10)$ & $  +16$ & $  -22$ & $   +1$ & $   -2$\\ 
&12--38	&$0.0257$&		${ 6.5}$ &$   5\,(   5)$ & $   9\,(   9)$ & $  +15$ & $  -21$ & $   -4$ & $  +10$\\ 
\hline
\end{tabular}

\label{table:cross-section-y-ranges-2}
\end{table*}

\begin{table*}[!ht]
\topcaption{The product of the $\PgU$(nS) acceptance-corrected production cross sections, $\sigma$, integrated and differential in $\pt^{\PgU}$, and the respective dimuon branching fraction, $\mathcal{B}$, measured for four polarization scenarios, in the helicity frame (HX) and Colins--Soper (CS) frame, each for $\lambda_{\theta}=1$ and $\lambda_{\theta}=-1$,  integrated over the rapidity range $0.8<|y^\PgU|<1.2$ The statistical uncertainty (stat.), the sum of the systematic uncertainties in quadrature $(\Sigma_\text{syst.})$, and the total uncertainty ($\Delta \sigma$; including stat., $\sum_\text{syst.}$, and the uncertainty in the integrated luminosity) are in percent. For the four polarization scenarios the fractional change to the central value of the cross section relative to the unpolarized value is given in percent. The numbers in parentheses are negative variations.}
\centering
$0.8<|y^\PgU|<1.2$\\
\begin{tabular}{c|c c c c c |  c c| c c} \hline
&\multicolumn{5}{c|}{} & \multicolumn{2}{c|}{$\rm HX$} & \multicolumn{2}{c}{$\rm CS$} \\
&$\pt (\rm \GeVc)$  & $\sigma \cdot {\cal B}~(\rm nb)$ &       $\frac{\rm stat.} {\sigma}$ &$\frac{\sum_{\rm{syst.}}}{\sigma}$ & $\frac{\Delta\sigma}{\sigma}$
& $\lambda_{\theta} = 1$ & $\lambda_{\theta} = -1$ & $\lambda_{\theta} = 1$ & $\lambda_{\theta} = -1$ \\ \hline
\multirow{8}{*}{$\PgUa$}
&0--2	&$0.198$&		${ 3.7}$ &$  11\,(   9)$ & $  12\,(  10)$ & $  +20$ & $  -25$ & $  +20$ & $  -24$\\ 
&2--4	&$0.426$&		${ 2.7}$ &$   9\,(   9)$ & $  10\,(  10)$ & $  +20$ & $  -26$ & $  +19$ & $  -25$\\ 
&4--6	&$0.331$&		${ 2.9}$ &$   7\,(   6)$ & $   8\,(   8)$ & $  +17$ & $  -25$ & $  +13$ & $  -21$\\ 
&6--8	&$0.217$&		${ 3.4}$ &$   7\,(   7)$ & $   9\,(   8)$ & $  +22$ & $  -18$ & $  +12$ & $   -7$\\ 
&8--11	&$0.174$&		${ 2.9}$ &$   5\,(   5)$ & $   7\,(   7)$ & $  +15$ & $  -21$ & $   -1$ & $   +2$\\ 
&11--15	&$0.0879$&		${ 3.4}$ &$   6\,(   5)$ & $   8\,(   8)$ & $  +15$ & $  -21$ & $   -4$ & $   +9$\\ 
&15--50	&$0.0482$&		${ 3.8}$ &$   5\,(   5)$ & $   8\,(   8)$ & $  +13$ & $  -19$ & $   -5$ & $  +12$\\ 
\hline
\multirow{6}{*}{$\PgUb$}
&0--3	& $0.098$ &		${ 6.2}$ &$  14\,(  13)$ & $  15\,(  15)$ & $  +18$ & $  -23$ & $  +17$ & $  -23$\\
&3--7	& $0.155$ &		${ 5.3}$ &$  11\,(   8)$ & $  13\,(  10)$ & $  +19$ & $  -22$ & $  +16$ & $  -19$\\
&7--11	& $0.0803$ &		${ 5.3}$ &$   8\,(   7)$ & $  10\,(  10)$ & $  +14$ & $  -20$ & $   +1$ & $   -2$\\
&11--15	& $0.0309$ &		${ 6.2}$ &$   6\,(   6)$ & $  10\,(   9)$ & $  +14$ & $  -20$ & $   -4$ & $   +9$\\
&15--42	& $0.0214$ &		${ 6.2}$ &$   6\,(   6)$ & $  10\,(   9)$ & $  +13$ & $  -18$ & $   -5$ & $  +11$\\
\hline
\multirow{4}{*}{$\PgUc$}
&0--7	&$0.109$&		${ 7.7}$  & $  15\,(  14)$ & $  17\,(  17)$ &  $+16 $&   $-22 $&   $+14 $&   $-20$\\ 
&7--12	&$0.0579$&		${ 6.9}$  & $   9\,(   9)$ & $  12\,(  12)$ &  $+13 $&   $-19 $&    $+2 $&    $-3$\\ 
&12--38	&$0.0250$&		${ 6.8}$  & $   6\,(   6)$ & $  10\,(  10)$ &  $+13 $&   $-19 $&    $-4 $&   $+10$\\ 
\hline
\end{tabular}

\label{table:cross-section-y-ranges-3}
\end{table*}
\clearpage

\begin{table*}[!ht]
\topcaption{The product of the $\PgU$(nS) acceptance-corrected production cross sections, $\sigma$, integrated and differential in $\pt^{\PgU}$, and the respective dimuon branching fraction, $\mathcal{B}$, measured for four polarization scenarios, in the helicity frame (HX) and Colins--Soper (CS) frame, each for $\lambda_{\theta}=1$ and $\lambda_{\theta}=-1$,  integrated over the rapidity range $1.2<|y^\PgU|<1.6$ The statistical uncertainty (stat.), the sum of the systematic uncertainties in quadrature $(\Sigma_\text{syst.})$, and the total uncertainty ($\Delta \sigma$; including stat., $\sum_\text{syst.}$, and the uncertainty in the integrated luminosity) are in percent. For the four polarization scenarios the fractional change to the central value of the cross section relative to the unpolarized value is given in percent. The numbers in parentheses are negative variations.}
\centering
$1.2<|y^\PgU|<1.6$\\
\begin{tabular}{c|c c c c c |  c c| c c} \hline
&\multicolumn{5}{c|}{} & \multicolumn{2}{c|}{$\rm HX$} & \multicolumn{2}{c}{$\rm CS$} \\
&$\pt (\rm \GeVc)$  & $\sigma \cdot {\cal B}~(\rm nb)$ &       $\frac{\rm stat.} {\sigma}$ &$\frac{\sum_{\rm{syst.}}}{\sigma}$ & $\frac{\Delta\sigma}{\sigma}$
& $\lambda_{\theta} = 1$ & $\lambda_{\theta} = -1$ & $\lambda_{\theta} = 1$ & $\lambda_{\theta} = -1$ \\ \hline
\multirow{8}{*}{$\PgUa$}
&0--2	&$0.203$&		${ 3.5}$ &$  10\,(   8)$ & $  11\,(   9)$ & $  +19$ & $  -23$ & $  +21$ & $  -22$\\ 
&2--4	&$0.416$&		${ 2.4}$ &$   7\,(   6)$ & $   8\,(   7)$ & $  +20$ & $  -26$ & $  +18$ & $  -25$\\ 
&4--6	&$0.315$&		${ 2.8}$ &$   6\,(   5)$ & $   8\,(   7)$ & $  +20$ & $  -24$ & $  +17$ & $  -21$\\ 
&6--8	&$0.215$&		${ 3.0}$ &$   5\,(   6)$ & $   7\,(   7)$ & $  +17$ & $  -22$ & $  +11$ & $  -15$\\ 
&8--11	&$0.165$&		${ 3.0}$ &$   5\,(   5)$ & $   7\,(   7)$ & $  +13$ & $  -19$ & $   +2$ & $   -4$\\ 
&11--15	&$0.0868$&		${ 3.3}$ &$   5\,(   5)$ & $   7\,(   7)$ & $  +13$ & $  -19$ & $   -3$ & $   +6$\\ 
&15--50	&$0.0485$&		${ 3.8}$ &$   4\,(   4)$ & $   7\,(   7)$ & $  +12$ & $  -18$ & $   -5$ & $  +11$\\ 
\hline
\multirow{6}{*}{$\PgUb$}
&0--3	& $0.099$ &		${ 6.2}$ &$  11\,(  11)$ & $  13\,(  13)$ & $  +18$ & $  -24$ & $  +18$ & $  -23$\\ 
&3--7	& $0.161$ &		${ 4.5}$ &$   7\,(   6)$ & $   9\,(   9)$ & $  +18$ & $  -23$ & $  +17$ & $  -22$\\ 
&7--11	& $0.0758$ &		${ 5.5}$ &$   7\,(   6)$ & $  10\,(   9)$ &   $+14 $&   $-19 $&    $+6 $&    $-9$\\
&11--15	& $0.0305$ &		${ 6.5}$ &$   5\,(   5)$ & $   9\,(   9)$ &   $+12 $&   $-18 $&    $-2 $&    $+5$\\
&15--42	& $0.0219$ &		${ 6.4}$ &$   5\,(   4)$ & $   9\,(   9)$ &   $+11 $&   $-17 $&    $-4 $&   $+10$\\
\hline
\multirow{4}{*}{$\PgUc$}
&0--7 &$0.113$&               ${ 6.8}$ &$   7\,(   7)$ & $  11\,(  10)$ & $  +17$ & $  -23$ & $  +16$ & $  -22$\\
&7--12 &$0.0480$&              ${ 7.7}$ &$  10\,(  10)$ & $  13\,(  13)$ & $  +13$ & $  -18$ & $   +5$ & $   -8$\\
&12--38 &$0.0276$&              ${ 6.7}$ &$   8\,(   7)$ & $  11\,(  11)$ & $  +11$ & $  -17$ & $   -3$ & $   +7$\\
\hline
\end{tabular}

\label{table:cross-section-y-ranges-4}
\end{table*}
\clearpage

\begin{table*}[!ht]
\topcaption{The product of the $\PgU$(nS) acceptance-corrected production cross sections, $\sigma$, integrated and differential in $\pt^{\PgU}$, and the respective dimuon branching fraction, $\mathcal{B}$, measured for four polarization scenarios, in the helicity frame (HX) and Colins--Soper (CS) frame, each for $\lambda_{\theta}=1$ and $\lambda_{\theta}=-1$,  integrated over the rapidity range $1.6<|y^\PgU|<2.0$. The statistical uncertainty (stat.), the sum of the systematic uncertainties in quadrature $(\Sigma_\text{syst.})$, and the total uncertainty ($\Delta \sigma$; including stat., $\sum_\text{syst.}$, and the uncertainty in the integrated luminosity) are in percent. For the four polarization scenarios the fractional change to the central value of the cross section relative to the unpolarized value is given in percent. The numbers in parentheses are negative variations.}
\centering
$1.6<|y^\PgU|<2.0$\\
\begin{tabular}{c|c c c c c |  c c| c c} \hline
&\multicolumn{5}{c|}{} & \multicolumn{2}{c|}{$\rm HX$} & \multicolumn{2}{c}{$\rm CS$} \\
&$\pt (\rm \GeVc)$  & $\sigma \cdot {\cal B}~(\rm nb)$ &       $\frac{\rm stat.} {\sigma}$ &$\frac{\sum_{\rm{syst.}}}{\sigma}$ & $\frac{\Delta\sigma}{\sigma}$
& $\lambda_{\theta} = 1$ & $\lambda_{\theta} = -1$ & $\lambda_{\theta} = 1$ & $\lambda_{\theta} = -1$ \\ \hline
\multirow{8}{*}{$\PgUa$}
&0--2	&$0.211$&		${ 3.1}$ &$   7\,(   6)$ & $   9\,(   8)$ & $  +22$ & $  -26$ & $  +23$ & $  -27$\\ 
&2--4	&$0.401$&		${ 2.5}$ &$   6\,(   4)$ & $   7\,(   6)$ & $  +22$ & $  -26$ & $  +24$ & $  -27$\\ 
&4--6	&$0.330$&		${ 2.6}$ &$   5\,(   4)$ & $   7\,(   6)$ & $  +20$ & $  -25$ & $  +23$ & $  -27$\\ 
&6--8	&$0.191$&		${ 3.2}$ &$   4\,(   4)$ & $   7\,(   6)$ & $  +18$ & $  -22$ & $  +20$ & $  -24$\\ 
&8--11	&$0.151$&		${ 3.4}$ &$   5\,(   4)$ & $   7\,(   7)$ & $  +15$ & $  -20$ & $  +15$ & $  -19$\\ 
&11--15	&$0.0746$&		${ 4.1}$ &$   5\,(   5)$ & $   8\,(   7)$ & $  +10$ & $  -16$ & $   +5$ & $   -9$\\ 
&15--50	&$0.0465$&		${ 4.4}$ &$   6\,(   6)$ & $   9\,(   8)$ & $   +9$ & $  -14$ & $   -1$ & $   +3$\\ 
\hline
\multirow{6}{*}{$\PgUb$}
&0--3	& $0.091$ &		${ 6.8}$ &$   8\,(   8)$ & $  11\,(  11)$ & $  +21$ & $  -25$ & $  +22$ & $  -27$\\
&3--7	& $0.175$ &		${ 4.4}$ &$   7\,(   6)$ & $   9\,(   9)$ & $  +19$ & $  -24$ & $  +23$ & $  -27$\\
&7--11	& $0.0608$ &		${ 7.0}$ &$  7\,(   6)$ & $  10\,(  10)$ &   $+14 $&   $-20 $&   $+17 $&   $-22$\\
&11--15	& $0.0263$ &		${ 8.3}$ &$  5\,(   5)$ & $  11\,(  10)$ &   $+10 $&   $-15 $&    $+7 $&   $-11$\\
&15--42	& $0.0205$ &		${ 8.1}$ &$  6\,(   5)$ & $  11\,(  11)$ &    $+8 $&   $-13 $&    $-0 $&    $+1$\\
\hline
\multirow{4}{*}{$\PgUc$}
&0--7 &$0.117$&               ${ 7.2}$ &$   9\,(   9)$ & $  13\,(  12)$ & $  +19$ & $  -23$ & $  +23$ & $  -26$\\
&7--12 &$0.0341$&              ${10.6}$ &$  18\,(  18)$ & $  22\,(  21)$ & $  +13$ & $  -19$ & $  +16$ & $  -21$\\
&12--38 &$0.0243$&              ${ 8.8}$ &$   6\,(   6)$ & $  11\,(  11)$ & $   +8$ & $  -13$ & $   +4$ & $   -6$\\
\hline
\end{tabular}

\label{table:cross-section-y-ranges-5}
\end{table*}
\clearpage

\begin{table*}[!ht]
\topcaption{The product of the $\PgU$(nS) acceptance-corrected production cross sections, $\sigma$, integrated and differential in $\pt^{\PgU}$, and the respective dimuon branching fraction, $\mathcal{B}$, measured for four polarization scenarios, in the helicity frame (HX) and Colins--Soper (CS) frame, each for $\lambda_{\theta}=1$ and $\lambda_{\theta}=-1$,  integrated over the rapidity range $2.0<|y^\PgU|<2.4$ . The statistical uncertainty (stat.), the sum of the systematic uncertainties in quadrature $(\Sigma_\text{syst.})$, and the total uncertainty ($\Delta \sigma$; including stat., $\sum_\text{syst.}$, and the uncertainty in the integrated luminosity) are in percent. For the four polarization scenarios the fractional change to the central value of the cross section relative to the unpolarized value is given in percent. The numbers in parentheses are negative variations.}
\centering
$2.0<|y^\PgU|<2.4$\\
\begin{tabular}{c|c c c c c |  c c| c c} \hline
&\multicolumn{5}{c|}{} & \multicolumn{2}{c|}{$\rm HX$} & \multicolumn{2}{c}{$\rm CS$} \\
&$\pt (\rm \GeVc)$  & $\sigma \cdot {\cal B}~(\rm nb)$ &       $\frac{\rm stat.} {\sigma}$ &$\frac{\sum_{\rm{syst.}}}{\sigma}$ & $\frac{\Delta\sigma}{\sigma}$
& $\lambda_{\theta} = 1$ & $\lambda_{\theta} = -1$ & $\lambda_{\theta} = 1$ & $\lambda_{\theta} = -1$ \\ \hline
\multirow{8}{*}{$\PgUa$}
&0--2	&$0.135$&		${ 5.3}$ &$   8\,(   7)$ & $  11\,(   9)$ & $  +30$ & $  -31$ & $  +32$ & $  -31$\\ 
&2--4	&$0.338$&		${ 2.8}$ &$   7\,(   6)$ & $   9\,(   8)$ & $  +23$ & $  -28$ & $  +30$ & $  -32$\\ 
&4--6	&$0.289$&		${ 4.1}$ &$   8\,(   6)$ & $  10\,(   9)$ & $  +27$ & $  -21$ & $  +38$ & $  -28$\\ 
&6--8	&$0.150$&		${ 5.8}$ &$   7\,(   6)$ & $  10\,(   9)$ & $  +23$ & $  -21$ & $  +34$ & $  -29$\\ 
&8--11	&$0.126$&		${ 5.8}$ &$   7\,(   5)$ & $  10\,(   9)$ & $  +18$ & $  -20$ & $  +31$ & $  -29$\\ 
&11--15	&$0.0667$&		${ 7.2}$ &$   8\,(   7)$ & $  12\,(  11)$ & $  +14$ & $  -18$ & $  +28$ & $  -29$\\ 
&15--50	&$0.0356$&		${ 9.1}$ &$   9\,(   9)$ & $  14\,(  13)$ & $   +6$ & $  -13$ & $  +18$ & $  -24$\\ 
\hline
\multirow{6}{*}{$\PgUb$}
&0--3	& $0.0475$ &		${13.2}$ &$   9\,(   8)$ & $  16\,(  16)$ & $  +27$ & $  -30$ & $  +30$ & $  -32$\\
&3--7	& $0.148$ &		${ 7.5}$ &$   7\,(   5)$ & $  11\,(  10)$ & $  +20$ & $  -25$ & $  +31$ & $  -32$\\
&7--11	& $0.0636$ &		${10.9}$ &$   8\,(   7)$ & $  14\,(  13)$&   $+16 $&   $-22 $&   $+29 $&   $-31$\\
&11--15	& $0.0224$ &		${14.7}$ &$   9\,(   8)$ & $  18\,(  17)$&   $+12 $&   $-18 $&   $+26 $&   $-29$\\
&15--42	& $0.0137$ &		${20.4}$ &$  11\,(  11)$ & $  24\,(  23)$&    $+9 $&   $-14 $&   $+20 $&   $-24$\\
\hline
\multirow{4}{*}{$\PgUc$}
&0--7 &$0.105$&               ${11.2}$ &$   8\,(   6)$ & $  14\,(  13)$ & $  +22$ & $  -28$ & $  +30$ & $  -32$\\
&7--12 &$0.0366$&              ${19.3}$ &$  11\,(   9)$ & $  22\,(  22)$ & $  +14$ & $  -20$ & $  +29$ & $  -31$\\
&12--38 &$0.0231$&              ${15.8}$ &$   9\,(   8)$ & $  19\,(  18)$ & $   +9$ & $  -14$ & $  +24$ & $  -27$\\
\hline
\end{tabular}

\label{table:cross-section-y-ranges-6}
\end{table*}

\begin{table*}[!ht]
\topcaption{The product of the $\PgU$(nS) acceptance-corrected production cross sections, $\sigma$, integrated and differential in $\pt^{\PgU}$, and the respective dimuon branching fraction, $\mathcal{B}$, measured for four polarization scenarios, in the helicity frame (HX) and Colins--Soper (CS) frame, each for $\lambda_{\theta}=1$ and $\lambda_{\theta}=-1$,  integrated over the \pt range  $\pt^\PgU<50\GeVc$. The statistical uncertainty (stat.), the sum of the systematic uncertainties in quadrature $(\Sigma_\text{syst.})$, and the total uncertainty ($\Delta \sigma$; including stat., $\sum_\text{syst.}$, and the uncertainty in the integrated luminosity) are in percent. For the four polarization scenarios the fractional change to the central value of the cross section relative to the unpolarized value is given in percent. The numbers in parentheses are negative variations.}
\centering
$\pt^\PgU<50\GeVc$\\
\begin{tabular}{c|c c c c c |  c c| c c} \hline
&\multicolumn{5}{c|}{} & \multicolumn{2}{c|}{$\rm HX$} & \multicolumn{2}{c}{$\rm CS$} \\ 
&$|y|$ & $\sigma  \cdot {\cal B}~(\rm nb)$ & 	$\frac{\rm stat.} {\sigma}$ &$\frac{\sum_{\rm{syst.}}}{\sigma}$ & $\frac{\Delta\sigma}{\sigma}$
& $\lambda_{\theta} = 1$ & $\lambda_{\theta} = -1$ & $\lambda_{\theta} = 1$ & $\lambda_{\theta} = -1$ \\ \hline
\multirow{7}{*}{$\PgUa$}
&0--0.2	&$0.770$&		${ 1.7}$ &$  10\,(   9)$ & $  11\,(  10)$ & $  +27$ & $  -16$ & $  +22$ & $   -9$\\ 
&0.2--0.4	&$0.777$&		${ 1.7}$ &$   9\,(   9)$ & $  10\,(  10)$ & $  +19$ & $  -25$ & $  +12$ & $  -16$\\ 
&0.4--0.6	&$0.770$&		${ 1.7}$ &$   9\,(   9)$ & $  10\,(  10)$ & $  +20$ & $  -18$ & $  +13$ & $  -15$\\ 
&0.6--0.8	&$0.795$&		${ 1.7}$ &$   8\,(   8)$ & $   9\,(   9)$ & $  +20$ & $  -24$ & $  +13$ & $  -15$\\ 
&0.8--1.0	&$0.761$&		${ 1.7}$ &$   9\,(   8)$ & $  10\,(   9)$ & $  +18$ & $  -24$ & $  +12$ & $  -15$\\ 
&1.0--1.2	&$0.738$&		${ 1.8}$ &$   6\,(   5)$ & $   7\,(   7)$ & $  +18$ & $  -24$ & $  +11$ & $  -16$\\ 
&1.2--1.4	&$0.714$&		${ 1.7}$ &$   7\,(   7)$ & $   8\,(   8)$ & $  +18$ & $  -23$ & $  +12$ & $  -16$\\ 
&1.4--1.6	&$0.738$&		${ 1.7}$ &$   5\,(   4)$ & $   7\,(   6)$ & $  +18$ & $  -23$ & $  +14$ & $  -18$\\ 
&1.6--1.8	&$0.721$&		${ 1.7}$ &$   6\,(   5)$ & $   7\,(   7)$ & $  +19$ & $  -23$ & $  +18$ & $  -21$\\ 
&1.8--2	&$0.708$&		${ 1.9}$ &$   6\,(   5)$ & $   7\,(   6)$ & $  +20$ & $  -24$ & $  +23$ & $  -26$\\ 
&2--2.2	&$0.636$&		${ 2.3}$ &$   6\,(   5)$ & $   8\,(   7)$ & $  +20$ & $  -24$ & $  +28$ & $  -30$\\ 
&2.2--2.4	&$0.528$&		${ 4.2}$ &$   7\,(   5)$ & $   9\,(   8)$ & $  +22$ & $  -26$ & $  +32$ & $  -32$\\
&0--2.4	&$8.66$&		${ 0.6}$ &$   7\,(   6)$ & $   8\,(   7)$ & $  +19$ & $  -24$ & $  +16$ & $  -19$\\  
\hline
\multirow{5}{*}{$\PgUb$}
&0--0.4	&$0.376$&		${ 2.5}$ &$  10\,(   9)$ & $  11\,(  11)$ & $  +16$ & $  -21$ & $  +10$ & $  -13$\\
&0.4--0.8	&$0.400$&		${ 2.5}$ &$   9\,(   8)$ & $  10\,(   9)$ & $  +17$ & $  -23$ & $  +10$ & $  -13$\\
&0.8--1.2	&$0.380$&		${ 2.8}$ &$   9\,(   8)$ & $  10\,(   9)$ & $  +16$ & $  -22$ & $  +10$ & $  -13$\\
&1.2--1.6	&$0.382$&		${ 2.7}$ &$   7\,(   6)$ & $   8\,(   8)$ & $  +16$ & $  -22$ & $  +12$ & $  -16$\\
&1.6--2	&$0.380$&		${ 2.9}$ &$   7\,(   6)$ & $   8\,(   8)$ & $  +17$ & $  -22$ & $  +19$ & $  -23$\\
&2--2.4	&$0.295$&		${ 5.1}$ &$   7\,(   5)$ & $   9\,(   8)$ & $  +19$ & $  -24$ & $  +29$ & $  -31$\\
&0--2.4	&$2.21$&		${ 1.3}$ &$   8\,(   7)$ & $   9\,(   8)$ & $  +26$ & $  -16$ & $  +23$ & $  -12$\\\hline
\multirow{3}{*}{$\PgUc$}
&0--0.4	&$0.185$&		${ 3.9}$ &$  11\,(  10)$ & $  12\,(  12)$ & $  +15$ & $  -20$ & $   +9$ & $  -11$\\
&0.4--0.8	&$0.202$&		${ 3.9}$ &$  10\,(   9)$ & $  11\,(  11)$ & $  +16$ & $  -22$ & $   +9$ & $  -11$\\
&0.8--1.4	&$0.282$&		${ 3.8}$ &$   9\,(   9)$ & $  11\,(  10)$ & $  +15$ & $  -21$ & $   +8$ & $  -11$\\
&1.4--2	&$0.290$&		${ 4.0}$ &$   7\,(   7)$ & $   9\,(   9)$ & $  +16$ & $  -21$ & $  +16$ & $  -20$\\
&2--2.4	&$0.164$&		${ 8.3}$ &$   7\,(   6)$ & $  12\,(  11)$ & $  +18$ & $  -23$ & $  +29$ & $  -31$\\
&0--2.4	&$1.12$&		${ 2.1}$ &$   8\,(   7)$ & $  10\,(   8)$ & $  +16$ & $  -21$ & $  +14$ & $  -17$\\\hline
\end{tabular}

\label{table:cross-section-rapdiff}
\end{table*}

\begin{table*}
  \centering
  \topcaption{The ratio of $\PgUc/\PgUa$ acceptance-corrected cross sections, and its \pt dependence, integrated over the rapidity range $|y^\PgU|<2.4$, for the unpolarized scenario, with statistical and systematic uncertainties. The numbers in parentheses are negative variations.}
  \begin{tabular}{cc} \hline 
\pt (\GeVc) & \PgUc/\PgUa  \\ \hline
0--2 &$0.108\pm0.007\,(0.007)\pm0.015\,(0.015)$\\ 
2--5 &$0.101\pm0.006\,(0.006)\pm0.015\,(0.012)$\\ 
5--8 &$0.122\pm0.007\,(0.007)\pm0.016\,(0.014)$\\ 
8--10 &$0.154\pm0.010\,(0.009)\pm0.022\,(0.017)$\\ 
10--13 &$0.197\pm0.012\,(0.011)\pm0.020\,(0.015)$\\ 
13--16 &$0.258\pm0.019\,(0.018)\pm0.021\,(0.019)$\\ 
16--18 &$0.305\pm0.034\,(0.032)\pm0.027\,(0.026)$\\ 
18--22 &$0.283\pm0.030\,(0.028)\pm0.022\,(0.027)$\\ 
22--38 &$0.323\pm0.039\,(0.036)\pm0.050\,(0.089)$\\ 
0--38 &$0.129\pm0.003\,(0.003)\pm0.018\,(0.015)$\\\hline 
 \end{tabular}

  \label{tab:xsec_ratio_1S}
\end{table*}

\begin{table*}
  \centering
  \topcaption{The ratio of $\PgUb/\PgUa$ acceptance-corrected cross sections, and its \pt dependence, integrated over the rapidity range $|y^\PgU|<2.4$, for the unpolarized scenario, with statistical and systematic uncertainties. The numbers in parentheses are negative variations.}
  \begin{tabular}{cc} \hline 
\pt (\GeVc) & \PgUb/\PgUa \\ \hline
0--2 &$0.221\pm0.010\,(0.009)\pm0.030\,(0.024)$\\ 
2--5 &$0.230\pm0.008\,(0.008)\pm0.028\,(0.030)$\\ 
5--8 &$0.266\pm0.011\,(0.010)\pm0.033\,(0.025)$\\ 
8--10 &$0.270\pm0.012\,(0.012)\pm0.034\,(0.029)$\\ 
10--13 &$0.330\pm0.016\,(0.015)\pm0.029\,(0.028)$\\ 
13--16 &$0.372\pm0.023\,(0.022)\pm0.025\,(0.028)$\\ 
16--18 &$0.502\pm0.046\,(0.043)\pm0.040\,(0.041)$\\ 
18--22 &$0.449\pm0.039\,(0.036)\pm0.032\,(0.048)$\\ 
22--38 &$0.466\pm0.047\,(0.044)\pm0.074\,(0.121)$\\ 
0--38 &$0.264\pm0.005\,(0.005)\pm0.032\,(0.028)$\\ 
\hline 
 \end{tabular}

  \label{tab:xsec_ratio_2S}
\end{table*}

\begin{table*}
  \centering
  \topcaption{The ratio of $\PgUc/\PgUb$ acceptance-corrected cross sections, and its \pt dependence, integrated over the rapidity range $|y^\PgU|<2.4$, for the unpolarized scenario, with statistical and systematic uncertainties. The numbers in parentheses are negative variations.}
  \begin{tabular}{cc} \hline 
\pt (\GeVc) & \PgUc/\PgUb \\ \hline
0--2 &$0.490\pm0.042\,(0.040)\pm0.067\,(0.070)$\\ 
2--5 &$0.440\pm0.032\,(0.030)\pm0.082\,(0.057)$\\ 
5--8 &$0.459\pm0.035\,(0.033)\pm0.054\,(0.053)$\\ 
8--10 &$0.569\pm0.047\,(0.044)\pm0.078\,(0.063)$\\ 
10--13 &$0.597\pm0.046\,(0.043)\pm0.062\,(0.044)$\\ 
13--16 &$0.693\pm0.063\,(0.059)\pm0.060\,(0.047)$\\ 
16--18 &$0.607\pm0.082\,(0.074)\pm0.058\,(0.051)$\\ 
18--22 &$0.630\pm0.079\,(0.071)\pm0.060\,(0.044)$\\ 
22--38 &$0.69\pm0.10\,(0.09)\pm0.15\,(0.14)$\\ 
0--38 &$0.490\pm0.015\,(0.015)\pm0.069\,(0.058)$\\ 
\hline 
 \end{tabular}

  \label{tab:xsec_ratio_3S}
\end{table*}

\begin{table*}[!ht]
\topcaption{The product of the \upsn acceptance-corrected production cross sections, $\sigma$, integrated and differential in $\pt^{\PgU}$, and the respective dimuon branching fraction, $\mathcal{B}$, integrated over the rapidity range $|y^{\PgU}| < 1.2$ and the $\pt^{\PgU}$ range from 10 to 50 \GeVc. The cross sections assume the \upsn are unpolarized. The statistical uncertainty (stat.), the sum in quadrature of the systematic uncertainties excluding the contribution from the polarization uncertainty ($\sum_\text{syst.}$), the systematic uncertainties from the polarization (syst.(pol)), and the total uncertainty ($\Delta\sigma$; including stat., $\sum_\text{syst.}$, syst.(pol), and the uncertainty in the integrated luminosity) are in percent. The numbers in parentheses are negative variations.}
  \centering
  $|y^\PgU|<1.2$\\
\begin{tabular}{c|c c c c c c} \hline
\multirow{7}{*}{$\Upsilon(\rm 1S)$}&\pt (\rm \GeVc)  & $\sigma \cdot {\cal B}~(\rm nb)$ & 	$\frac{\rm stat.}{\sigma}$ & $\frac{\sum_{\rm{syst.}}}{\sigma}$ & $\frac{\rm syst.(\rm pol.)}{\sigma}$ & $\frac{\Delta\sigma}{\sigma}$\\ \hline
&10--11     &$0.139$&               ${ 2.8}$ &$   8\,(   7)$ & $   5\,(   1)$ & $  10\,(   9)$\\
&11--12     &$0.0949$&              ${ 3.4}$ &$   5\,(   4)$ & $   4\,(   1)$ & $   8\,(   7)$\\
&12--13     &$0.0779$&              ${ 3.6}$ &$   6\,(   5)$ & $   3\,(   1)$ & $   8\,(   8)$\\
&13--14     &$0.0540$&              ${ 4.1}$ &$   5\,(   4)$ & $   3\,(   1)$ & $   8\,(   7)$\\
&14--15     &$0.0436$&              ${ 4.4}$ &$   5\,(   5)$ & $   3\,(   1)$ & $   9\,(   8)$\\
&15--16     &$0.0322$&              ${ 5.0}$ &$   5\,(   5)$ & $   3\,(   1)$ & $   9\,(   8)$\\
&16--18     &$0.0415$&              ${ 4.2}$ &$   5\,(   4)$ & $   2\,(   2)$ & $   8\,(   8)$\\
&18--20     &$0.0286$&              ${ 4.9}$ &$   5\,(   4)$ & $   2\,(   2)$ & $   8\,(   8)$\\
&20--22     &$0.0164$&              ${ 6.4}$ &$   4\,(   4)$ & $   2\,(   2)$ & $   9\,(   9)$\\
&22--25     &$0.0113$&              ${ 7.6}$ &$   5\,(   4)$ & $   2\,(   2)$ & $  10\,(  10)$\\
&25--30     &$0.0105$&              ${ 7.8}$ &$   4\,(   4)$ & $   1\,(   2)$ & $  10\,(  10)$\\
&30--50     &$0.00720$&             ${ 9.4}$ &$   5\,(   5)$ & $   3\,(   2)$ & $  12\,(  12)$\\
&10--50     &$0.558$&               ${ 1.3}$ &$   6\,(   5)$ & $   4\,(   2)$ & $   8\,(   7)$\\
\hline
\multirow{5}{*}{$\Upsilon(\rm 2S)$}
&10--11.5     &$0.0661$&              ${ 4.9}$ &$   8\,(   8)$ & $   9\,(   3)$ & $  13\,(  10)$\\
&11.5--13     &$0.0400$&              ${ 5.7}$ &$   6\,(   6)$ & $   8\,(   4)$ & $  12\,(  10)$\\
&13--14.5     &$0.0287$&              ${ 6.3}$ &$   6\,(   5)$ & $   8\,(   4)$ & $  12\,(  10)$\\
&14.5--16     &$0.0206$&              ${ 6.9}$ &$   6\,(   6)$ & $   7\,(   4)$ & $  12\,(  11)$\\
&16--18     &$0.0215$&              ${ 6.4}$ &$   9\,(   9)$ & $   6\,(   3)$ & $  13\,(  12)$\\
&18--19.5     &$0.0091$&              ${ 9.5}$ &$   6\,(   5)$ & $   5\,(   3)$ & $  13\,(  12)$\\
&19.5--22     &$0.0108$&              ${ 8.5}$ &$   7\,(   7)$ & $   6\,(   4)$ & $  13\,(  12)$\\
&22--26     &$0.0089$&              ${ 9.3}$ &$   9\,(   8)$ & $   7\,(   4)$ & $  15\,(  14)$\\
&26--42     &$0.00737$&             ${ 9.7}$ &$   6\,(   6)$ & $   6\,(   4)$ & $  14\,(  13)$\\
&10--42     &$0.213$&               ${ 2.4}$ &$   5\,(   5)$ & $   7\,(   3)$ & $  10\,(   8)$\\\hline
\multirow{3}{*}{$\Upsilon(\rm 3S)$}
&10--13     &$0.0591$&              ${ 5.2}$ &$   7\,(   6)$ & $   5\,(   3)$ & $  11\,(  10)$\\
&13--16     &$0.0318$&              ${ 6.3}$ &$   6\,(   5)$ & $   8\,(   4)$ & $  12\,(  10)$\\
&16--18     &$0.0129$&              ${ 9.1}$ &$  10\,(  10)$ & $   9\,(   5)$ & $  17\,(  15)$\\
&18--22     &$0.0128$&              ${ 8.4}$ &$   7\,(   7)$ & $   8\,(   4)$ & $  14\,(  12)$\\
&22--38     &$0.0104$&              ${ 8.9}$ &$   6\,(   6)$ & $   6\,(   4)$ & $  13\,(  12)$\\
&10--38     &$0.127$&               ${ 3.2}$ &$   7\,(   5)$ & $   7\,(   3)$ & $  11\,(   8)$\\
\hline
\end{tabular}

  \label{table:cross-section-cmspol-ns}
\end{table*}

}
\cleardoublepage \section{The CMS Collaboration \label{app:collab}}\begin{sloppypar}\hyphenpenalty=5000\widowpenalty=500\clubpenalty=5000\textbf{Yerevan Physics Institute,  Yerevan,  Armenia}\\*[0pt]
S.~Chatrchyan, V.~Khachatryan, A.M.~Sirunyan, A.~Tumasyan
\vskip\cmsinstskip
\textbf{Institut f\"{u}r Hochenergiephysik der OeAW,  Wien,  Austria}\\*[0pt]
W.~Adam, E.~Aguilo, T.~Bergauer, M.~Dragicevic, J.~Er\"{o}, C.~Fabjan\cmsAuthorMark{1}, M.~Friedl, R.~Fr\"{u}hwirth\cmsAuthorMark{1}, V.M.~Ghete, N.~H\"{o}rmann, J.~Hrubec, M.~Jeitler\cmsAuthorMark{1}, W.~Kiesenhofer, V.~Kn\"{u}nz, M.~Krammer\cmsAuthorMark{1}, I.~Kr\"{a}tschmer, D.~Liko, I.~Mikulec, M.~Pernicka$^{\textrm{\dag}}$, D.~Rabady\cmsAuthorMark{2}, B.~Rahbaran, C.~Rohringer, H.~Rohringer, R.~Sch\"{o}fbeck, J.~Strauss, A.~Taurok, W.~Waltenberger, C.-E.~Wulz\cmsAuthorMark{1}
% ...
\vskip\cmsinstskip
\dag:~Deceased\\
1:~~Also at Vienna University of Technology, Vienna, Austria\\
2:~~Also at CERN, European Organization for Nuclear Research, Geneva, Switzerland\\
% ...

\end{sloppypar}
\end{document}